\pdfoutput=1
\documentclass{JINST}

\graphicspath{{./img/}}
\DeclareGraphicsExtensions{.pdf,.jpeg}

\title{Charge-sensitive front-end electronics with operational amplifiers for CdZnTe detectors}

\author{P. F\"odisch$^a$\thanks{Corresponding author.}, M. Berthel$^b$, B. Lange$^a$, T. Kirschke$^a$, W. Enghardt$^{b,c,d}$ and P. Kaever$^a$\\
\llap{$^a$}Helmholtz-Zentrum Dresden - Rossendorf, Department of Research Technology,\\
Bautzner Landstr. 400, 01328 Dresden, Germany\\
\llap{$^b$}OncoRay - National Center for Radiation Research in Oncology, Faculty of Medicine and University Hospital Carl Gustav Carus, Technische Universit\"at Dresden,\\
Fetscherstr. 74, PF 41, 01307 Dresden, Germany\\
\llap{$^c$}Helmholtz-Zentrum Dresden - Rossendorf, Institute of Radiooncology\\
Bautzner Landstr. 400, 01328 Dresden, Germany\\
\llap{$^d$}German Cancer Consortium (DKTK) and German Cancer Research Center (DKFZ)\\
Im Neuenheimer Feld 280, 69120 Heidelberg, Germany\\

  E-mail: \email{p.foedisch@hzdr.de}}

\abstract{Cadmium zinc telluride (CdZnTe, CZT) radiation detectors are suitable for a variety of applications, due to their high spatial resolution and spectroscopic energy performance at room temperature. However, state-of-the-art detector systems require high-performance readout electronics. Though an application-specific integrated circuit (ASIC) is an adequate solution for the readout, requirements of high dynamic range and high throughput are not available in any commercial circuit. Consequently, the present study develops the analog front-end electronics with operational amplifiers for an $8\,\times\,8$ pixelated CZT detector. For this purpose, we modeled an electrical equivalent circuit of the CZT detector with the associated charge-sensitive amplifier (CSA). Based on a detailed network analysis, the circuit design is completed by numerical values for various features such as ballistic deficit, charge-to-voltage gain, rise time, and noise level. A verification of the performance is carried out by synthetic detector signals and a pixel detector. The experimental results with the pixel detector assembly and a ${}^{22}\mathrm{Na}$ radioactive source emphasize the depth dependence of the measured energy. After pulse processing with depth correction based on the fit of the weighting potential, the energy resolution is $2.2\,\%$ (FWHM) for the $511\,\mathrm{keV}$ photopeak.}

\newcommand{\cj}{\mathrm{j}}
\usepackage{amsmath}
\begin{document}

\section{Introduction}
Cadmium zinc telluride (CdZnTe, CZT) is a room-temperature semiconductor material for radiation detectors~\protect\cite{delsordo}. It is available in compact detector units with highly segmented pixel layouts and is ideally suited for high-resolution gamma-ray spectroscopy~\protect\cite{vergerbuffet} and 3D imaging~\protect\cite{polaris}. As has been previously reported, CZT detectors have been used in Compton camera systems~\protect\cite{kormoll,fernando,wlee}. With regard to our investigations, CZT detectors are potentially useful in imaging systems for proton therapy~\protect\cite{McCleskey,polf,taya,golnikczt}.
State-of-the-art readout systems for highly segmented CZT detectors are conventionally built with an application-specific integrated circuit (ASIC)~\protect\cite{he_asic,gan}. The ASICs are optimized for gamma-ray spectroscopy. Low energy range, usually up to $2\,\mathrm{MeV}$~\protect\cite{McCleskey,gan,foedisch_rena}, limited count rate capability, and poor availability and product life cycle are unsolved challenges of an ASIC-based readout system for an imaging system in proton therapy. In this environment, high energies up to $7\,\mathrm{MeV}$ and count rates up to $1\,\mathrm{Mcps}$ have to be handled~\protect\cite{schumann,fernando2,pausch}. Instead of using an ASIC for the readout electronics, commercial off-the-shelf (COTS) operational amplifiers have been used for the front-end electronics \protect\cite{ramachers}. Along with space-saving multi-channel analog-to-digital converters (ADC) and a field-programmable gate array (FPGA), all tasks related to the signal acquisition and processing can be done with a COTS system. A programmable digital system benefits from its versatility, which is needed for the evaluation of a detector system for new applications like proton therapy. Even for applications in a fixed installation, a system made of COTS components provides the advantages of proven reliability and life-cycle support.

In general the front-end electronics are the key element of the overall performance of the detector system. Our goal is to maximize the dynamic range of the front-end electronics since the CZT is exposed to high-energy gamma rays, but also has to detect low-energy scatter events used for the Compton imaging. As this is the main goal, which cannot be solved with a state-of-the-art ASIC, the timing information of an interaction must be preserved by the readout system. Most ASICs merely include simple analog signal processing (e.g.\ leading-edge trigger for timing and peak-hold circuit for energy information), making pulse shape analysis or advanced timing difficult or even impossible. As the design of the front-end electronics is a tradeoff between bandwidth, noise, complexity, size, and costs, the best solution must be driven by the application. For the prototype of a Compton camera, we investigate a space-saving and simple circuit design with minimal components. The system must include at least 65 analog readout channels, set up with COTS voltage feedback operational amplifiers.

\section{Basic design considerations}
\subsection{CZT detector assembly}
\begin{figure}[ht]
\centering
\includegraphics[width=0.49\textwidth]{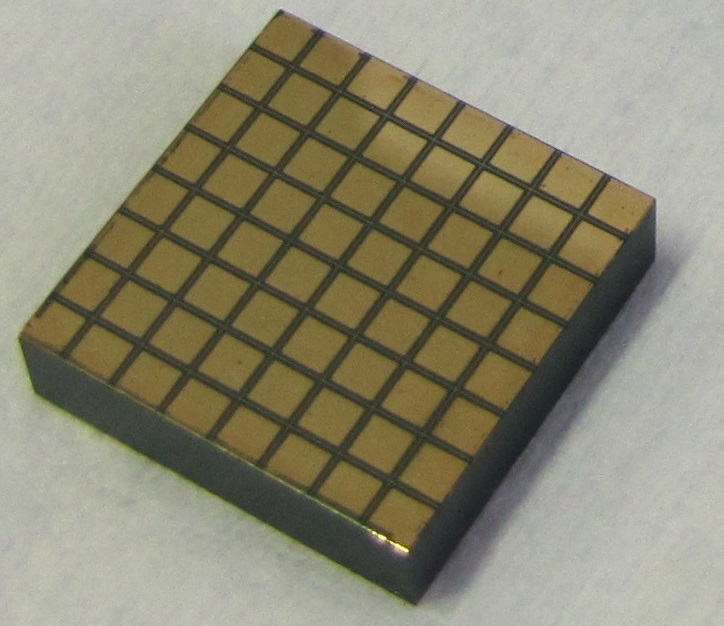}
\includegraphics[width=0.49\textwidth]{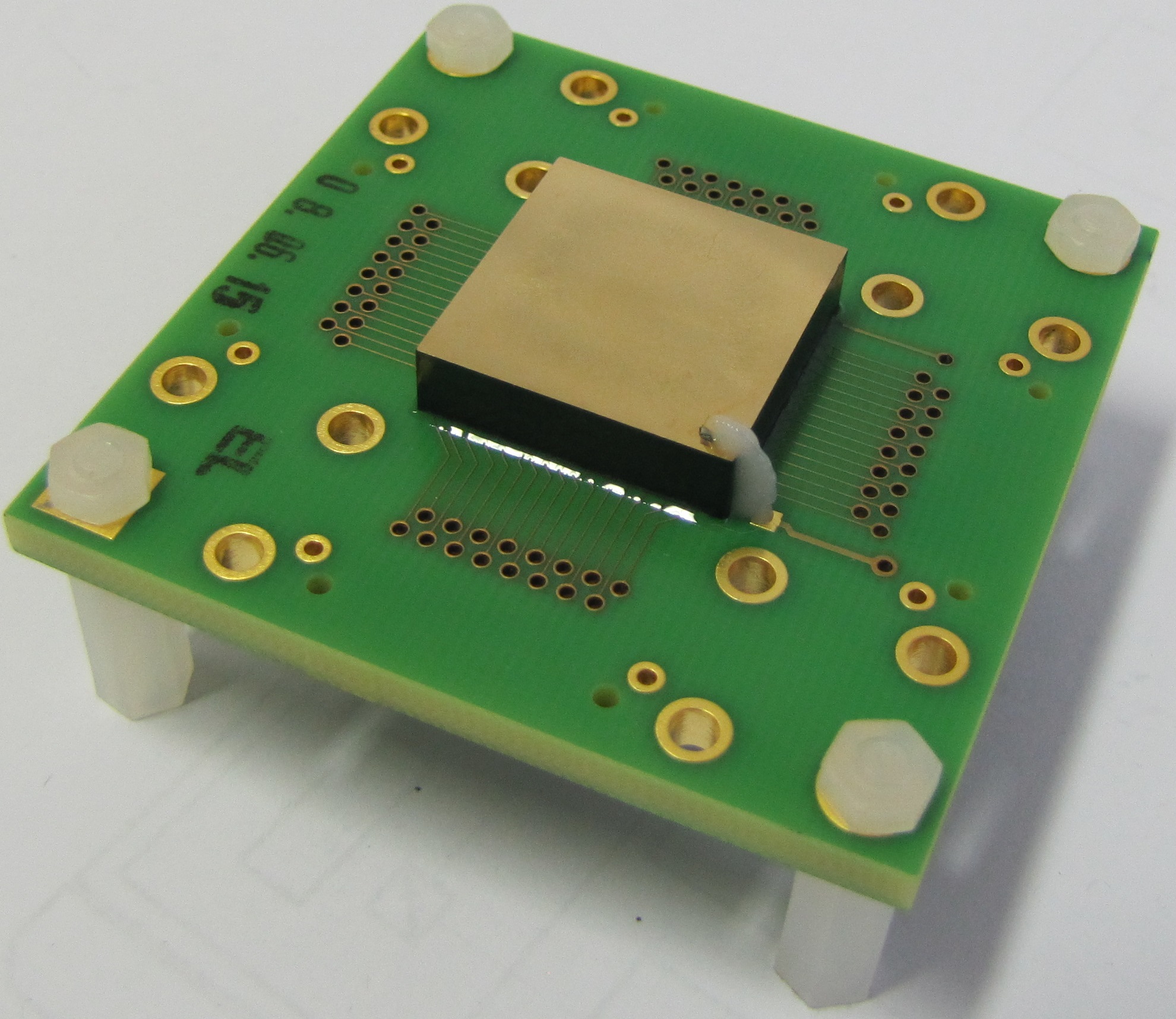}
\caption{A $5\,\mathrm{mm}$ thick pixelated CZT detector from Redlen Technologies~\protect\cite{redlen_datasheet} with $8\,\times\,8$ electrodes (each $2.2\,\mathrm{mm}\,\times\,2.2\,\mathrm{mm}$) on the front side (left) and a continuous planar electrode ($19.42\,\mathrm{mm}\,\times\,19.42\,\mathrm{mm}$) on the back side (right). The detector is mounted on a $3.2\,\mathrm{mm}$-thick carrier board with rugged connectors on the bottom side.}
\label{fig_czt}
\end{figure}

For a medical imaging application, we use a CZT detector as the scattering layer in a Compton camera. In our case, this is a pixelated CZT radiation detector from Redlen Technologies~\protect\cite{redlen_datasheet}. The detector size is $19.42\,\times\,19.42\,\mathrm{{mm}^{2}}$ with a thickness of $5\,\mathrm{mm}$. Towards the continuous planar electrode on the back side, there are 64 pixel electrodes aligned in an $8\,\times\,8$ array on the front side. The size of a pixel pad is $2.2 \times 2.2\,\mathrm{{mm}^{2}}$ for all pixels except the corner pixels with $1.98\,\times\,1.98\,\mathrm{{mm}^{2}}$. In addition, a steering grid surrounds all pixels. The inter-pixel space is $0.26\,\mathrm{mm}$. The bulk device and an assembled detector are shown in figure~\protect\ref{fig_czt}.
The detector is mounted on a printed circuit board (PCB) with the continuous planar electrode on top. A bond wire is attached to a copper pad on the carrier PCB. Furthermore, a conductive adhesive on each pixel pad ensures the electrical connection to the pixel array on the PCB. An underfill between the pixel array and the PCB supports the adhesive connection and improves the mechanical stability. We also decided to use a $3.2\,\mathrm{mm}$-thick FR-4 material for the PCB to enhance the robustness of the assembly. The electrodes of the detector are accessible via rugged high-speed connectors on the bottom side. To shield the detector against visible light, a 3D-printed cap is attached above the detector on the top of the carrier board. Our front-end electronics are designed to work with this type of detector assembly, and the readout boards are plugged into the side faces of the detector assembly (see figure~\protect\ref{fig_hardware}) so a stacked system with arbitrary depth, as required for the evaluation of the Compton camera, can be easily constructed. Further investigations on ruggedization of CZT detectors and detector assemblies have been presented in~\protect\cite{phlu}.

\subsection{Electrical characteristics of a CZT pixel detector}
\begin{figure}[ht]
\centering
\includegraphics[width=0.4\textwidth]{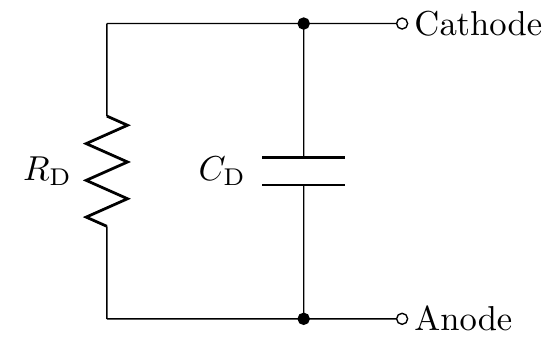}
\caption{An electrical equivalent circuit of a CZT detector. The bulk resistivity is modeled with the resistor $R_\mathrm{D}$, which is typically of several tens of $\mathrm{G}\Omega$. The capacitor $C_\mathrm{D}$ represents the parallel-plate geometry of the electrodes. Its value can be estimated with eq.~\protect\ref{eqn_platecapacitor}.}
\label{fig_detector}
\end{figure}

From the electrical point of view, a CZT detector can be modeled with the equivalent circuit shown in figure~\protect\ref{fig_detector}. With an external operating voltage at the electrodes of the detector, the terminals are referred to as cathode and anode in accordance with the applied polarity. Usually, the continuous electrode is biased with a negative potential and the pixel electrodes are at ground potential. For an ideal detector material, this would force the negative charge carriers (electrons) to move towards the anode and the positive charge carriers (holes) to move towards the cathode. As a consequence of charge trapping due to structural defects, impurities, and irregularities of the material \cite{awadalla}, the mobility and lifetime of the holes in CZT are very poor compared to the electrons \cite{spieler}. Only the moving electrons induce a signal on the electrodes, while the portion of the signal due to the holes may be neglected. Thus, if the generated electrons move to the position-sensitive side of the detector, the overall detection performance is improved. As the readout electronics are directly connected to the electrodes, the electrical characteristics of the detector influence the dynamic behavior of the entire circuit. Finally, the network model for the readout electronics must include the electrical equivalent circuit of the detector.
In general, a very simple equivalent circuit is adequate to model the properties of the detector. As summarized in figure~\protect\ref{fig_detector}, it is a passive two-terminal component with a permittivity and a resistance. A capacitor represents the permittivity of the material and the conductance is modeled as a resistor. For the evaluated pixelated CZT detector, the capacitance can be roughly approximated by the model of the parallel-plate capacitor with an electrode area $A$ separated by the distance $d$. That capacitance $C$ is calculated by
\begin{equation}
C = \epsilon_\mathrm{0}\epsilon_\mathrm{r}\frac{A}{d} \, ,
\label{eqn_platecapacitor}
\end{equation}
where $\epsilon_\mathrm{0}$ is the vacuum permittivity and $\epsilon_\mathrm{r}$ is the relative permittivity of CZT. For the detector in this study, the values are $A = 377\,\mathrm{{mm}^2}$, $d = 5\,\mathrm{mm}$. In practical terms, the bias voltage does not influence the relative permittivity in the applicable voltage range up to $-600\,\mathrm{V}$. The capacitance of the detector is therefore largely independent of the bias voltage~\protect\cite{garson}. The value of $\epsilon_\mathrm{r}$ depends on the manufacturing process, but ranges from 10 to 11~\protect\cite{spieler,antonis}. Thus, the entire bulk capacitance is in the range from $6.6\,\mathrm{pF}$ to $7.4\,\mathrm{pF}$. The capacitance of a single pixel can also be calculated by eq.~\protect\ref{eqn_platecapacitor}, where the size of the pixel determines the value of $A$~\protect\cite{rossi}. With the same assumption for $\epsilon_\mathrm{r}$, the pixel capacitance is in the range from $69\,\mathrm{fF}$ to $95\,\mathrm{fF}$, including the smaller corner pixels. Besides the estimation of capacitance, the bulk resistivity of the detector is needed to model the electrical characteristics. We measured the leakage current of the assembled CZT detector from figure~\protect\ref{fig_czt} with a precision high-voltage source with current monitor (Iseg SHQ series)~\cite{iseg}. This device reports a current of $10\,\mathrm{nA} \pm 1\,\mathrm{nA}$ with a detector bias voltage of $-500\,\mathrm{V}$. For a homogenous material, the resistance is defined as:
\begin{equation}
R=\rho\frac{d}{A}
\label{eqn_resistor}
\end{equation}
where $\rho$ is the resistivity of the material. Our measurement corresponds to a resistor of $50\,\mathrm{G\Omega}$ for the two-terminal equivalent circuit or a resistivity of $4\cdot10^{11}\Omega\mathrm{cm}$. This is in accordance with the values from the datasheet ($\rho > 10^{11}\Omega\mathrm{cm}$) from Redlen~\protect\cite{redlen_datasheet}.

Finally, the electrical characteristics of a CZT crystal mainly depend on the manufacturing process. If they cannot be experimentally verified, the values for the components of the electrical equivalent circuit can be estimated with the geometry of the detector and the constants from the literature by eqn.~\protect\ref{eqn_platecapacitor} and \ref{eqn_resistor}.
The equivalent circuit shown in figure~\protect\ref{fig_detector} is the simplest electrical representation of the detector unit. It does not model a frequency dependency with a complex permittivity. Additionally, the stray capacitances introduced by the traces and the carrier board itself, cross-coupling between pixels, and any inductivities of the connectors are ignored. However, a well-designed PCB layout can minimize these effects.

\subsection{High voltage biasing and grounding}
\begin{figure}[ht]
\centering
\includegraphics[width=\textwidth]{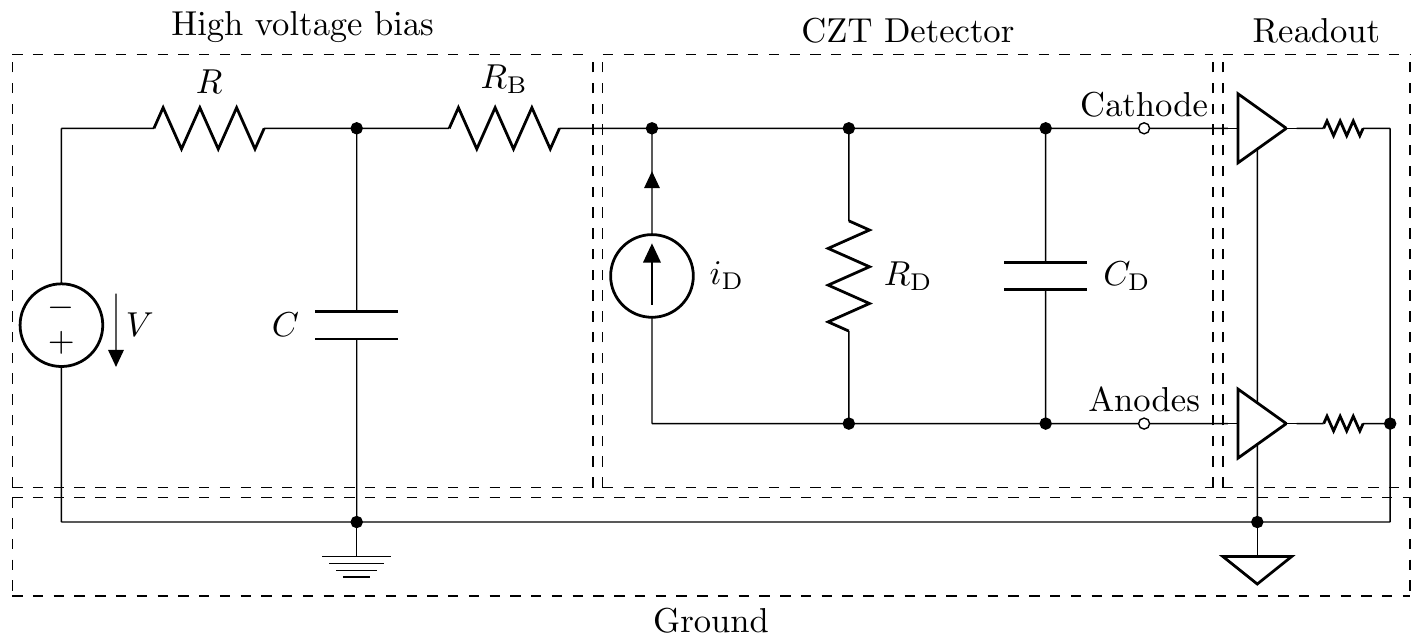}
\caption{Basic connection scheme of the CZT detector. A high voltage supply with an $RC$ low-pass filter and a resistor $R_\mathrm{B}$ are used to generate the bias voltage for the cathode. The readout electronics are directly connected to the cathode and anodes of the detector. The anodes are biased by ground signal of the readout electronics, and consequently both ground potentials must be tied together.}
\label{fig_hv_detector_readout}
\end{figure}

A fundamental operating condition for a CZT detector is the presence of an electric field between the electrodes. Thus, the charge carriers generated by incident radiation move towards the electrodes. Typical electric field strengths for CZT detectors are in the range of $1\,\mathrm{\frac{kV}{cm}}$~\protect\cite{redlen_datasheet}. As the continuous electrode of the detector is biased with a negative voltage, the ground potential is connected to the pixelated electrodes on the opposite side. In general, the high voltage with low ripple is generated by an external power supply connected to the detector via a cable or PCB traces. To reduce any pickup noise related to electromagnetic interference, we put a high-voltage filter close to the detector electrode. This is in the simplest case a passive first-order low-pass filter ($RC$ network shown in figure~\protect\ref{fig_hv_detector_readout}).
The value of the resistor $R_\mathrm{B}$ should be chosen to minimize the voltage drop across the high-voltage filter and maximize the voltage across the detector. The value of the filter capacitor $C$ should be as high as possible to achieve the best noise filtering. According to eqs.~\protect\ref{eqn_platecapacitor} and \ref{eqn_resistor}, the capacitance is inversely proportional to the resistance. Therefore, the capacitor $C$ must be chosen, such that the insulation resistance of the dielectric material is much higher than the resistance of the detector. Commercial capacitors of class~1 have an insulation resistance of more than $100\,\mathrm{G\Omega}$ with a maximum capacitance of $10\,\mathrm{nF}$. Thus, a passive first-order low-pass filter with a cutoff frequency below $1\,\mathrm{Hz}$ is possible (e.g.\ $R=47\,\mathrm{M\Omega}$, $C=10\,\mathrm{nF}$). As a noise-filtered voltage is the output of the $RC$ circuit, it cannot be directly connected to the electrode in order to bias the detector. One reason for this is that the filter capacitor $C$ would be in parallel with the capacitance $C_\mathrm{D}$ of the detector. A bias resistor $R_\mathrm{B}$ of $47\,\mathrm{M\Omega}$ separates the filter network from the detector. Another point that has to be taken into account is the path of current flow generated by the detector. The current should not flow into the high-voltage source. This can be ensured by choosing a high resistance for biasing the detector, so that the time constant $R_\mathrm{B}C$ is much larger than the time constant of the readout electronics~\protect\cite{grupen}. The active components of the readout electronics have their own power supply, which is separated from the high-voltage supply. Further, the anodes are biased with the ground potential of the readout electronics (signal ground in figure~\protect\ref{fig_hv_detector_readout}). Both grounds have to be at the same potential and must be tied together. The electric field between the cathode and the anodes is therefore referenced to a known potential.

\subsection{Signal formation in CZT detectors}
\begin{figure}[ht]
\centering
\includegraphics[width=0.49\textwidth]{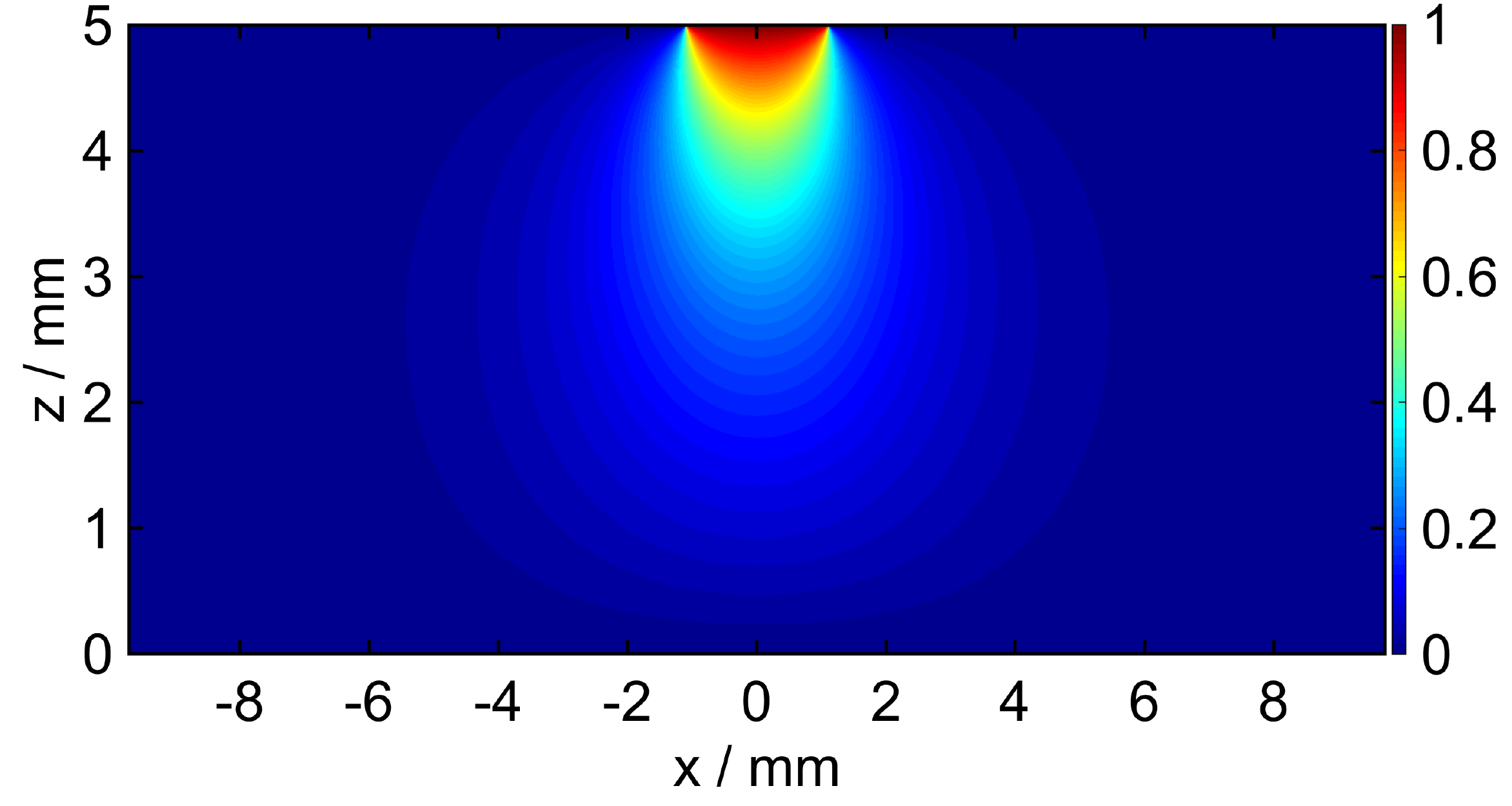}
\includegraphics[width=0.49\textwidth]{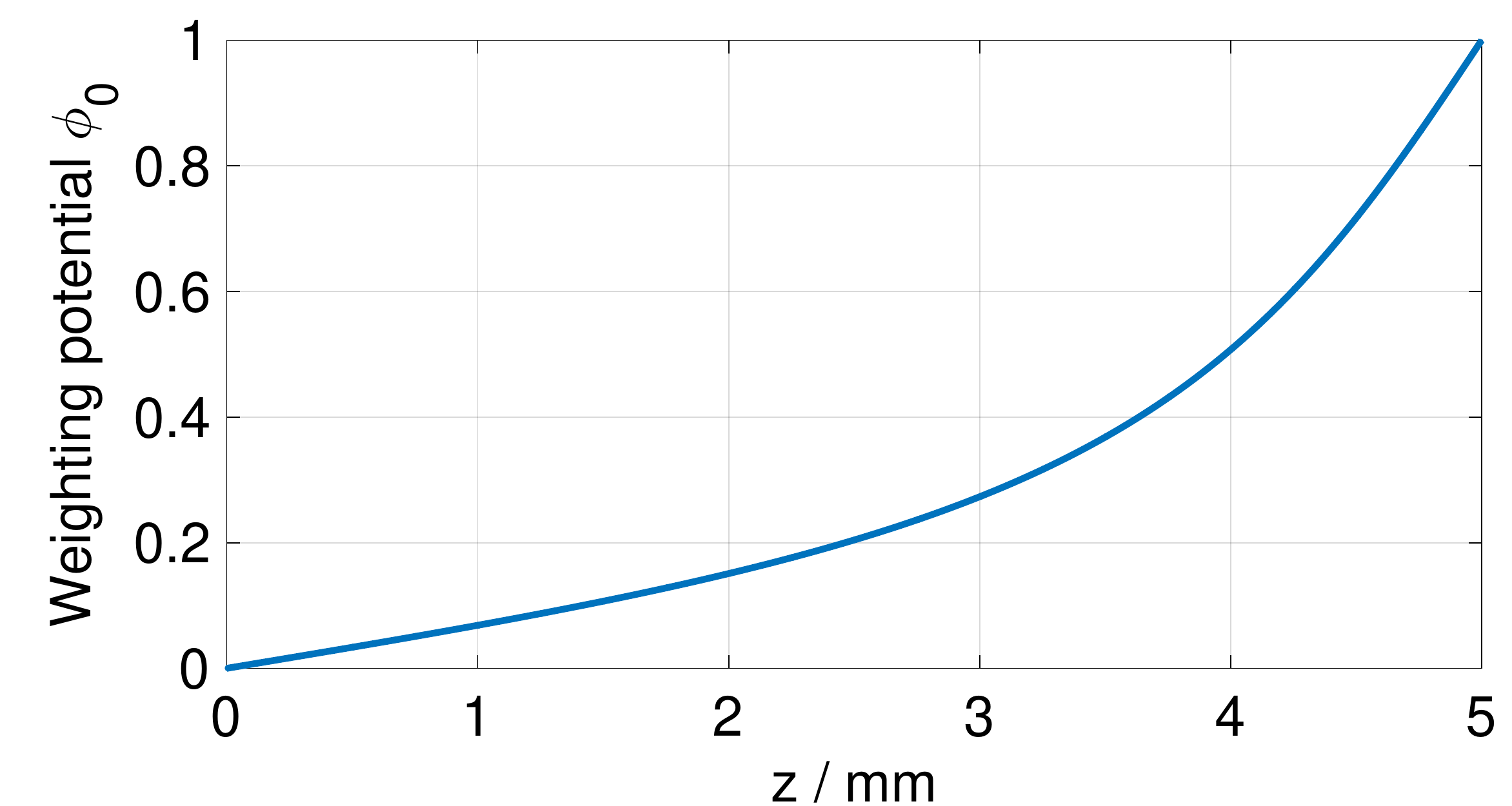}
\includegraphics[width=0.49\textwidth]{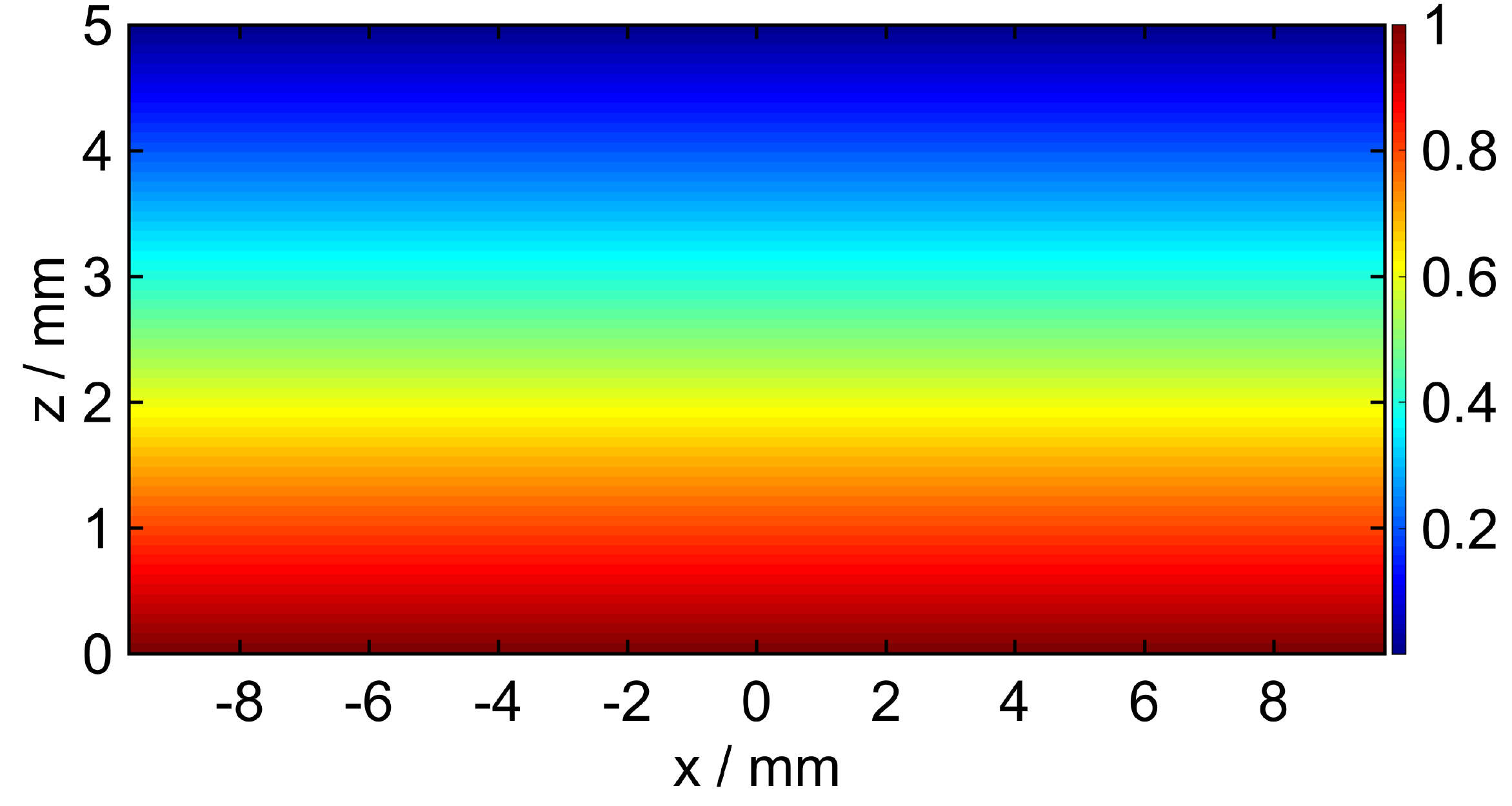}
\includegraphics[width=0.49\textwidth]{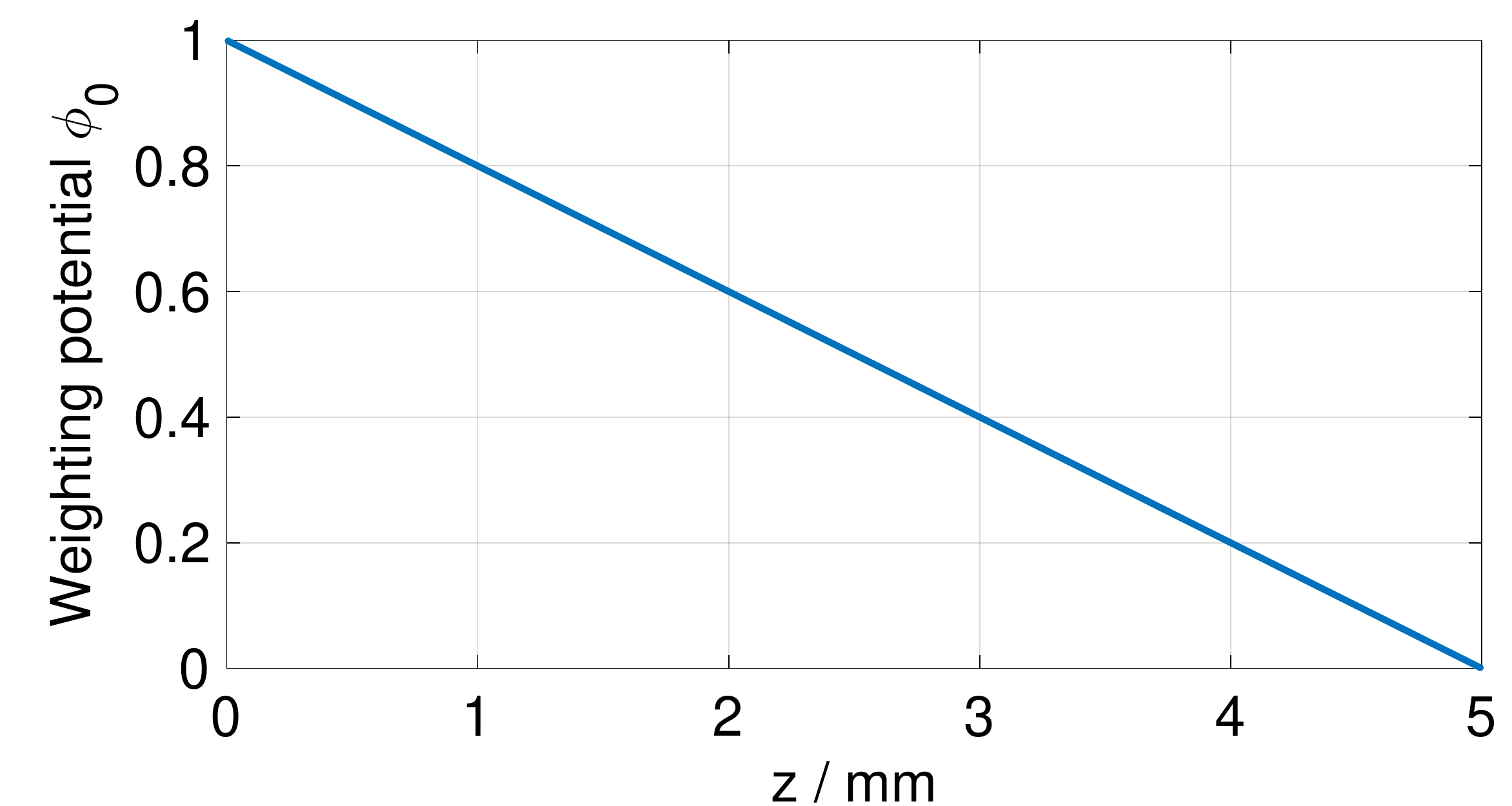}
\caption{The weighting potentials of the cathode (bottom) and pixel anode (top) of the $5\,\mathrm{mm}$-thick detector. The cathode is located at position $z = 0\,\mathrm{mm}$. The right plots show the weighting potential under the collecting electrode ($x=y=0\,\mathrm{mm}$).}
\label{fig_weighting_potentials}
\end{figure}

\begin{figure}[ht]
\centering
\includegraphics[width=0.49\textwidth]{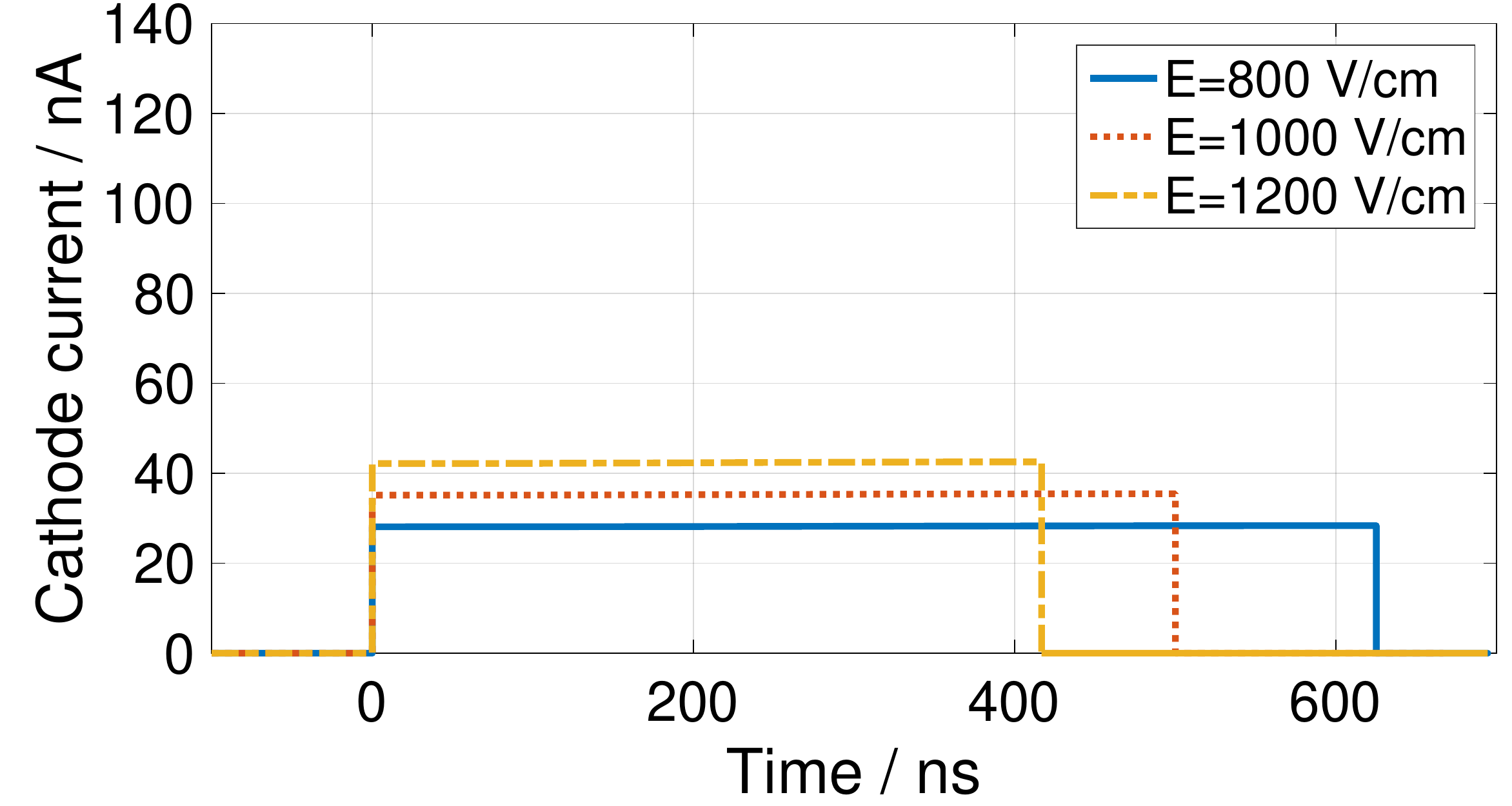}
\includegraphics[width=0.49\textwidth]{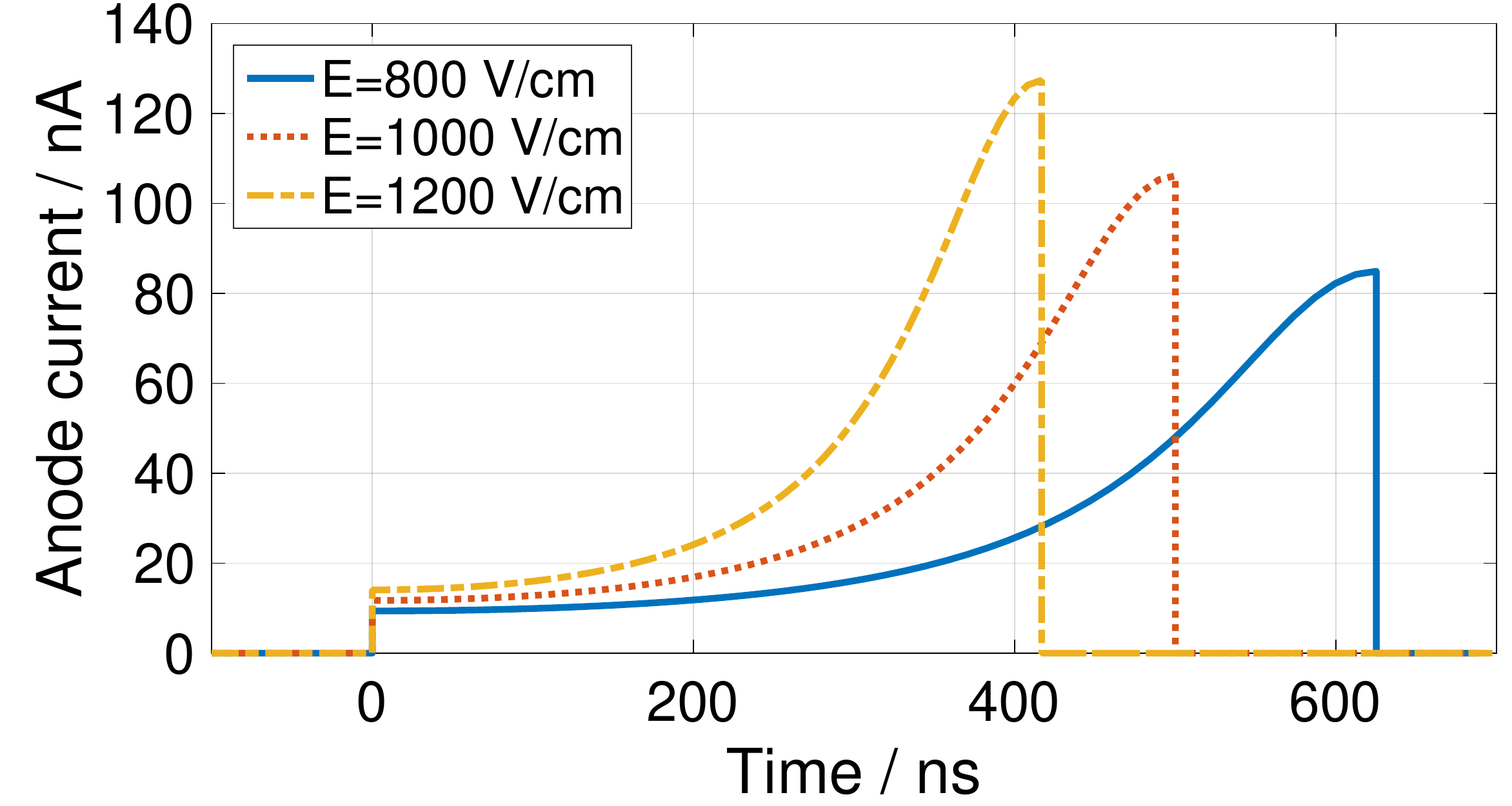}
\caption{Induced current $i$ on the detector electrodes for incident radiation with an energy of $511\,\mathrm{keV}$. The left plot shows the current induced on the cathode and the right plot shows the current induced on an anode pixel. The signals are calculated with eq.~\protect\ref{eqn_current}. An increased electric field strength $E$ causes a higher current $i$ with shorter drift times $t_\mathrm{D}$. The generated charge remains constant.}
\label{fig_current}
\end{figure}

As implied, incident radiation hitting the detector generates free charge carriers. These electrons and holes move towards the electrodes because of the applied electric field. However, the generated charge is proportional to the incident gamma-ray energy and the signal of the detector is an electric current. The induced current through an electrode is defined as
\begin{equation}
i = q\stackrel{\rightarrow}{v}\stackrel{\rightarrow}{E_\mathrm{0}}(\mathbf{x}) \, ,
\label{eqn_current}
\end{equation}
where $q$ is the moving charge, $\stackrel{\rightarrow}{v}$ is the instantaneous velocity of the charge $q$ and $\stackrel{\rightarrow}{E_\mathrm{0}}(\mathbf{x})$ is the weighting field associated with the electrode at the position $\mathbf{x}$ of the charge~\protect\cite{heramo}. The weighting potential $\varphi_\mathrm{0}(\mathbf{x})$ is defined as
\begin{equation}
\stackrel{\rightarrow}{E_\mathrm{0}}(\mathbf{x}) = -\nabla\stackrel{\rightarrow}{\varphi_\mathrm{0}}(\mathbf{x}) \, ,
\label{eqn_weightingpotential}
\end{equation}
by setting one electrode to unit potential and all others to zero. For a parallel-plate geometry of a detector, where the widths in $x$ and $y$ dimensions of the electrodes are much larger than the thickness $z$, the electric field inside the detector is distributed homogeneously with a constant field strength. Therefore, the weighting field $\stackrel{\rightarrow}{E_\mathrm{0}}(\mathbf{x})$ is equal to the electric field, and, solving eq.~\protect\ref{eqn_weightingpotential} in the $z$-dimension, the normalized weighting potential $\varphi_\mathrm{0_\mathrm{C}}(z)$ of the cathode is a linear function~\protect\cite{heramo}:
\begin{equation}
\varphi_\mathrm{0_\mathrm{C}}(z) = z, 0 \leq z \leq 1 \, .
\end{equation}
The solution for the Poisson equation in two dimensions was given by~\protect\cite{rossi,wermes}. These authors presented an equation for the calculation of the weighting potential for a detector with a segmented electrode layout. By setting one dimension of the pixel area to zero and normalizing the thickness $z$ of the detector to 1, the weighting potential $\varphi_\mathrm{0_\mathrm{A}}$ of a single pixel with the normalized width $a$ can be calculated as
\begin{equation}
\varphi_\mathrm{0_\mathrm{A}} (x,z) = \frac{1}{\pi}\mathrm{arctan}\left(\frac{\mathrm{sin}(\pi z) \mathrm{sinh}(\pi \frac{a}{2})}{\mathrm{cosh}(\pi x)-\mathrm{cos}(\pi z)\mathrm{cosh}(\pi \frac{a}{2})}\right) \, ,
\label{eqn_weightingpixel}
\end{equation}
where $x$ and $z$ are the coordinates and $0 \leq z \leq 1$. The weighting potentials are shown in figure~\protect\ref{fig_weighting_potentials}.
The weighting field $E_\mathrm{0}(z)$ under the collecting electrode ($x=0$, $y=0$) can be calculated by solving eq.~\protect\ref{eqn_weightingpotential} with eq.~\protect\ref{eqn_weightingpixel}. This results in the following equation:
\begin{equation}
E_\mathrm{0}(z) = \frac{\mathrm{sinh}(\pi \frac{a}{2})}{\mathrm{cos}(\pi z)-\mathrm{cosh}(\pi \frac{a}{2})} \, .
\label{eqn_weighting_field}
\end{equation}
As we can estimate the weighting field of the detector with eq.~\protect\ref{eqn_weighting_field}, the next step is to calculate the expected electric current to simulate the transient behavior of the detector signals. With the assumption that the electric field $E$ is constant and homogenous across the detector because all anodes are at the same potential, the electric field is calculated by
\begin{equation}
E = \frac{V}{d} \, ,
\label{eqn_efield}
\end{equation}
where $V$ is the applied bias voltage and $d$ is the thickness of the detector. Then, the velocity $v$ of a moving charge $q$ in the detector is calculated by
\begin{equation}
v = \mu_\mathrm{e} E \, ,
\label{eqn_velocity}
\end{equation}
where $\mu$ is the mobility of the charge carriers (electrons). A typical value of $\mu_\mathrm{e}$ for a CZT is about $1000\,\frac{\mathrm{{cm}^2}}{\mathrm{Vs}}$~\protect\cite{heknoll,cho}. With the ionization energy $\mathrm{E}_\mathrm{i} = 4.64\,\mathrm{eV}$ for CZT~\protect\cite{spieler} and the incident radiation $\mathrm{E}_\mathrm{\gamma}$, the generated moving charge $q$ is defined as
\begin{equation}
q = e\frac{\mathrm{E}_\gamma}{\mathrm{E}_i}
\label{eqn_moving_charge}
\end{equation}
where $e$ is the elementary charge. By inserting the eqs.~\protect\ref{eqn_weighting_field}, \ref{eqn_efield}, \ref{eqn_velocity}, \ref{eqn_moving_charge} into eq.~\protect\ref{eqn_current}, the detector current can be numerically estimated. An example of incident radiation of $\mathrm{E}_\gamma = 511\,\mathrm{keV}$ is shown in figure~\ref{fig_current}.

Figure~\protect\ref{fig_current} shows the currents flowing through the electrodes for an interaction at the cathode side of the detector. If the depth of interaction is closer to the anode side, the total charge collection is incomplete, as the fraction of charge from the holes cannot be measured. This circumstance introduces a depth dependence and requires a correction for spectroscopic applications.

\subsection{Readout concepts}
\label{sec_readout}
\begin{figure}[ht]
\centering
\includegraphics[width=0.329\textwidth]{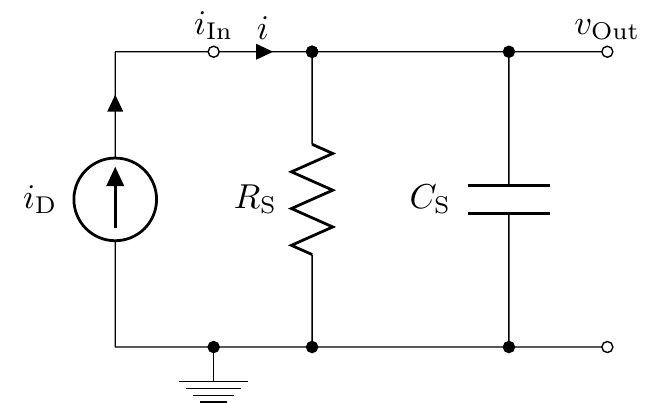}
\includegraphics[width=0.329\textwidth]{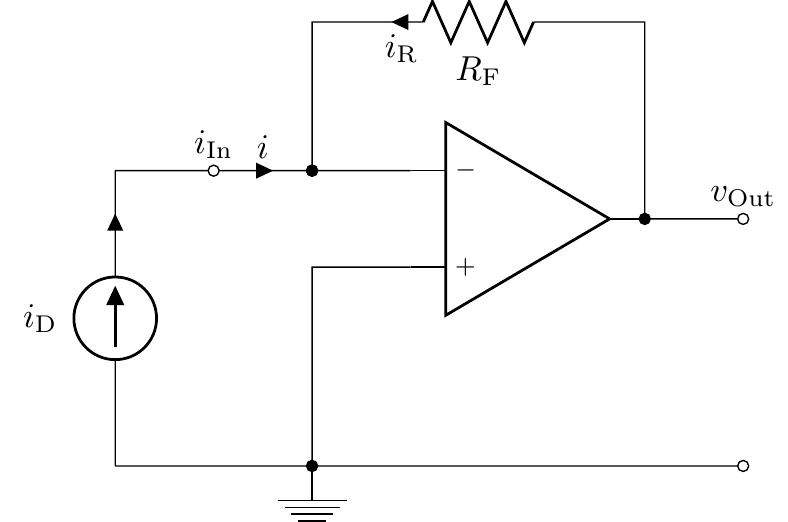}
\includegraphics[width=0.329\textwidth]{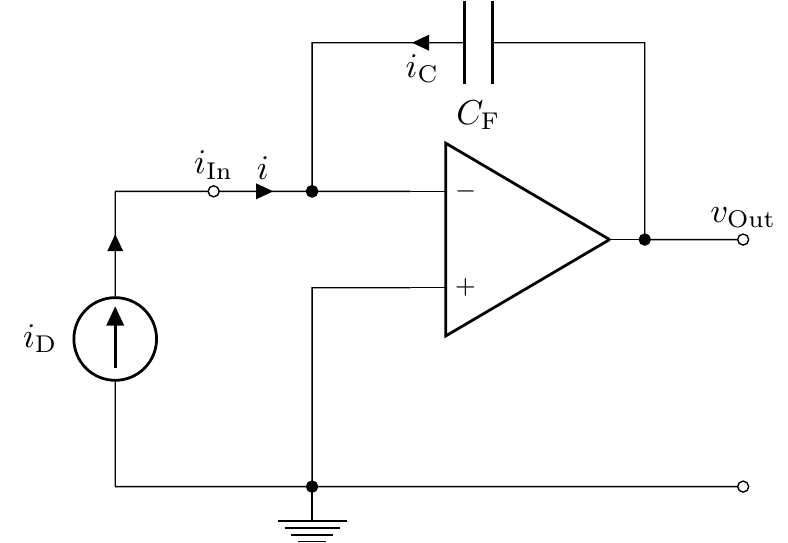}
\caption{Three different readout circuits for the conversion and amplification of the detector current $i_\mathrm{D}$ to a voltage $v_\mathrm{Out}$. The left circuit converts the current to a voltage by means of a shunt resistor. The transimpedance amplifier (middle) is also used as a current-to-voltage converter, whereas the charge-sensitive amplifier (right) is used for a charge-to-voltage conversion.}
\label{fig_readout_concept}
\end{figure}

The results from figure~\protect\ref{fig_current} give an estimate of the required sensitivity for the readout electronics. The signals are in the range of several $\mathrm{nA}$ (e.g.\ $8\,\mathrm{nA}$ for $100\,\mathrm{keV}$ at the cathode with an electric field of $1200\,\frac{\mathrm{V}}{\mathrm{cm}}$). In addition, the drift time $t_\mathrm{D}$ can be as short as $40\,\mathrm{ns}$ if the interaction takes place in the last $10\,\%$ of the detector volume at the anode side. As the detector signals are electric currents, the simplest way to acquire these signals is to use a shunt resistor and a measurement of the voltage drop across this resistor. Finally, the measured transient voltage signal could be fed to an arbitrary signal processing system. In fact, a single resistor would be the simplest, smallest, and cheapest solution for the front-end electronics. With a resistor in the range of some $\mathrm{M}\Omega$, i.e.\ is small enough to force the detector current to flow into it, a voltage drop in the range of some $\mathrm{mV}$ is generated. This is sufficient for most signal processing systems. However, this concept suffers from a lack of bandwidth. As the left circuit in figure~\protect\ref{fig_readout_concept} shows, the induced current will flow through $R_\mathrm{S}$, but the $RC$ time constant of this two-terminal circuit determines the bandwidth and therefore the rise time of this circuit. Moreover, the detector adds its own capacitance to the shunting elements. The $-3\,\mathrm{dB}$ bandwidth of the output signal is given by
\begin{equation}
f_{\mathrm{-3db}} = \frac{1}{2\pi R_\mathrm{S} (C_\mathrm{D} + C_\mathrm{S})} \, ,
\end{equation}
where $C_\mathrm{D}$ is the capacitance of the detector and $C_\mathrm{S}$ is the shunting capacitance of the readout electronics. Even with an undersized value of $1\,\mathrm{pF}$ for $C_\mathrm{D} + C_\mathrm{S}$ and a shunt resistor of $1\,\mathrm{M\Omega}$, the resulting bandwidth is only $159\,\mathrm{kHz}$. However, the rise time $t_\mathrm{r}$ of the output pulse is proportional to the cutoff frequency $f_\mathrm{c}$ of a low-pass filter. For a first-order low-pass filter, the rise time from $10\,\%$ to $90\,\%$ of the step response is calculated by
\begin{equation}
\label{eqn_risetime}
t_\mathrm{r} = \frac{\mathrm{ln}(0.9) - \mathrm{ln}(0.1)}{2 \pi f_\mathrm{c}} \, ,
\end{equation}
where $f_\mathrm{c}$ is the $-3\,\mathrm{dB}$ cutoff frequency. In other words, the rise time for the given example is about $2.2\,\mathrm{\mu s}$. Most of the current pulses will no longer be detectable.

\begin{figure}[ht]
\centering
\includegraphics[width=0.4\textwidth]{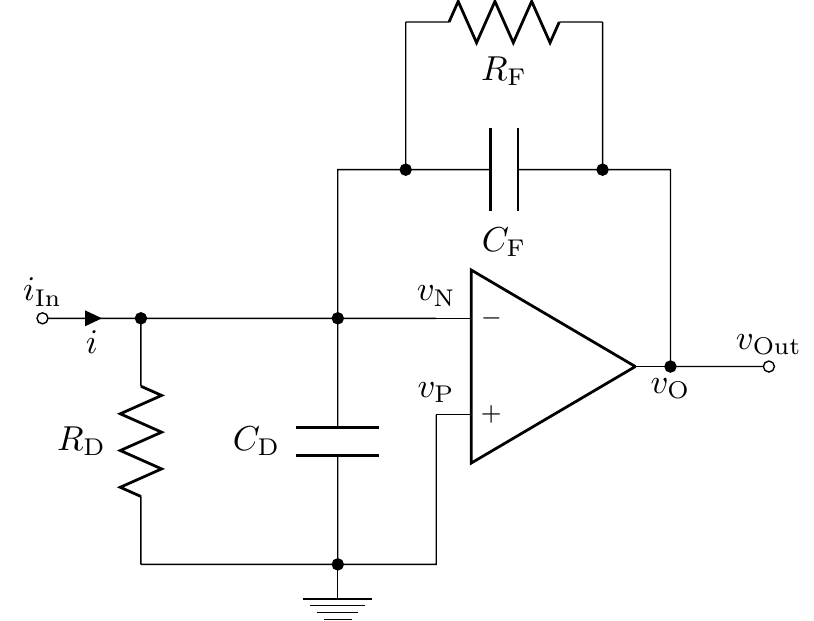}
\caption{A real configuration of a readout circuit as charge-sensitive amplifier. The shunting impedance cannot be eliminated, but the effects are reduced by the operational amplifier. The resistor $R_\mathrm{F}$ from the feedback path also has the parasitic capacitance $C_\mathrm{F}$. This circuit will be used for the further analysis.}
\label{fig_opa_csa}
\end{figure}

To increase the bandwidth for such a current-to-voltage converter, a transimpedance amplifier (TIA) is an appropriate solution (figure~\protect\ref{fig_readout_concept}, middle circuit). This configuration forces the generated current to flow into the negative terminal (virtual ground) of the amplifier. That means $C_\mathrm{D}$ and $C_\mathrm{S}$ are still present, but do not have the same impact on the time constant as the shunting readout circuit. Instead, the rise time and current-to-voltage gain of an ideal TIA are determined by the feedback network. Similarly to the example with the shunting resistor, the TIA is assumed to have a current-to-voltage gain of $1\,\mathrm{M\Omega}$, but the bandwidth is now limited by the parasitic capacitance of the feedback resistor $R_\mathrm{F}$. The resulting bandwidth of the TIA is about $1.6\,\mathrm{MHz}$ with a rise time of $220\,\mathrm{ns}$, if the parasitic capacitance of the resistor is around $100\,\mathrm{fF}$. This is sufficient to detect events near the cathode with drift times longer than the rise time of the TIA, but most of the events will suffer from a significant pulse amplitude loss. To further increase the bandwidth of the TIA, several methods can be applied~\protect\cite{ltjournal}, but, nevertheless, the achievable gain-bandwidth product is too low in relation to the rise times of the anode signal and an adequate signal-to-noise ratio (SNR). Preserving the pulse shape by means of a current-to-voltage converter is quite challenging; however, the detector generates a charge, which can be measured with a modification of the feedback network. A single capacitor $C_\mathrm{F}$ in the feedback loop of the amplifier results in a current-integrating circuit (figure~\protect\ref{fig_readout_concept}, right). Finally, the current $i_\mathrm{c}$ through the capacitor is
\begin{equation}
i_\mathrm{D} = -i_\mathrm{C} = C_F\frac{d v_\mathrm{Out}}{dt} \, .
\end{equation}
Therefore, the voltage $v_\mathrm{Out}$ at the output becomes
\begin{equation}
v_\mathrm{Out} =\frac{-1}{C_\mathrm{F}}\int{i_d dt} = \frac{-q}{C_\mathrm{F}} + v_0\, .
\label{eqn_charge_to_voltage}
\end{equation}
This results in a voltage output whose amplitude is proportional to the moving charge $q$ generated by the detector. This circuit is referred to as the charge-sensitive amplifier (CSA). 
Both the TIA and the CSA are basic negative-voltage-feedback operational amplifier circuits. As the gain of the TIA is designed with a resistor in its feedback path, the gain of the CSA is determined by its feedback capacitance. Nevertheless, a resistor has a parasitic capacitance and a capacitance also has a parasitic resistance. Thus, both circuits are adequately modeled by an operational amplifier with an $RC$ feedback network. Equally importantly, an operational amplifier also introduces a shunting resistor and capacitor. These are in parallel with the impedance of the detector and abstracted by the resistor $R_\mathrm{D}$ and capacitor $C_\mathrm{D}$ in figure~\protect\ref{fig_opa_csa}. On the whole, the model of a CSA is a mixture of the three circuits shown in figure~\protect\ref{fig_readout_concept}.

\begin{figure}[ht]
\centering
\includegraphics[width=0.49\textwidth]{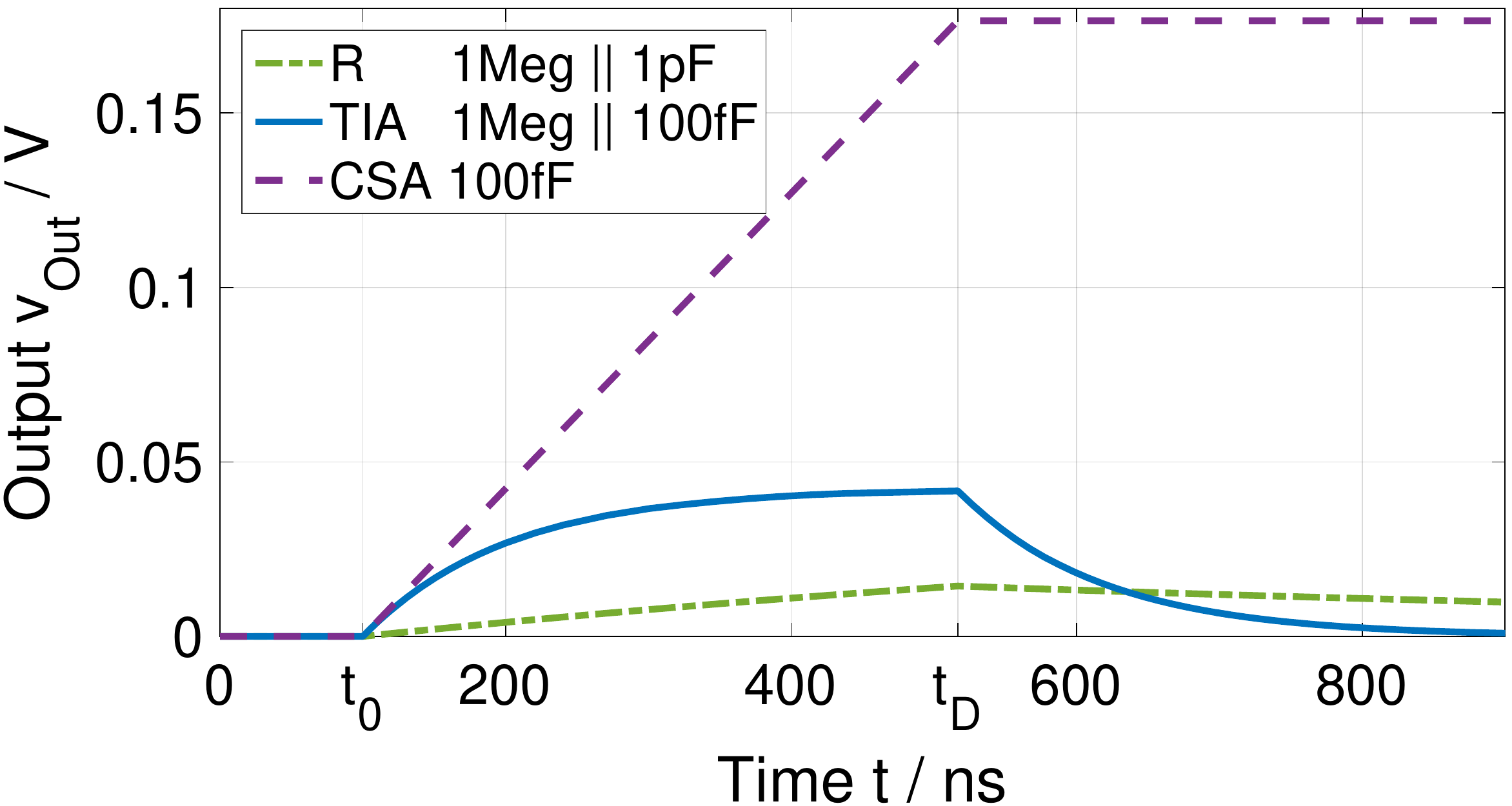}
\includegraphics[width=0.49\textwidth]{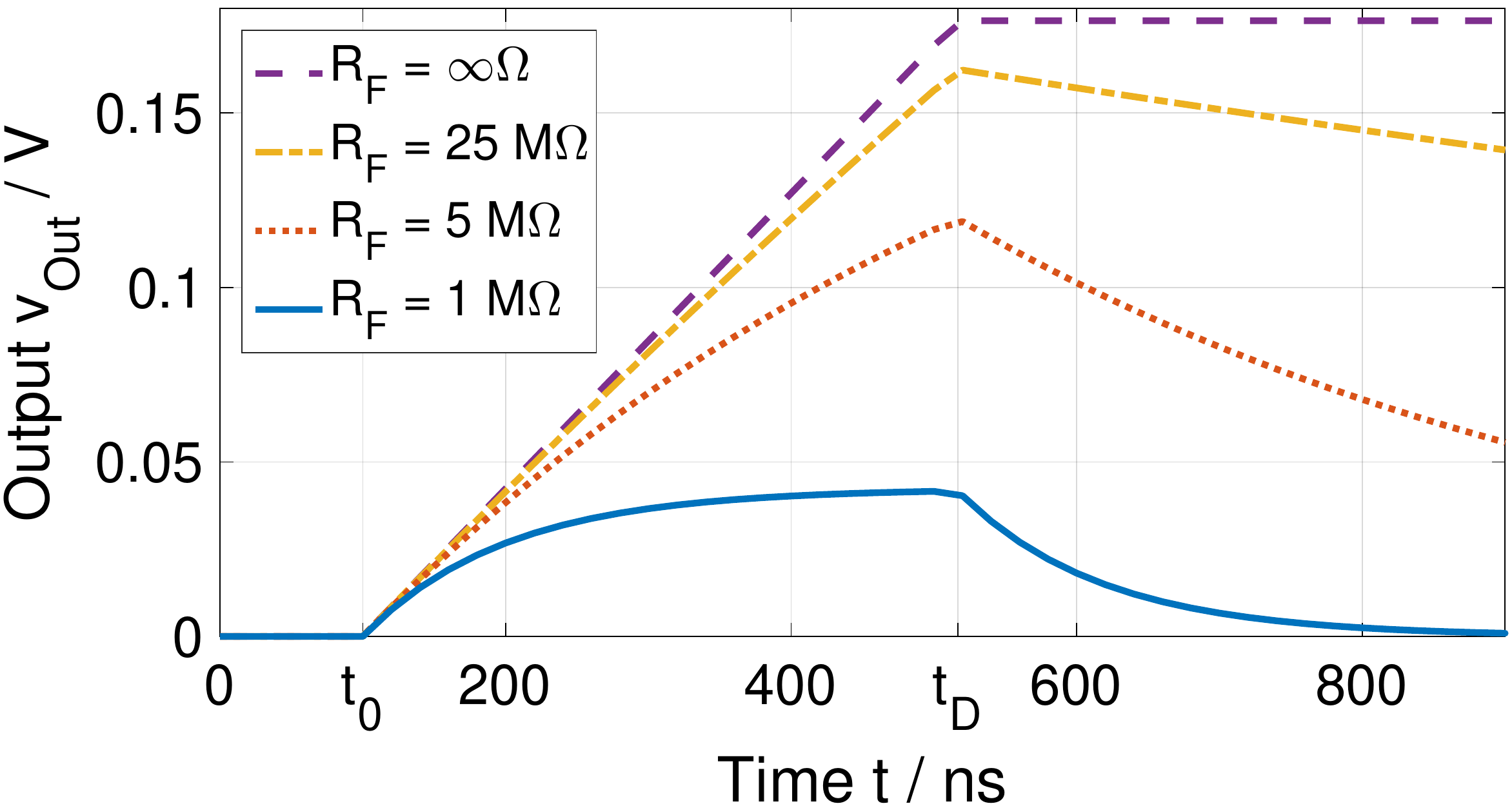}
\caption{Output signals from the three different readout circuits (left). The resistor-based current-to-voltage conversion has poor gain and bandwidth. These are improved by the transimpedance amplifier (TIA). The largest gain is achieved with the charge-sensitive amplifier (CSA). With a constant parasitic capacitance $C_\mathrm{F} = 100\,\mathrm{fF}$ and an increased feedback resistor $R_\mathrm{F}$, the TIA becomes a CSA (right).}
\label{fig_voltage_signals}
\end{figure}

For illustration, the simulated signal waveforms of the three readout circuits with typical values are shown in figure~\protect\ref{fig_voltage_signals}. With this example, it is clearly visible that the CSA achieves the highest amplitude. As the voltage output of the CSA begins to increase at time $t=t_0$ and reaches its peak amplitude $v_{\mathrm{peak}}$ at time $t_\mathrm{D}$ when the current flow stops, the resulting output voltage for an input current pulse with rectangle shape and constant amplitude $i_\mathrm{D}$ is
\begin{equation}
v_{\mathrm{peak}}\left( R_\mathrm{F} \right) = i_\mathrm{D} R_\mathrm{F} \left(1 - {\operatorname{e}}^{\frac{-t_\mathrm{D}}{R_\mathrm{F} C_\mathrm{F}}}\right) \, .
\end{equation}
Therefore, the maximum output voltage $v_{\mathrm{max}}$ for the same input signal is calculated by replacing $R_\mathrm{F}$ with a resistor $R = \infty$. Thus, we can define $v_{\mathrm{peak}}$ over $v_{\mathrm{max}}$ as the peak ratio $P$ of the charge-sensitive preamplifier with the time constant $\tau = R_\mathrm{F} C_\mathrm{F}$ as
\begin{eqnarray}
\frac{v_{\mathrm{peak}}\left( R_\mathrm{F} \right)}{v_{\mathrm{max}}\left( R \right)}  &=& \frac{i_\mathrm{D} R_\mathrm{F} \left(1 - {\operatorname{e}}^{\frac{-{t}_{\mathrm{D}}}{R_\mathrm{F} C_\mathrm{F}}}\right)}{i_\mathrm{D} R \left(1 - {\operatorname{e}}^{\frac{-{t}_{\mathrm{D}}}{R C_\mathrm{F}}}\right)}\\
P = \lim\limits_{R \to \infty} \frac{v_{\mathrm{peak}}\left( R_\mathrm{F} \right)}{v_{\mathrm{max}}\left( R \right)} &=& \frac{\tau}{t_\mathrm{D}} \left( 1 - {\operatorname{e}}^{\frac{-{t}_{\mathrm{D}}}{\tau}}\right) \, .\label{eqn_ballistic_deficit_p}
\end{eqnarray}
Eq.~\protect\ref{eqn_ballistic_deficit_p} describes the attenuation of the peak amplitude. Referring to~\protect\cite{knoll}, the degree of which the infinite time constant amplitude has been decreased is called the ballistic deficit $B$. With eq.~\protect\ref{eqn_ballistic_deficit_p}, a numerical expression of $B$ can be defined as follows:
\begin{equation}
B = 1 - P \, .
\label{eqn_ballistic_deficit}
\end{equation}
\begin{figure}[ht]
\centering
\includegraphics[width=0.49\textwidth]{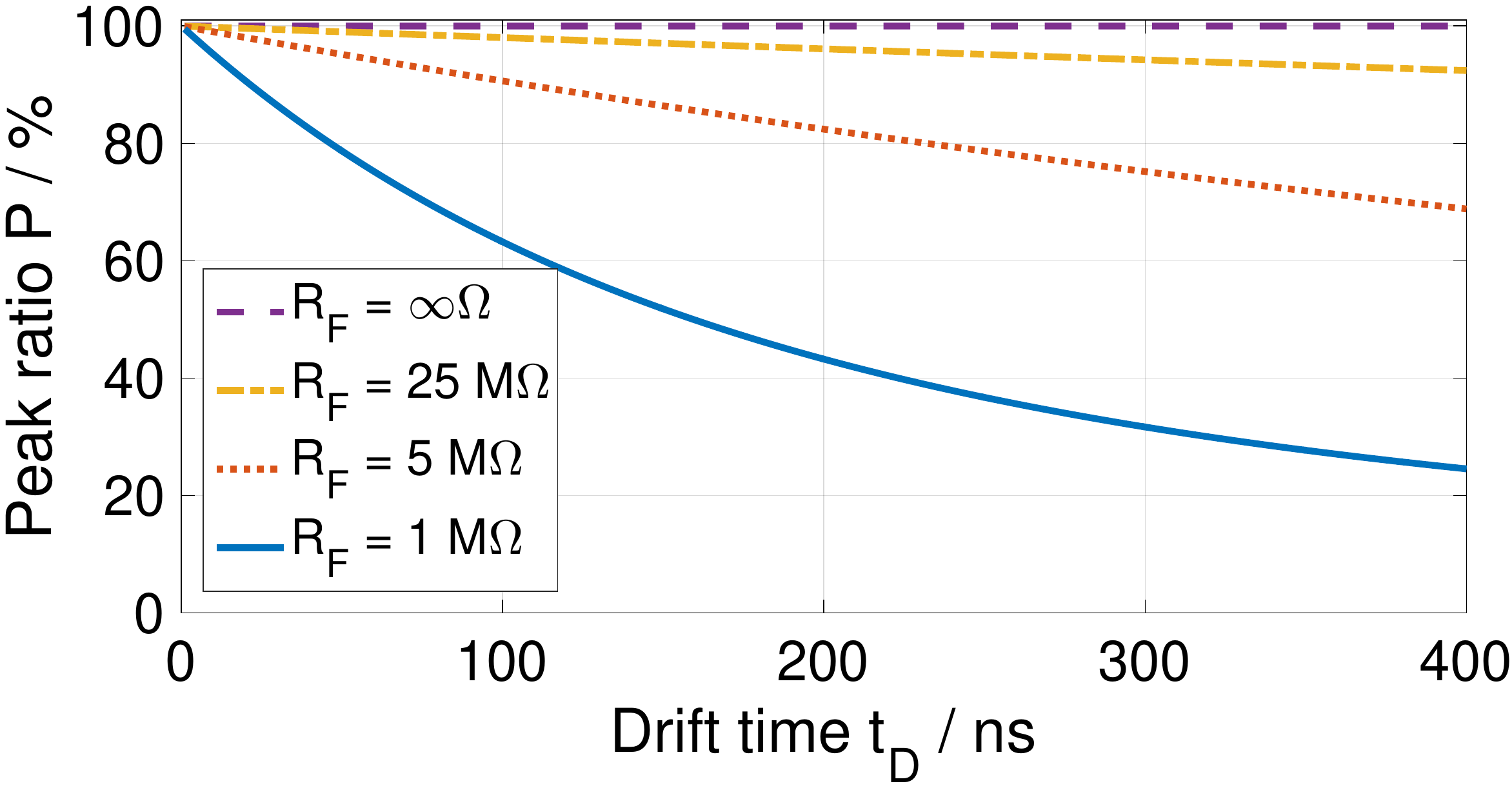}
\includegraphics[width=0.49\textwidth]{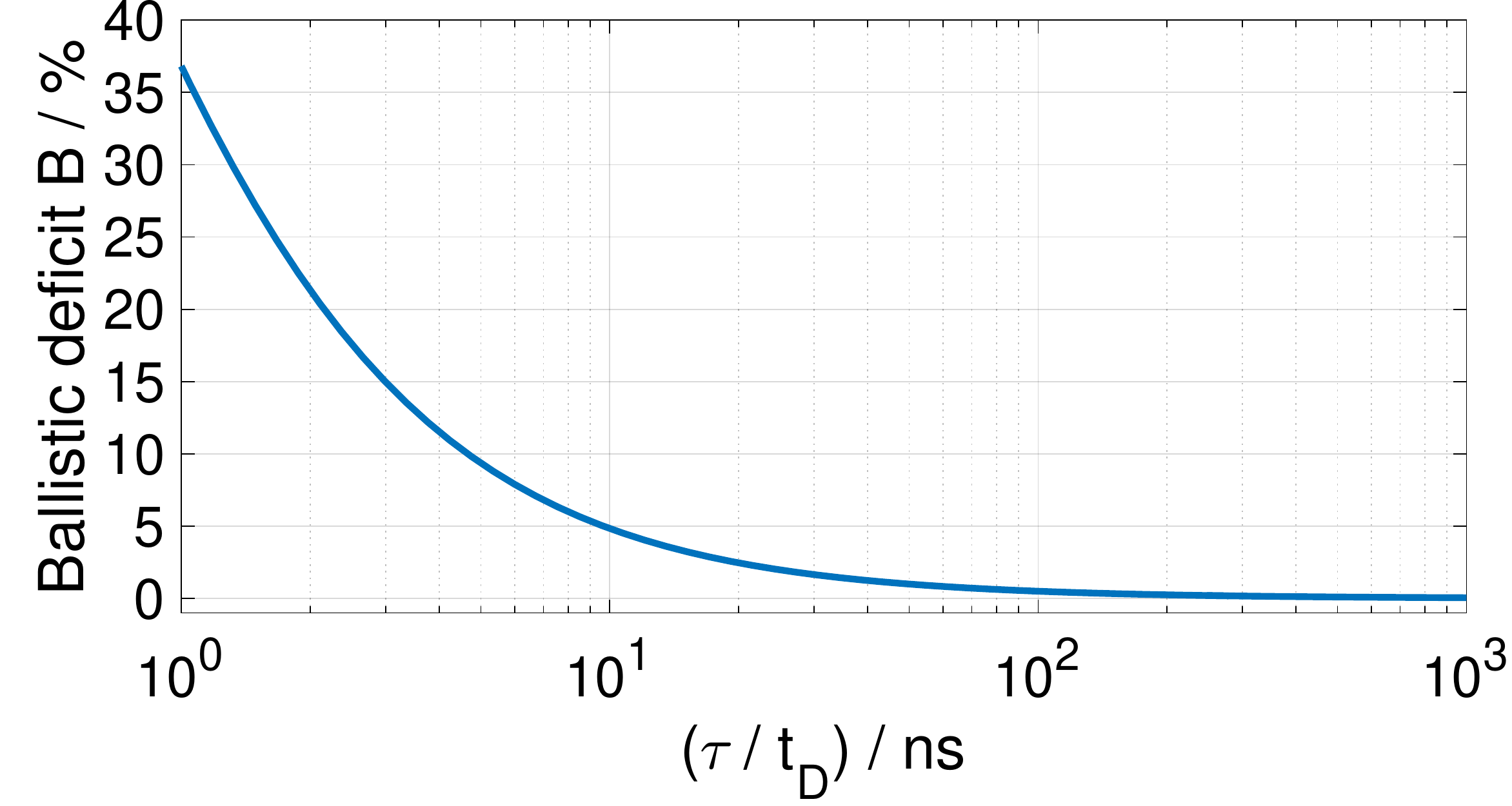}
\caption{The peak amplitude of the output signal from the charge-sensitive amplifier decreases with an increasing time constant $\tau = R_\mathrm{F} C_\mathrm{F}$ (left). Therefore, the ballistic deficit $B$ according to eq.~\protect\ref{eqn_ballistic_deficit} depends on the time constant $\tau$ and the drift time $t_\mathrm{D}$ of the moving charge (right).}
\label{fig_ballistic}
\end{figure}
The calculations are based on the assumption that the input current pulse has a rectangular shape and a constant value during the drift time $t_\mathrm{D}$, as it is seen by the cathode. If we consider a pixelated detector, where the size of the pixel is small compared to the continuous electrode and detector thickness, the current pulse as shown in figure~\protect\ref{fig_current} only rises when the moving charge reaches the pixel electrode. However, the drift time $t_\mathrm{D}$ is the same as for the cathode current, but most of the charge is deposited at the end of the pulse. Thus, the anode current can be simply imagined as a rectangular pulse with a shorter drift time and higher current than the cathode signal. Consequently, the ratio of $\frac{\tau}{t_\mathrm{D}}$ is larger than the ratio of the continuous electrode, if the same CSA is used. Thus the ballistic deficit has a greater impact on the cathode signals than the anode signals.

Eq.~\protect\ref{eqn_ballistic_deficit} is essential to select an optimized feedback time constant dependent on the characteristics of the detector. Usually, the charge-to-voltage gain is a requirement imposed by the lowest measurement range of the application and is set by a carefully selected feedback capacitance (see eq.~\protect\ref{eqn_charge_to_voltage}). One parameter which can be optimized is the value of the feedback resistor. Of course, the best choice is a value close to infinity, since it matches the ideal CSA. However, this is practically not useful, as every current pulse from the detector forces the amplifier to integrate the charge onto its steady state voltage level. After a while, the amplifier overflows, when the output voltage reaches the level of the supply voltage (or even far below this level). Then, it cannot process another event until the feedback capacitor has discharged. A conventional method for the discharge of the capacitor uses a switch, which is added to the feedback network. This requires an additional reset logic and active control, as featured by an ASIC. A more prevalent practice is to reset the amplifier by making an appropriate choice of the feedback resistance $R_F$. This feedback resistor discharges the capacitor $C_\mathrm{F}$ with the characteristic time constant $\tau$. An optimally adjusted value of the resistor is a tradeoff between the ballistic deficit, count rate, and frequency response of the amplifier.
\subsection{Operational amplifier}
\begin{figure}[ht]
\centering
\includegraphics[width=0.4\textwidth]{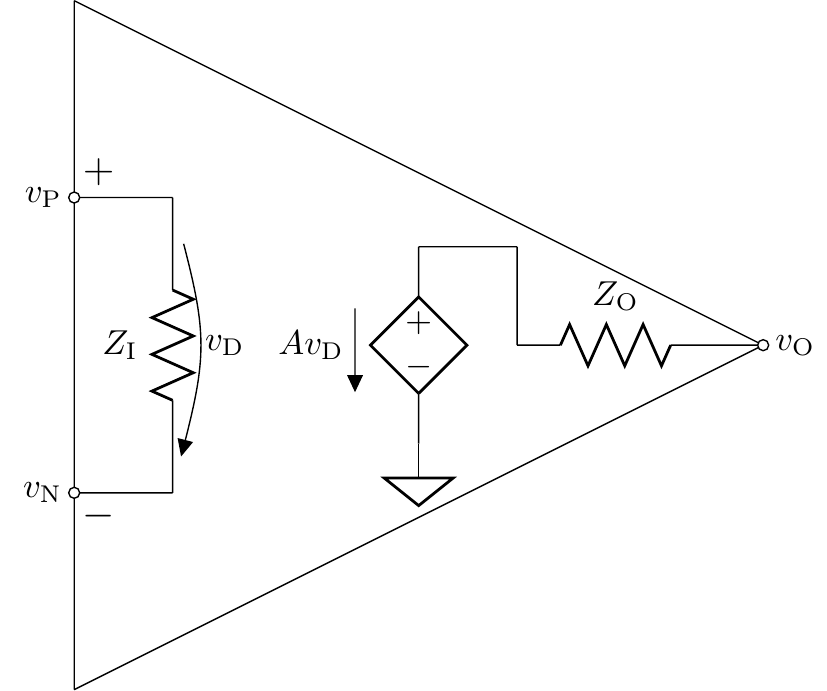}
\caption{A model of a realistic operational amplifier~\protect\cite{hor}. The voltage difference $v_\mathrm{D}$ between the two terminals is amplified by the bandwidth-limited factor $A$. The input impedance $Z_\mathrm{I}$ should be as high as possible, whereas the output impedance $Z_\mathrm{O}$ should be close to zero.}
\label{fig_real_opa}
\end{figure}

For the design and investigation of a CSA with an operational amplifier, we will analyze the circuits in time and frequency domains. Our notation for a complex number in the frequency domain is
\begin{equation}
s = \sigma + \cj\omega \, ,
\end{equation}
where $\cj$ is the imaginary unit, $\sigma$ is the real part, and $\omega$ is the imaginary part in the range of real positive values. For many circuit designs, it is useful and sufficient to analyze a network containing an operational amplifier with ideal constraints, which means the device has an infinitely high input impedance, infinite open-loop voltage gain, etc. For a realistic and detailed analysis, an operational amplifier can be modeled as shown in figure~\protect\ref{fig_real_opa}.
In contrast to the ideal operational amplifier, the input impedance $Z_\mathrm{I}$ is not infinite. The basic function of an operational amplifier is the amplification of the voltage drop across its positive and negative input terminals. The output voltage is therefore given by
\begin{equation}
\label{eqn_vo}
v_\mathrm{O} = (v_\mathrm{P} - v_\mathrm{N})A \, ,
\end{equation}
where $v_\mathrm{P}$ is the potential at the positive terminal and $v_\mathrm{N}$ at the negative terminal. For an ideal operational amplifier, the transfer function $A(s)$ has no frequency dependencies, so that the corresponding frequency response
\begin{equation}
H_\mathrm{A} ( f ) = A(0) = A_\mathrm{OL}
\end{equation}
is constant. $A_\mathrm{OL}$ is the zero-frequency open-loop gain. For a realistic operational amplifier, the frequency response of the open-loop gain has the shape of a low-pass filter. Consequently, we model the transfer function $A(s)$ of the operational amplifier as a first-order low-pass filter in the frequency domain using
\begin{equation}
A(s) = \frac{A_\mathrm{OL}}{1 + s \tau} \,
\label{eqn_As}
\end{equation}
with the finite open-loop voltage gain $A_\mathrm{OL}$ and the $-3_\mathrm{dB}$ cutoff frequency
\begin{equation}
f_c = \frac{1}{2 \pi \tau} \,.
\label{eqn_opa_fc}
\end{equation}
A characteristic parameter which simplifies the frequency response of an operational amplifier is the gain-bandwidth product (GBP). At frequencies larger than $f_c$, the GBP is constant for the first-order low-pass filter frequency response. At the frequency $f_\mathrm{GBP}$, the open-loop gain equals the unity gain. With the maximum open-loop voltage gain $A_\mathrm{OL}$, the functional relationship is given by
\begin{equation}
f_c  {A_\mathrm{OL}} = f_\mathrm{GBP} \,.
\end{equation}
With the given parameters $A_\mathrm{OL}$ and $f_\mathrm{GBP}$ of an operational amplifier, we can set
\begin{equation}
\tau = \frac{A_\mathrm{OL}}{2 \pi f_\mathrm{GBP}}
\end{equation}
and by setting $s=\cj \omega$ with $\omega = 2 \pi f$, the transfer function of the open-loop gain for periodic sinusoidal signals is
\begin{equation}
A(\cj 2 \pi f) = \frac{A_\mathrm{OL}}{1+\cj\frac{f}{f_{\mathrm{GBP}}}A_\mathrm{OL}} \, .
\label{eqn_transferfunction_opa}
\end{equation}
The frequency-dependent gain $G_\mathrm{A}(f)$ and phase shift $\Phi_\mathrm{A}(f)$ are derived from Eq.~\protect\ref{eqn_transferfunction_opa}.
\begin{eqnarray}
G_\mathrm{A} ( f )    & = & |A(\cj 2 \pi f)| = \frac{A_\mathrm{OL}}{\sqrt{1+{\left(\frac{f}{f_{\mathrm{GBP}}}A_\mathrm{OL}\right)}^2}}\\
\Phi_\mathrm{A} ( f ) & = & \angle {A(\cj 2 \pi f)} =  -{\operatorname{tan}}^{-1}\left( \frac{f}{f_\mathrm{GBP}} A_\mathrm{OL} \right)
\end{eqnarray}
\begin{figure}[ht]
\centering
\includegraphics[width=0.49\textwidth]{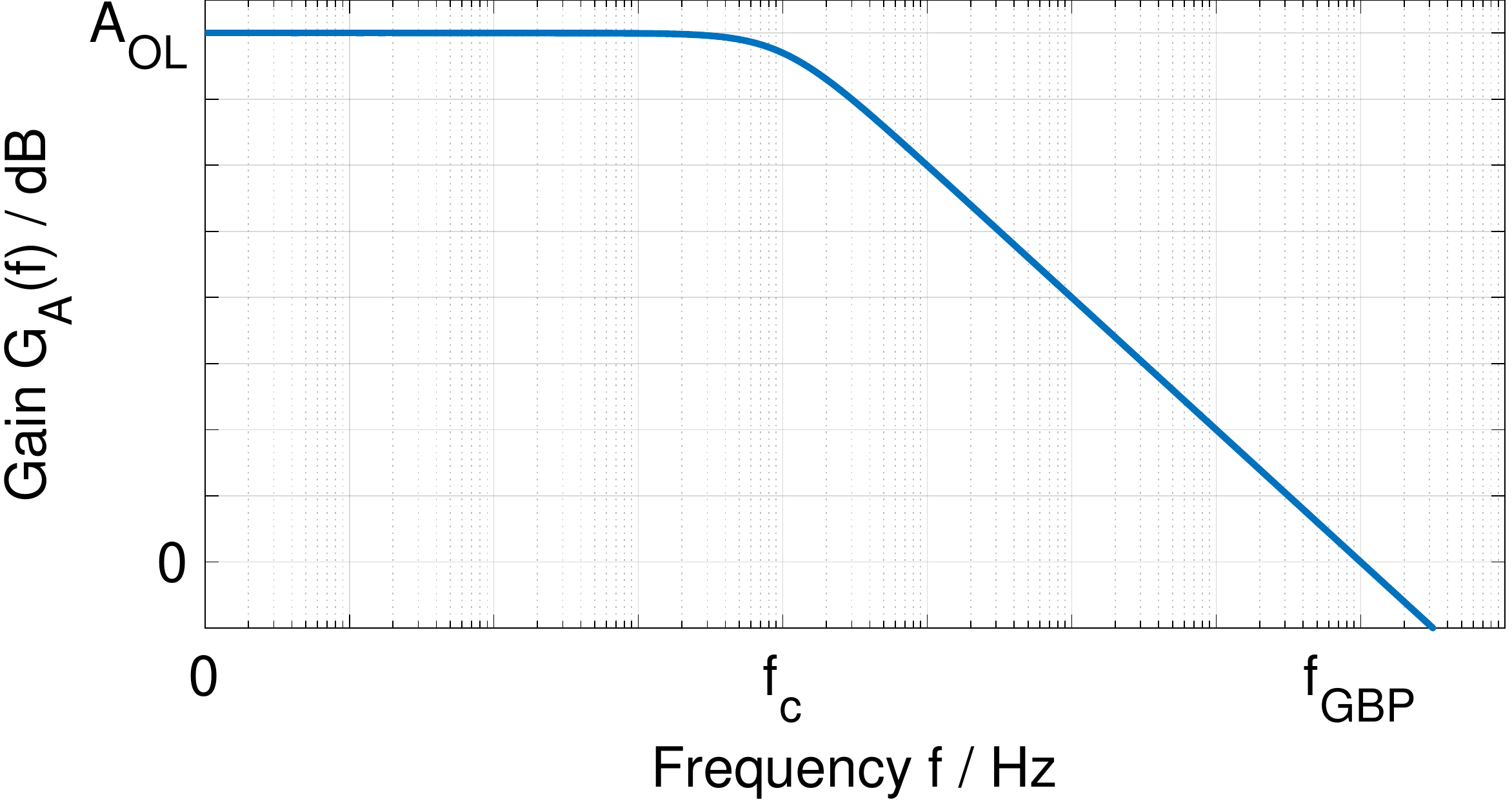}
\includegraphics[width=0.49\textwidth]{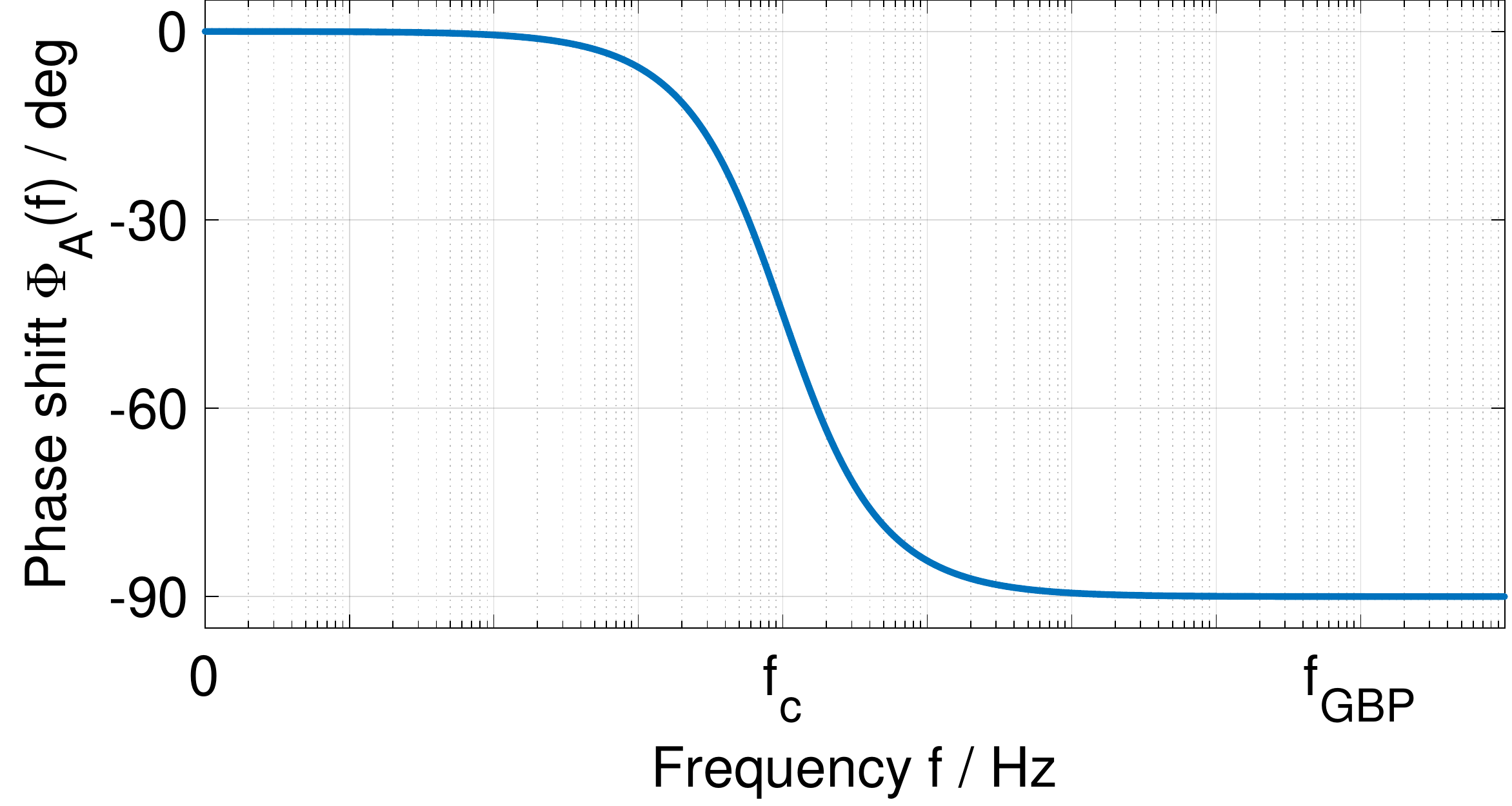}
\caption{The gain $G_\mathrm{A} ( f )$ and the phase shift ${\Phi}_\mathrm{A} ( f )$ of the frequency response of an operational amplifier with zero-frequency gain $A_\mathrm{OL}$. The frequency response is modeled as a first-order low-pass filter with the cutoff frequency $f_c = \frac{f_\mathrm{GBP}}{A_\mathrm{OL}}$. The attenuation of $A_\mathrm{OL}$ is $-20\,\mathrm{dB}\,/\,\mathrm{decade}$ above $f_\mathrm{c}$.}
\label{fig_opampbode}
\end{figure}

\section{Circuit design of a charge-sensitive preamplifier}
\begin{figure}[ht]
\centering
\includegraphics[width=0.4\textwidth]{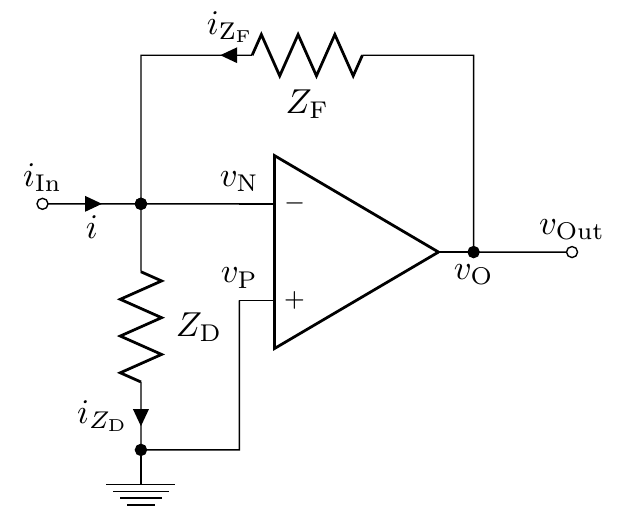}
\caption{An operational amplifier with a current input and a voltage feedback connected to the negative input terminal. With $Z_\mathrm{F} = C_\mathrm{F} || R_\mathrm{F}$, this configuration is used for the charge-to-voltage amplification. $Z_\mathrm{D}$ represents all impedance connected between the input $i_\mathrm{In}$ and ground (e.g.\ capacitance of the detector and parasitic capacitance of the negative input terminal).}
\label{fig_csa_opa}
\end{figure}

As briefly described in sec.~\protect\ref{sec_readout}, an operational amplifier integrates a charge, as there is a capacitor $C_\mathrm{F}$ in the feedback and the time constant of the $R_\mathrm{F}C_\mathrm{F}$ network is large compared to the drift time of the moving charge at its input. In general, a two-terminal equivalent circuit adequately represents the feedback circuit. Thus, the $R_\mathrm{F}C_\mathrm{F}$ network is summarized with the impedance $Z_\mathrm{F}$. In accordance with figure~\protect\ref{fig_detector}, the impedance of the detector is modeled with the two-terminal impedance $Z_\mathrm{D}$. Both impedances are connected to the inverting terminal of the operational amplifier. In addition, there are parasitic shunt impedances between the negative and the positive input terminal of the operational amplifier. As they are in parallel with the impedance of the detector, all additional shunt impedances are absorbed by the model of $Z_\mathrm{D}$.
For a simplified circuit analysis, the positive terminal of the operational amplifier is grounded. If the operational amplifier requires a single supply operation, the positive terminal is biased towards the desired potential for the virtual ground. For a circuit analysis, this is negligible. Therefore, the initial circuit for the analysis is shown in figure~\protect\ref{fig_csa_opa}.
\subsection{Circuit analysis}
\label{circuitanalysis}
\begin{figure}[ht]
\centering
\includegraphics[width=0.8\textwidth]{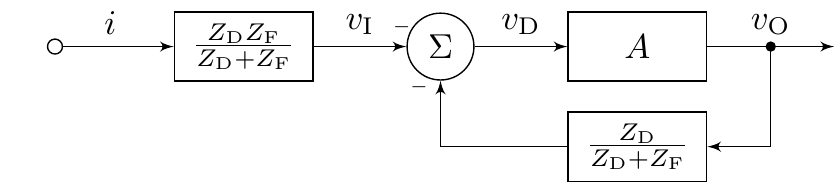}
\caption{Block diagram of the closed-loop transfer function. This structure shows the voltage feedback network and loop gain, which are essential for a stability analysis.}
\label{fig_csablock}
\end{figure}

Kirchhoff's current law is used for the fundamental circuit analysis. For the node at the negative terminal of the operational amplifier, the sum of all currents must be zero.
\begin{equation}
0 = i - i_{Z_\mathrm{D}} + i_{Z_\mathrm{F}}
\label{eqn_kirchhoff_sum}
\end{equation}
Further, the currents can be expressed in terms of the feedback impedance $Z_\mathrm{F}$ and shunt impedance $Z_\mathrm{D}$. 
\begin{eqnarray}
i_{Z_\mathrm{F}} &=& \frac{(v_\mathrm{O}-v_\mathrm{N})}{Z_\mathrm{F}}\label{eqn_izf}\\
i_{Z_\mathrm{D}} &=& \frac{v_\mathrm{N}}{Z_\mathrm{D}}\label{eqn_izd}
\end{eqnarray}
According to eq.~\protect\ref{eqn_vo}, and setting $v_\mathrm{P} = 0$, the voltage $v_\mathrm{N}$ in eqs.~\protect\ref{eqn_izf} and \ref{eqn_izd} can be substituted with
\begin{equation}
v_\mathrm{N} = \frac{-v_\mathrm{O}}{A} \, .
\label{eqn_vn}
\end{equation}
The solution of eq.~\protect\ref{eqn_kirchhoff_sum} with eqs.~\protect\ref{eqn_izf}-\ref{eqn_vn} results in the current-to-voltage transfer function
\begin{equation}
\label{eqn_opa}
\frac{-v_\mathrm{O}}{i} = \frac{AZ_\mathrm{D}Z_\mathrm{F}}{Z_\mathrm{D}(A+1) + Z_\mathrm{F}} \, ,
\end{equation}
where A is the transfer function of the operational amplifier according to eq.~\protect\ref{eqn_As}.
Another useful method for the circuit analysis is the principle of superposition. Because the system can be described with linear equations, a superposition of all current and voltage sources is possible. That means that first, if the current source for $i$ is turned off (replaced with an open circuit), the voltage source for $v_\mathrm{O}$ is acting alone and the resulting voltage at the negative input terminal of the operational amplifier is
\begin{equation}
{\left.v_{\mathrm{N}}\right|}_{i=0} = v_\mathrm{O} \frac{Z_\mathrm{D}}{Z_\mathrm{D} + Z_\mathrm{F}} \, .
\label{eqn_superpos1}
\end{equation}
Second, if the voltage source $v_\mathrm{O}$ is turned off (replaced with a short), the current source is acting alone and the voltage at the negative input terminal of the operational amplifier is
\begin{equation}
{\left.v_{\mathrm{N}}\right|}_{v_{\mathrm{O}}=0} = i\frac{Z_\mathrm{D}Z_\mathrm{F}}{Z_\mathrm{D}+Z_\mathrm{F}}
\label{eqn_superpos2}
\end{equation} 
Finally, the superposition for the voltage node $v_\mathrm{N}$ is the sum of eqs.~\protect\ref{eqn_superpos1} and \ref{eqn_superpos2}
\begin{equation}
v_\mathrm{N} = {\left.v_{\mathrm{N}}\right|}_{i=0}  + {\left.v_{\mathrm{N}}\right|}_{v_{\mathrm{O}}=0} \, .
\label{eqn_superpos}
\end{equation}
By inserting eqs.~\protect\ref{eqn_vn}, \ref{eqn_superpos1} and \ref{eqn_superpos2} into eq.~\protect\ref{eqn_superpos}, the output voltage $v_\mathrm{O}$ can be rewritten as
\begin{equation}
\label{eqn_superposition}
v_{\mathrm{O}} = -A\left( v_\mathrm{O}\frac{Z_\mathrm{D}}{Z_\mathrm{D} + Z_\mathrm{F}} + i\frac{Z_\mathrm{D}Z_\mathrm{F}}{Z_\mathrm{D}+Z_\mathrm{F}} \right) \, .
\end{equation}
The derived eq.~\protect\ref{eqn_superposition} is the same as eq.~\protect\ref{eqn_opa}, but shows the terms for the basic block structure shown in figure~\protect\ref{fig_csablock} in an intuitive and easily readable form.

The basic block structure is used to identify the voltage feedback and the stability of the loop. It is obvious that the feedback network $\beta$ is
\begin{equation}
\label{eqn_feedbackb}
\beta = \frac{Z_D}{Z_D+Z_F}
\end{equation}
and the closed-loop voltage gain $G$ is only~\protect\cite{hor}
\begin{equation}
\label{eqn_voltage_gain}
G = \frac{A}{1+A\beta} = -\frac{v_\mathrm{O}}{v_\mathrm{I}} \, .
\end{equation}
The closed-loop gain $G$ becomes $\frac{1}{\beta}$ if the loop gain $A\beta$ is much greater than one. On the contrary, the closed-loop gain becomes infinite if the loop gain $A\beta = -1$. At this frequency, the system tends to be unstable, and oscillates. As $A$ and $\beta$ are defined to have positive real values, the case where the loop gain becomes $-1$ only occurs if the loop gain is $1$ and a phase shift of $180^{\circ}$ is introduced by the frequency response of the $A\beta$. A phase shift of a sinusoidal signal with $180^{\circ}$ is equal to a multiplication with $-1$. To make the feedback circuit stable, the phase shift therefore has to be less than $180^{\circ}$ at the frequency $f_\mathrm{i}$, where the loop gain is $1$. A sufficient phase margin at the frequency $f_\mathrm{i}$ is required to ensure a stable operation over the entire temperature range, and also to cover tolerances of the used integrated circuits. 
\begin{figure}[ht]
\centering
\includegraphics[width=\textwidth]{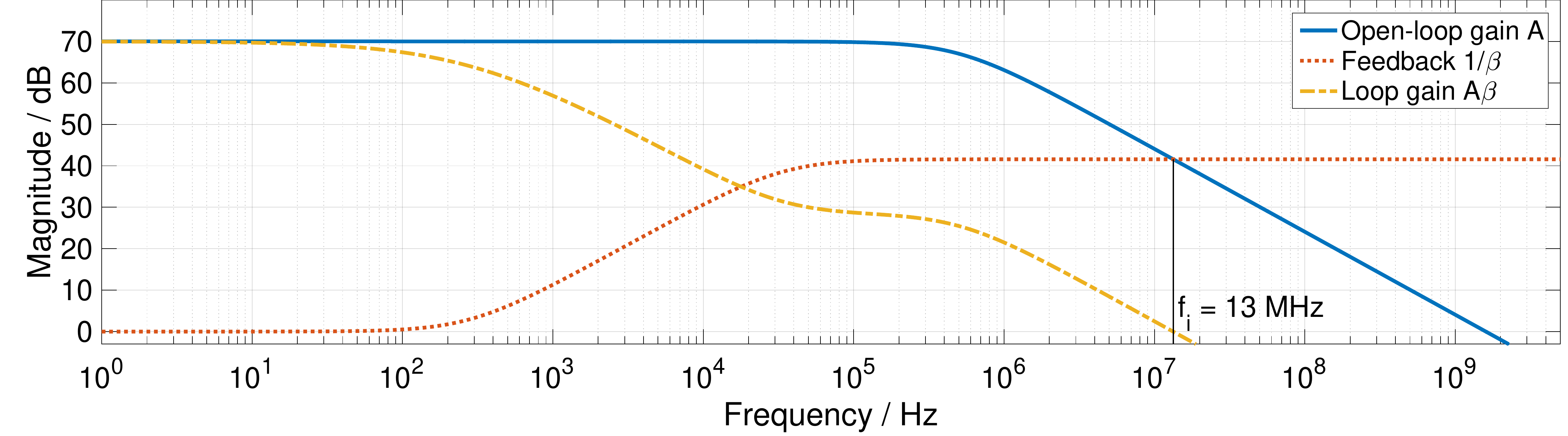}
\includegraphics[width=\textwidth]{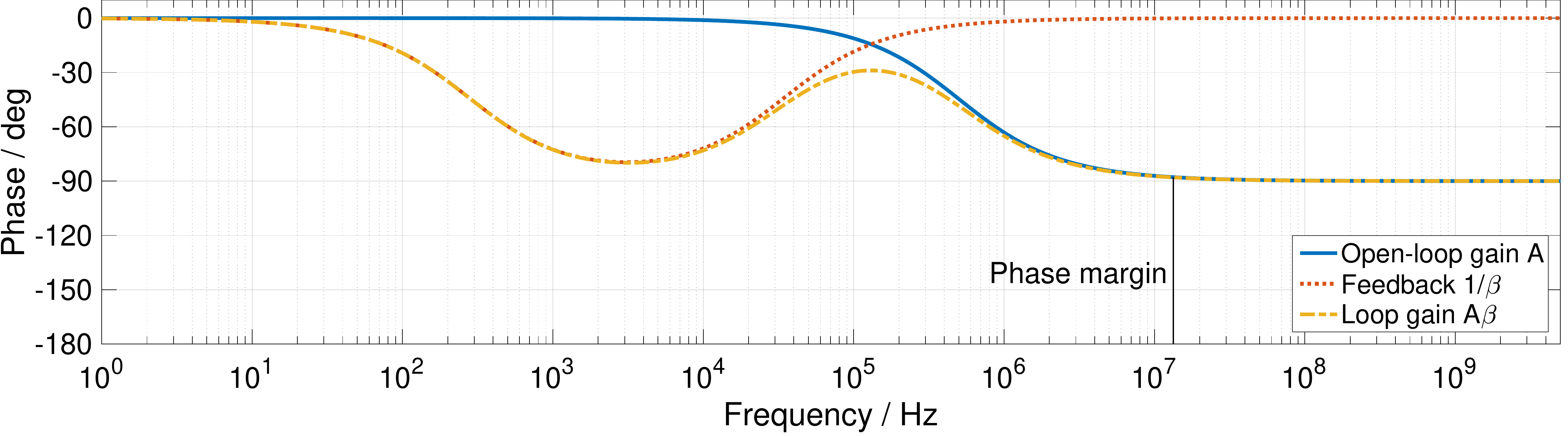}
\caption{The Bode plot of the closed-loop transfer function. The loop gain is 1 at the frequency $f_\mathrm{i}$ with a phase margin of roughly $90^{\circ}$. This is sufficient for a stable operation. See text for numerical values of the example.}
\label{fig_csastability}
\end{figure}
The example in figure~\protect\ref{fig_csastability} is calculated with the parameters from table~\protect\ref{tab_opas} for the OPA657. The feedback impedance was chosen to be $100\,\mathrm{fF}$ in parallel with $25\,\mathrm{M\Omega}$. Further, the detector capacitance of $6.7\,\mathrm{pF}$ in parallel with a $50\,\mathrm{G\Omega}$ resistor and the parasitic input capacitance of $5.2\,\mathrm{pF}$ are the shunting impedance $Z_\mathrm{D}$. The stability is investigated at the frequency, where the loop gain is 1. Mathematically, this intersection can be calculated by solving
\begin{equation}
\lvert A(\cj 2 \pi f) \beta(\cj 2 \pi f) \rvert = 1 \, .
\end{equation}
From the Bode plot shown in figure~\protect\ref{fig_csastability}, this point can be found at
\begin{equation}
\operatorname{log}(\lvert A \rvert) - \operatorname{log}(\lvert \frac{1}{\beta} \rvert ) = 0\,.
\end{equation}
In this plot, the reciprocal of the feedback network $\beta$ intersects the open-loop gain at approximately $13\,\mathrm{MHz}$. At this frequency, the phase shift of the loop gain is $88^{\circ}$, resulting in a phase margin $\phi$ of $92^{\circ}$, which is sufficient for a stable operation. The phase margin $\phi$ is calculated by
\begin{equation} 
\phi = 180^{\circ} - \left( \angle A(2 \pi f_i) + \angle \beta(2 \pi f_i) \right) \, .
\end{equation}

Besides the stability of the circuit, the effective input impedance also depends on the open-loop voltage gain of the amplifier, and is changed by the feedback network~\protect\cite{hor}. The effective input impedance of the CSA should be very low, to make sure that all current flows into the amplification circuit. Regarding figure~\protect\ref{fig_csa_opa}, the effective input impedance from the CSA is defined by the ratio of the voltage at the input terminal $i_\mathrm{In}$ and the input current $i$. The voltage at the input terminal is $v_\mathrm{N}$, so the effective input impedance $Z^{*}_\mathrm{I}$ is
\begin{equation}
Z^{*}_\mathrm{I} = \frac{v_\mathrm{N}}{i} \, .
\end{equation}
Solving eq.~\protect\ref{eqn_kirchhoff_sum} for $v_\mathrm{N}$ with eqs.~\protect\ref{eqn_izf}-\ref{eqn_vn}, $Z^{*}_\mathrm{I}$ can be expressed as
\begin{equation}
\label{eqn_zi}
Z^{*}_\mathrm{I} = \frac{Z_\mathrm{D} Z_\mathrm{F}}{Z_\mathrm{D} (A+1) + Z_\mathrm{F}} \, .
\end{equation}
As $A$ becomes infinitely large, as we assume for an ideal operational amplifier, the effective input impedance is zero. This satisfies the principle of the virtual ground. If we assume an ideal detector without a parasitic impedance from the CSA, $Z_\mathrm{D}$ is infinite. For this case, the effective input impedance
\begin{equation}
\lim_{Z_\mathrm{D} \to \infty}{Z^{*}_\mathrm{I}} = \frac{Z_\mathrm{F}}{A + 1}
\label{eqn_impedance_ideal}
\end{equation}
is determined by the open-loop gain and the feedback network $Z_\mathrm{F}$. To force the CSA, so that all current is integrated on the feedback capacitor, the impedance from eq.~\protect\ref{eqn_impedance_ideal} must be small compared to the shunting impedance $Z_\mathrm{D}$.

\subsection{Charge-to-voltage transfer function}
\begin{figure}[ht]
\centering
\includegraphics[width=0.48\textwidth]{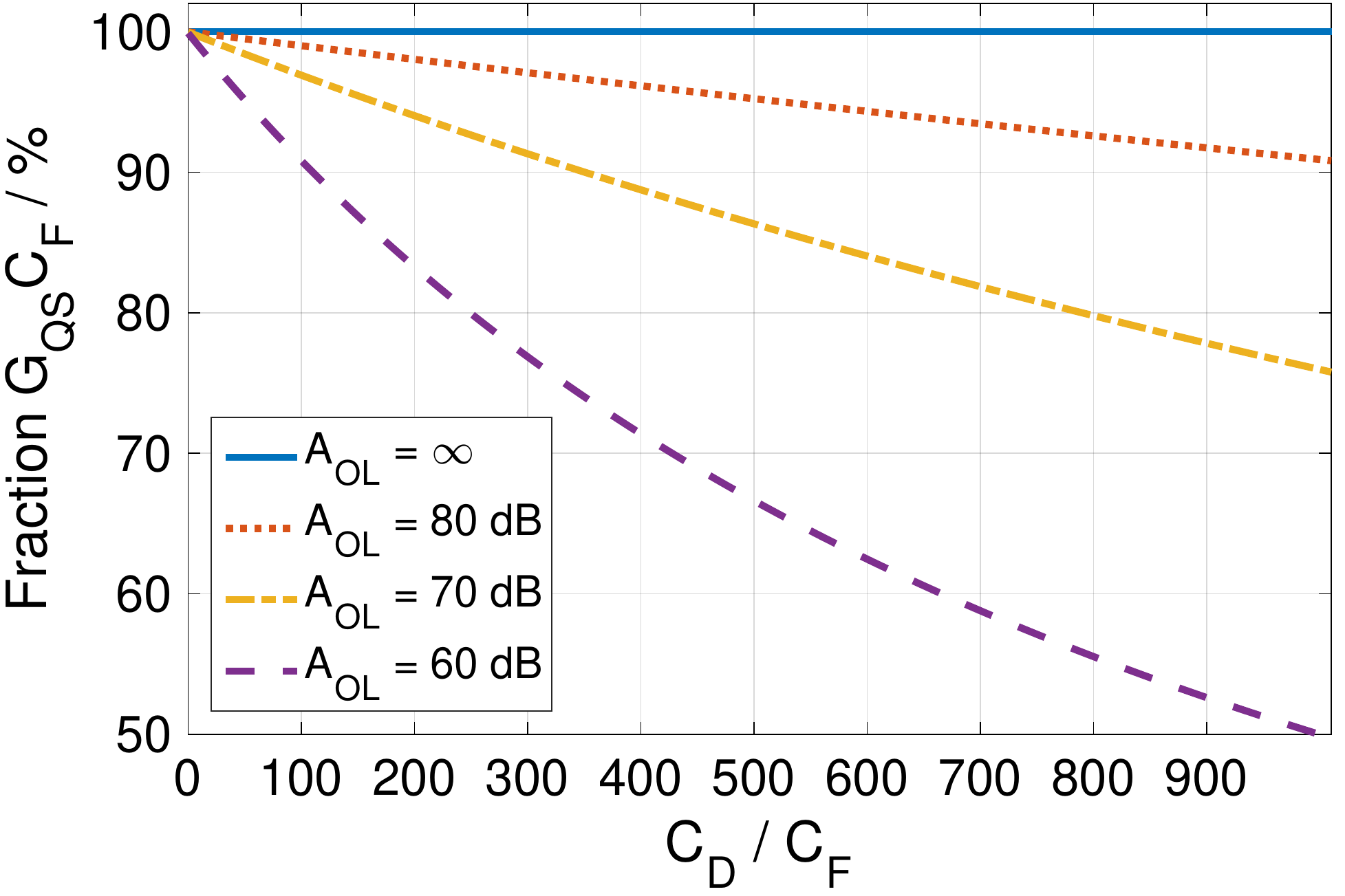}
\caption{The measured fraction of charge dependent on the ratio of input capacitance $C_\mathrm{D}$ over feedback capacitance $C_\mathrm{F}$. The fraction is increased with an increased zero-frequency open-loop voltage gain $A_\mathrm{OL}$ of the operational amplifier.}
\label{fig_charge_fraction}
\end{figure}

The CSA is designed to measure a charge with a voltage output signal, where the peak amplitude is proportional to the charge at the input. The relation of the output voltage $v_\mathrm{O}$ to the input charge $Q$ is the charge-to-voltage transfer function
\begin{equation}
\frac{v_\mathrm{O}}{Q} = H_\mathrm{Q} \, .
\label{eqn_chargetovoltage}
\end{equation}
As the charge is defined to be
\begin{equation}
\label{eqn_dqi}
\frac{dQ}{dt} = i\, ,
\end{equation}
where $i$ is the electrical current, the corresponding Laplace transformation of eq.~\protect\ref{eqn_dqi} is
\begin{equation}
\mathcal{L} \left\{ Q'(t) \right\} = s Q(s) = i(s) \, .
\label{eqn_q_laplace}
\end{equation}
If we replace $i$ in the current-to-voltage transfer function from eq.~\protect\ref{eqn_opa} by eq.~\protect\ref{eqn_q_laplace} and set the feedback network $Z_\mathrm{F} = \frac{1}{sC_\mathrm{F}}$ and parasitic impedance $Z_\mathrm{D} = \frac{1}{sC_\mathrm{D}}$ to single capacitors, then the charge-to-voltage transfer function is given by
\begin{equation}
H_Q = \frac{-A}{C_\mathrm{D}+C_\mathrm{F} (A+1) } \, .
\label{eqn_hq}
\end{equation}
The simplification of the impedances to single capacitors is valid, as we want to investigate the frequency response of the charge-to-voltage conversion. If the feedback time constant is chosen appropriately according to eq.~\protect\ref{eqn_ballistic_deficit}, there is no significant peak amplitude loss in the voltage signal. The peak amplitude of the voltage signal is therefore independent of the feedback resistor $R_\mathrm{F}$. The impedance $Z_\mathrm{D}$ can also be simplified for this analysis, as the equivalent input resistance of the amplifier and detector is much larger than the effective input impedance according to eq.~\protect\ref{eqn_zi}. For an ideal operational amplifier with infinite open-loop gain A, $H_\mathrm{Q}$ from eq.~\protect\ref{eqn_hq} becomes $\frac{-1}{C_\mathrm{F}}$ over the entire frequency domain. Since the open-loop voltage gain of an operational amplifier is not independent of frequency, a more realistic charge-to-voltage transfer function is obtained by replacing $A$ from eq.~\protect\ref{eqn_hq} by eq.~\protect\ref{eqn_As}
\begin{equation}
H_Q(s) = \frac{-A_\mathrm{OL}}{C_\mathrm{D} + C_\mathrm{F}(A_\mathrm{OL} + 1) + s\tau(C_\mathrm{D} + C_\mathrm{F})} \, .
\label{eqn_hqs}
\end{equation}
The charge-to-voltage gain is then given by
\begin{equation}
\label{eqn_hq_freq}
|H_Q(\cj 2 \pi f)| = G_\mathrm{Q}(f) =  \frac{A_\mathrm{OL}}{\sqrt{{\left(C_\mathrm{D} + C_\mathrm{F}(A_\mathrm{OL}+1)\right)}^2 + {\left(\frac{f}{f_{\mathrm{GBP}}}A_\mathrm{OL}(C_\mathrm{D}+C_\mathrm{F})\right)}^2}} \, .
\end{equation}
The steady state of the system is derived by calculating the zero-frequency gain:
\begin{equation}
G_\mathrm{QS} = \lim_{f \to 0}{G_\mathrm{Q}(f)} = \frac{A_\mathrm{OL}}{C_\mathrm{D} + C_\mathrm{F}(A_\mathrm{OL}+1)} \label{eqn_steadystate} \, .
\end{equation}
Eq.~\protect\ref{eqn_steadystate} shows that a finite open-loop gain $A_\mathrm{OL}$ attenuates the measured peak voltage. The measured fraction of charge as a ratio of $G_\mathrm{QS}$ over the ideal charge-to-voltage gain $\frac{1}{C_\mathrm{F}}$ is given by
\begin{equation}
G_\mathrm{QS} C_\mathrm{F} = \frac{A_\mathrm{OL}}{\frac{C_\mathrm{D}}{C_\mathrm{F}} + A_\mathrm{OL}+1}\label{eqn_chargefraction} \, .
\end{equation}
It is obvious that the measured peak amplitude is decreased by an increased detector capacitance~$C_\mathrm{D}$ or a reduced open-loop voltage gain $A_\mathrm{OL}$. Thus, the operational amplifier should provide a high and stable open-loop voltage gain for an improved system performance as illustrated in fig~\ref{fig_charge_fraction}.

\begin{figure}[ht]
\centering
\includegraphics[width=0.48\textwidth]{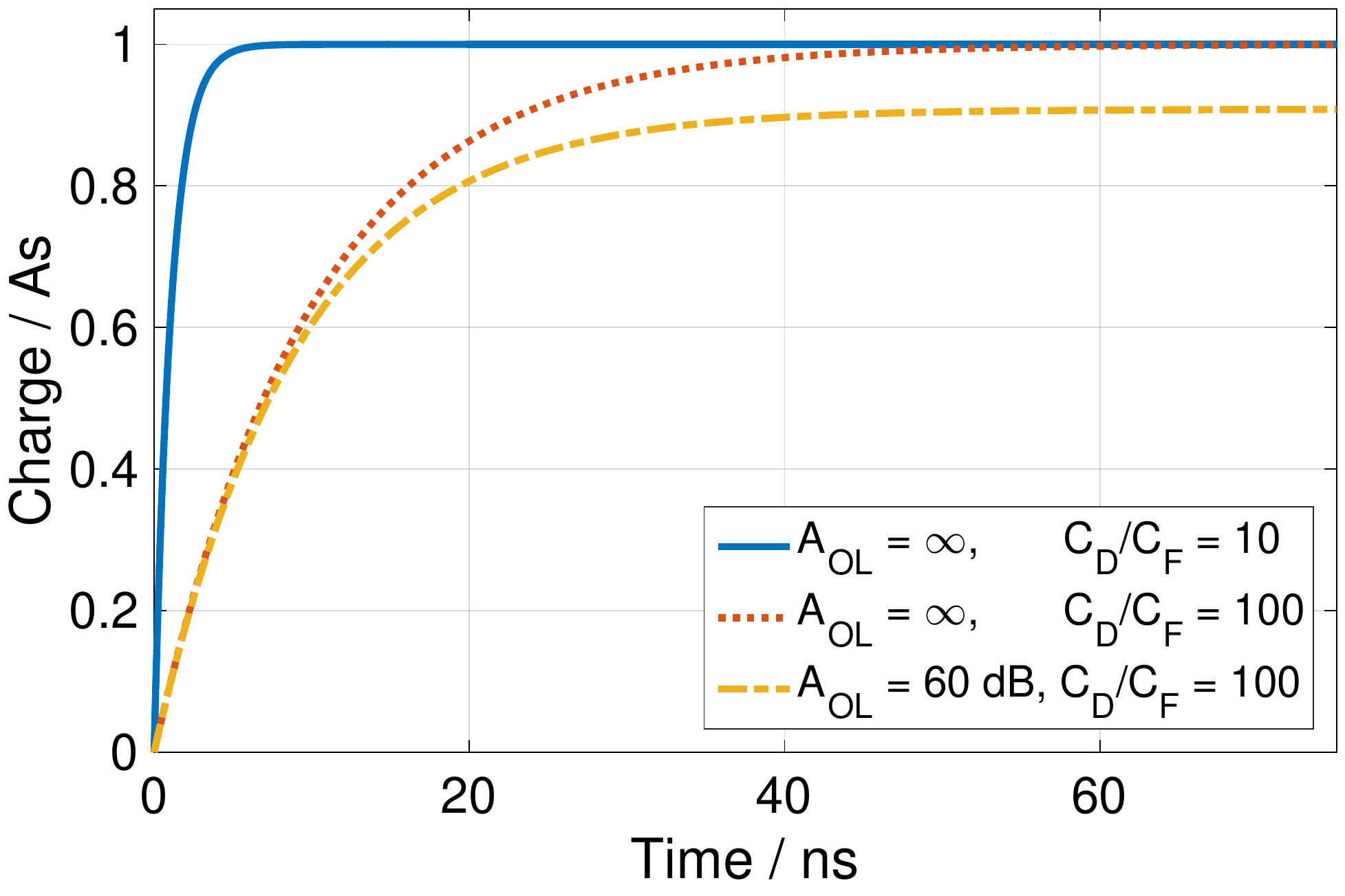}
\caption{The output signal of the charge-sensitive amplifier with a unity step at its input. The rise time is increased with an increasing ratio of $\frac{C_\mathrm{D}}{C_\mathrm{F}}$. The gain-bandwidth product of the operational amplifier was assumed to be $f_\mathrm{GBP}=1\,\mathrm{GHz}$. In this example, the rise times are $17\,\mathrm{ns}$ (solid line), $32\,\mathrm{ns}$ (dashed line) and $35\,\mathrm{ns}$ (dotted line). The decreased rise time related to a smaller open-loop gain $A_\mathrm{OL}$ is caused by the attenuation of the peak amplitude.}
\label{fig_csastepresponse}
\end{figure}

An equally important parameter of the CSA is the rise time of the output voltage as a reaction of a charge step at its input. As shown by eq.~\protect\ref{eqn_risetime}, the rise time is proportional to the cutoff frequency of the system. Therefore, to determine the rise time, we have to calculate its cutoff frequency, which is derived using eq.~\protect\ref{eqn_hq_freq}
\begin{eqnarray}
\frac{|H_\mathrm{Q}(\cj 2 \pi f_\mathrm{c})|}{|H_\mathrm{Q}(0)|} &=& \frac{1}{\sqrt{2}}\\
f_\mathrm{c}                               &=&  f_{\mathrm{GBP}} \frac{\frac{C_\mathrm{D}}{C_\mathrm{F}} + (A_\mathrm{OL}+1)}{A_\mathrm{OL}(\frac{C_\mathrm{D}}{C_\mathrm{F}} + 1)}\label{eqn_csabandwith} \, .
\end{eqnarray}
Eq.~\protect\ref{eqn_csabandwith} shows, that the upper bandwidth limit is lowered with an increased fraction of the shunting capacitance $C_\mathrm{D}$ over $C_\mathrm{F}$. The bandwidth is extended with a lower value of $A_\mathrm{OL}$, but this will reduce the measured fraction of charge, as shown in figure~\protect\ref{fig_charge_fraction}. This also means that the signal-to-noise ratio is decreased and therefore, the resolution of the charge measurement is also decreased. $A_\mathrm{OL}$ should therefore be as large as possible. For this assumption, the bandwidth of the CSA is
\begin{equation}
\lim_{A_\mathrm{OL} \to \infty}{f_\mathrm{c}} = \frac{f_{\mathrm{GBP}}}{\frac{C_\mathrm{D}}{C_\mathrm{F}} + 1} \, .
\label{eqn_csa_risetime}
\end{equation}
Eq.~\protect\ref{eqn_csa_risetime} shows that the cutoff frequency and therefore the rise time of the CSA are directly proportional to the gain-bandwidth product of the amplifier. The response of the transfer function from eq.~\protect\ref{eqn_hq} to a unit step is shown in figure~\protect\ref{fig_csastepresponse}.

\subsection{Input coupling of the CSA}
\begin{figure}[ht]
\centering
\includegraphics[width=0.6\textwidth]{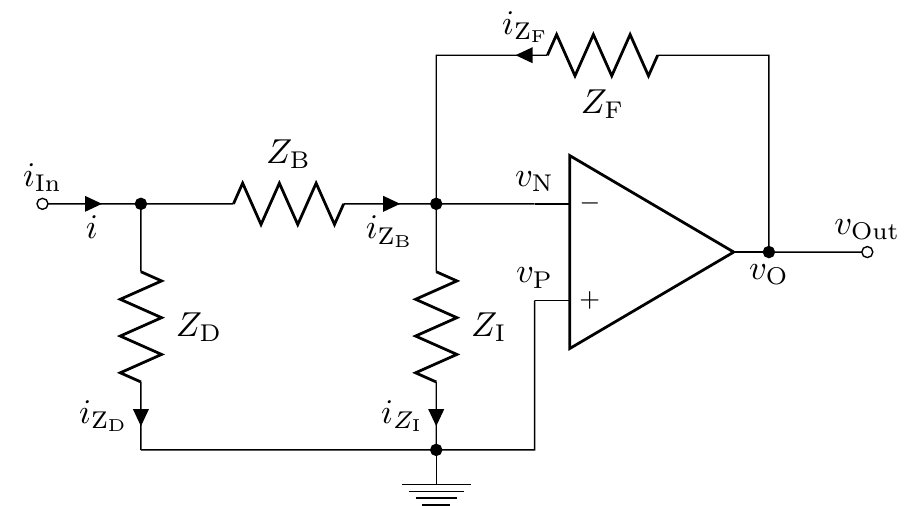}
\caption{A functionally equivalent configuration to figure~\protect\ref{fig_csa_opa} for the charge-to-voltage amplification. The input impedance at the terminal $i_\mathrm{In}$ is a Pi network instead of the single impedance $Z_\mathrm{D}$. $Z_\mathrm{B}$ is a high-voltage coupling capacitor to protect the low-voltage input terminals of the operational amplifier.}
\label{fig_blockingc}
\end{figure}

As shown in figure~\protect\ref{fig_hv_detector_readout}, at least one of the electrodes is biased at a negative high voltage. Thus, the amplifier at the cathode must be protected against the bias voltage, since it is not common for integrated circuits to operate at high input voltages in the range of several hundreds of volts. A capacitor in series to the amplifier input blocks the bias voltage of the detector, but allows the detector current to flow. This electrical circuit is shown in figure~\protect\ref{fig_blockingc}, where the coupling capacitor $C_\mathrm{B}$ is represented by the impedance $Z_\mathrm{B}$. In this configuration, the parasitic impedance $Z_\mathrm{I}$ and the detector impedance $Z_\mathrm{D}$ are separated by the impedance $Z_\mathrm{B}$. 
As the influence of the coupling capacitor is not apparent at first glance throughout the equation for the current-to-voltage transfer function, a calculation with an infinite loop gain turns out the simplified equation
\begin{equation}
\frac{-v_\mathrm{O}}{i} = \frac{Z_\mathrm{D} Z_\mathrm{F}}{Z_\mathrm{B} + Z_\mathrm{D}} \, .
\end{equation}
To eliminate the influence of the impedance $Z_\mathrm{B}$, it must be much smaller than $Z_\mathrm{D}$. Where all impedances are represented by a single capacitor, the steady-state charge-to-voltage gain is ultimately given by
\begin{equation}
\label{eqn_steadystateblocking}
\frac{v_\mathrm{O}}{Q} = \frac{1}{C_\mathrm{F} \left( 1 + \frac{C_\mathrm{D}}{C_\mathrm{B}} \right)} \, .
\end{equation}
Eq.~\protect\ref{eqn_steadystateblocking} shows that the coupling capacitor must be much larger than the detector capacitance to avoid a peak amplitude loss, thus reducing the signal-to-noise ratio.

\subsection{Noise}
\begin{figure}[ht]
\centering
\includegraphics[width=0.6\textwidth]{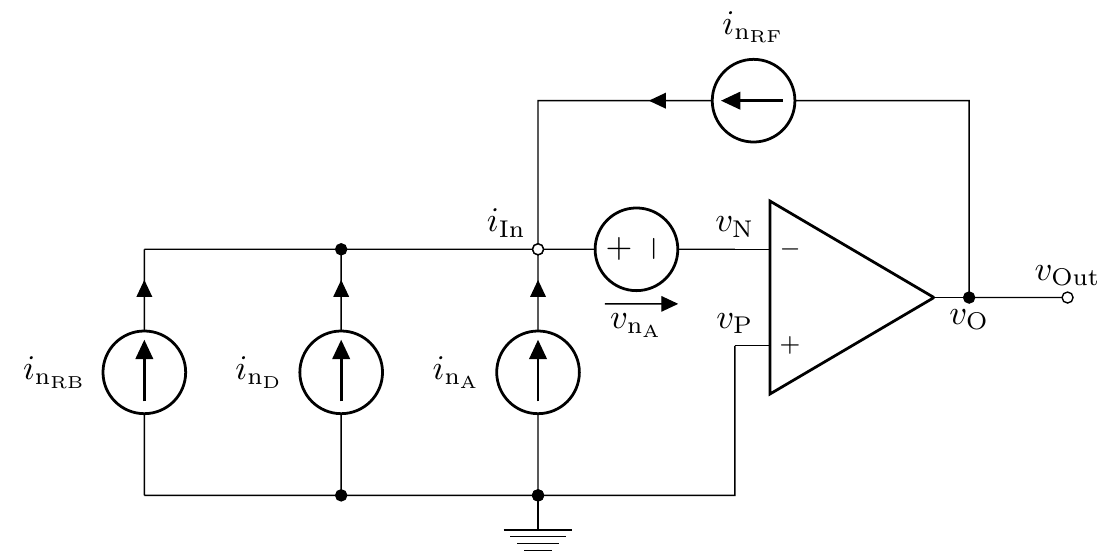}
\caption{The charge-sensitive amplifier configuration with its dominant noise sources at the input terminal $i_\mathrm{In}$. For the noise analysis, the current noise sources from the bias resistor ($i_\mathrm{n_{RB}}$), equivalent resistor of the detector ($i_\mathrm{n_{D}}$), feedback resistor ($i_\mathrm{n_{RF}}$), and amplifier ($i_\mathrm{n_{A}}$) can be substituted by a single source, as they act in parallel. Finally, the noise analysis is made with a current noise source $i_\mathrm{n}$ in parallel (parallel noise) and a voltage noise source $v_\mathrm{n}$ in series (series noise) to the input terminal.}
\label{fig_opa_noise}
\end{figure}

As has been pointed out, the noise caused by the electronics and the corresponding signal-to-noise ratio (SNR) characterize the quality of the CSA with respect to the achievable energy resolution. The precision of the charge measurement and the achievable timing are directly related to the SNR. Regarding both values, the front-end electronics should not limit the intrinsic resolution of the detector. Thus, to reduce the electronics noise, the noise sources of the detector system must be identified in the first step. The main source for the electronics noise is the operational amplifier. But the passive components of an electronic circuit also generate noise. The resulting noise at the output is the sum of all noise sources at the input, amplified by the noise gain. The essential noise sources of the electronics circuit are shown in figure~\protect\ref{fig_opa_noise}.
The noise contribution of the operational amplifier is simplified to a model, where its noise is characterized by an equivalent voltage noise source $v_\mathrm{n_\mathrm{A}}$ and current noise source $i_\mathrm{n_\mathrm{A}}$ at the inverting input terminal. Additionally, all resistors in the system contribute to the total noise by adding thermal noise. The equivalent noise current of a resistor at temperature $T$ is given by
\begin{equation}
i_\mathrm{n_\mathrm{R}} = \frac{4kT} {R}
\label{eqn_resistor_noise}
\end{equation}
where k is the Boltzmann constant \protect\cite{spieler}. The equivalent resistor $R_\mathrm{D}$ of the detector, the biasing resistor $R_\mathrm{B}$, and the equivalent resistor $R_\mathrm{F}$ of the feedback impedance contribute to the total noise following eq.~\protect\ref{eqn_resistor_noise}. Each is represented by a current source in figure~\protect\ref{fig_opa_noise}. As there are multiple noise sources in the system, they can all be absorbed into a single source for the voltage noise $v_\mathrm{n}$ and a single source for the current noise $i_\mathrm{n}$ with
\begin{equation}
i_\mathrm{n} = i_\mathrm{n_\mathrm{RD}} + i_\mathrm{n_\mathrm{RB}} + i_\mathrm{n_\mathrm{RF}} + i_\mathrm{n_\mathrm{A}}\, ,
\label{eqn_in_sum}
\end{equation}
as they are connected in parallel. Voltage noise sources can be summed, as they are connected in series. As we assume that all noise sources have a flat frequency spectrum (white noise), the resulting noise spectral density at the output is shaped by the noise transfer function (noise gain).

The noise spectral density is usually expressed in units of $\frac{\mathrm{nV}}{\sqrt{\mathrm{Hz}}}$. With regard to the block diagram of the closed-loop transfer function from figure~\ref{fig_csablock}, the current noise source is amplified by the current-to-voltage transfer function, whereas the voltage noise source is amplified by the closed-loop voltage gain.
\begin{figure}[ht]
\centering
\includegraphics[width=0.8\textwidth]{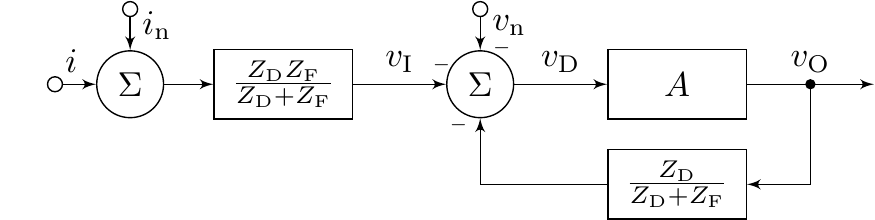}
\caption{Block diagram of the closed-loop transfer function with additional current noise source $i_\mathrm{n}$ and voltage noise source $v_\mathrm{n}$. Both sources act at the negative input terminal of the operational amplifier but have different transfer functions.}
\label{fig_opablocknoise}
\end{figure} 
As illustrated in figure~\protect\ref{fig_opablocknoise}, the input noise current $i_\mathrm{n}$ to output noise voltage $v_\mathrm{On}$ transfer function $G_\mathrm{in}$ is given by eq.~\protect\ref{eqn_opa} and the input noise voltage $v_\mathrm{n}$ to $v_\mathrm{On}$ transfer function $G_\mathrm{vn}$ is given by eq.~\protect\ref{eqn_voltage_gain}.

\begin{figure}[ht]
\centering
\includegraphics[width=0.58\textwidth]{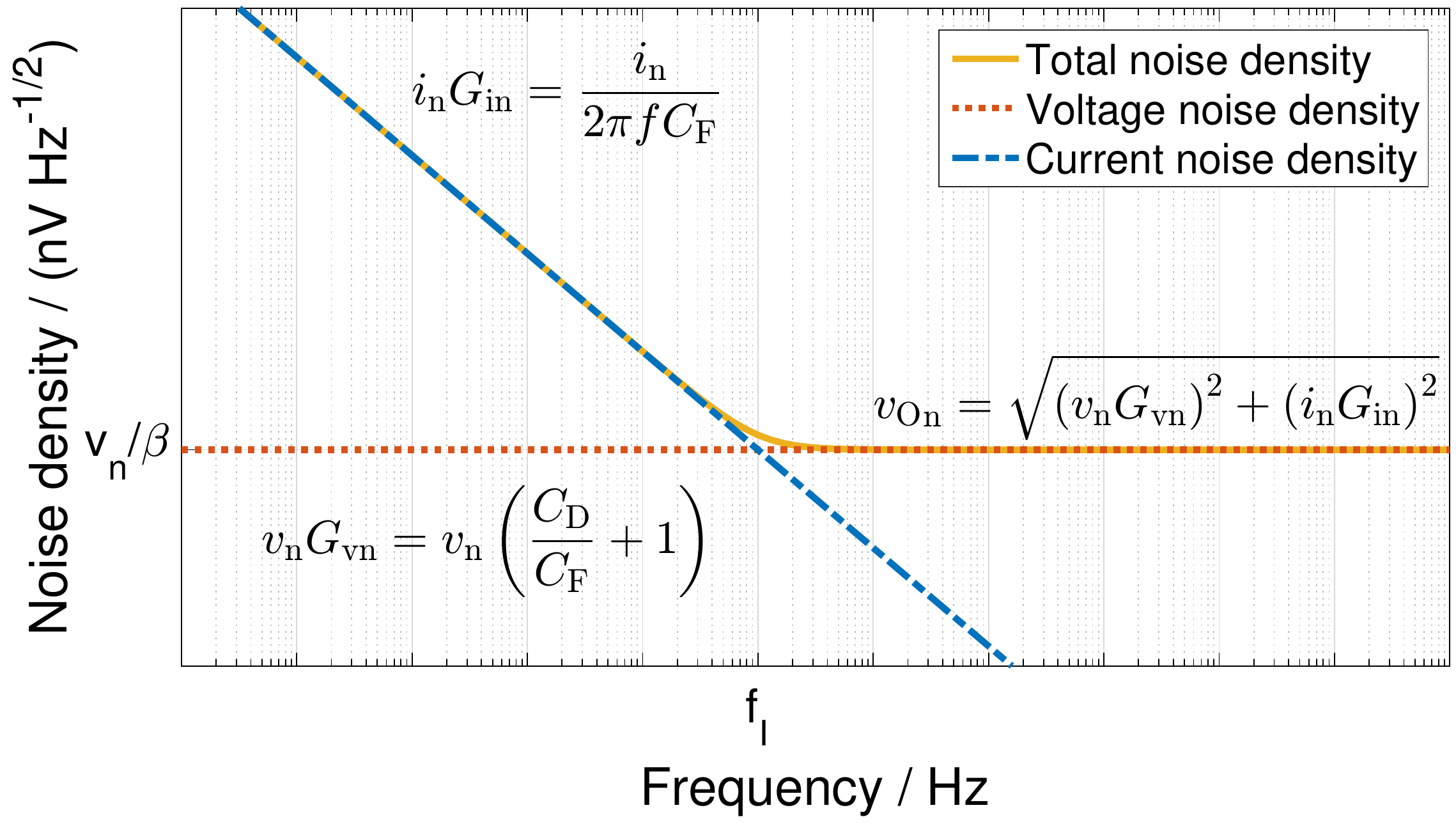}
\caption{The noise spectral density log-log plot for the ideal charge-sensitive amplifier with the capacitor $C_\mathrm{F}$ in the feedback network and the capacitor $C_\mathrm{D}$ at its input terminal. In the lower frequency range, the noise density is dominated by the current noise. Above the frequency $f_\mathrm{I}$, the portion of the voltage noise dominates the total noise density ${v_\mathrm{O}}_\mathrm{n}$.}
\label{fig_noise_density}
\end{figure}
\begin{figure}[ht]
\centering
\includegraphics[width=0.58\textwidth]{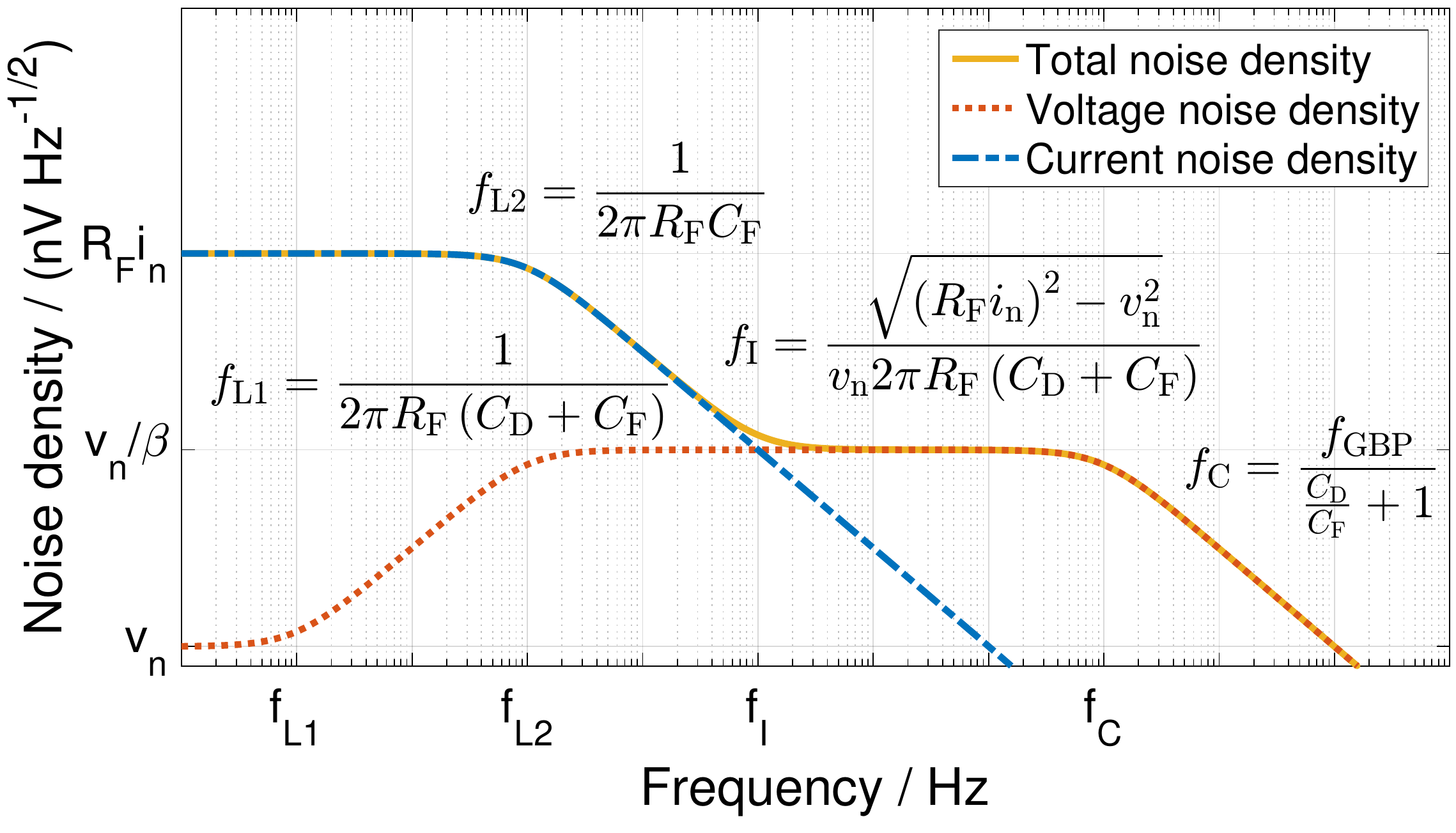}
\caption{The noise spectral density log-log plot for the charge-sensitive amplifier with the resistor $R_\mathrm{F}$ and capacitor $C_\mathrm{F}$ in the feedback network and the capacitor $C_\mathrm{D}$ at its input terminal. The frequency response of the current noise density has the shape of a low-pass filter, whereas the voltage noise density has the response of a band-pass filter. The time constant $\tau = R_\mathrm{F} C_\mathrm{F}$ determines the low cutoff frequency $f_\mathrm{L2}$ for both responses. The voltage noise density is limited by the corner frequency $f_\mathrm{c}$. Thus, the total noise density is bounded over the whole frequency range.}
\label{fig_noise_density_r}
\end{figure}

For a clear view on the components of the resulting spectral noise density in figure~\protect\ref{fig_noise_density}, the calculations are based on the ideal model of the CSA. Figure~\protect\ref{fig_noise_density} shows the contribution of the input current noise, which has a typical $\frac{1}{f}$ shape and dominates the low frequency range (referred to as flicker noise). The noise density in the upper frequency range is dominated by the input voltage noise but remains flat. If the open-loop gain is sufficiently large, the voltage noise is amplified by the factor $\frac{1}{\beta}$. A more realistic noise spectral density is shown in figure~\protect\ref{fig_noise_density_r}.
This illustration emphasizes the impact of the feedback resistor $R_\mathrm{F}$ and the gain-bandwidth product $f_\mathrm{GBP}$. The density of the flicker noise is limited in its upper value. It has the shape of a typical low-pass filter, which is determined by the time constant $R_\mathrm{F}C_\mathrm{F}$ of the feedback impedance. The component related to the voltage noise has the spectral density of a band-pass filter, where the upper-corner frequency $f_\mathrm{c}$ is limited by the gain-bandwidth product and the detector capacitance $C_\mathrm{D}$. At zero frequency, the voltage noise is limited to the value of $v_\mathrm{n}$.

The noise spectral density determines the root-mean-square amplitude of the output noise (rms~noise) over a given bandwidth. As the noise spectral density is bounded over the entire frequency range, the total rms~noise $V_\mathrm{rms}$ is calculated by
\begin{equation}
V_\mathrm{rms} = \sqrt{\int_{0}^{\infty}{{\left( v_\mathrm{n} G_\mathrm{vn} \right)}^2 + {\left( i_\mathrm{n} G_\mathrm{in} \right)}^2 df}}
\end{equation}
An additional signal processing with filters (pulse shapers) must be adapted in accordance to the noise spectral density.

\section{Implementation}
\begin{figure}[ht]
\centering
\includegraphics[width=0.45\textwidth]{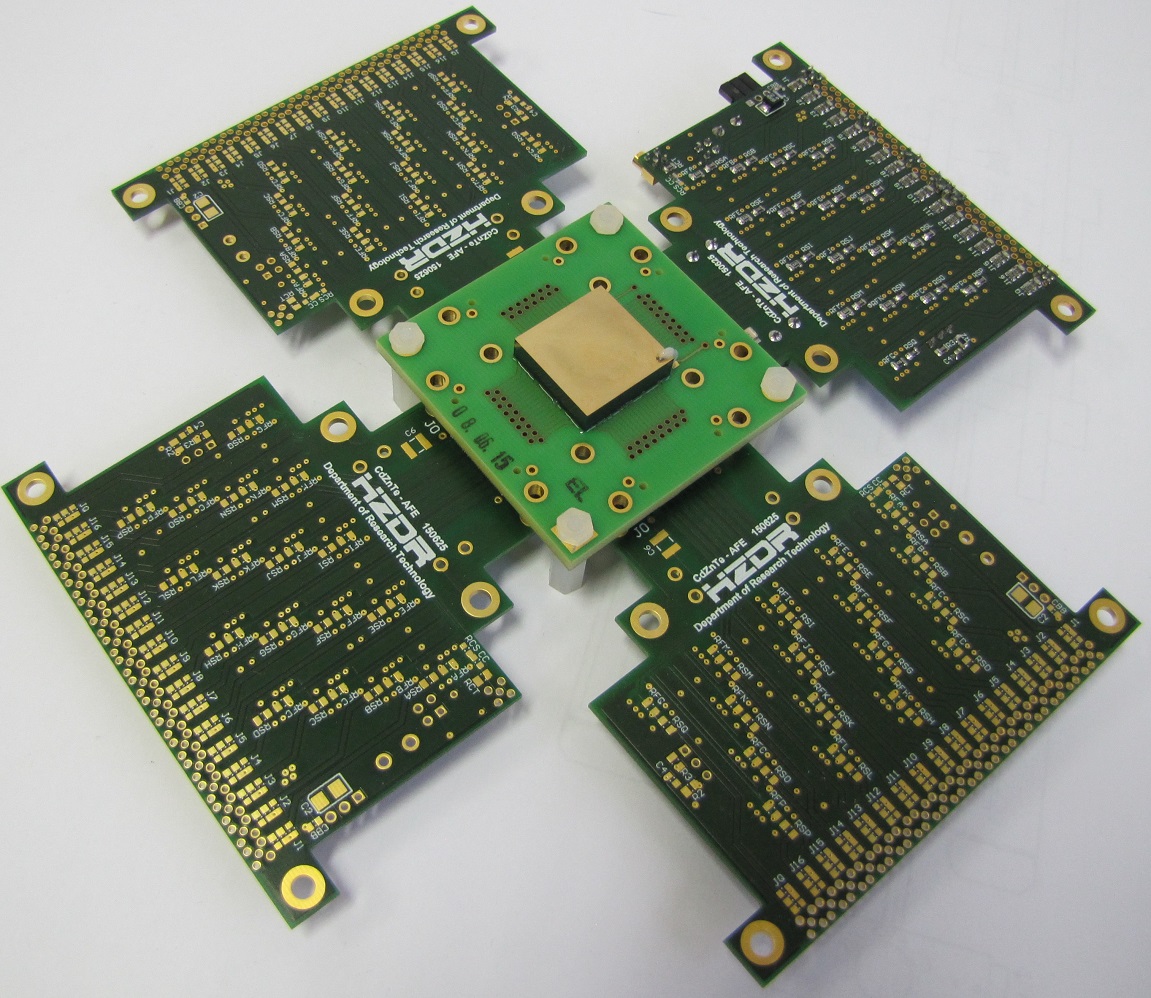}
\includegraphics[width=0.49\textwidth]{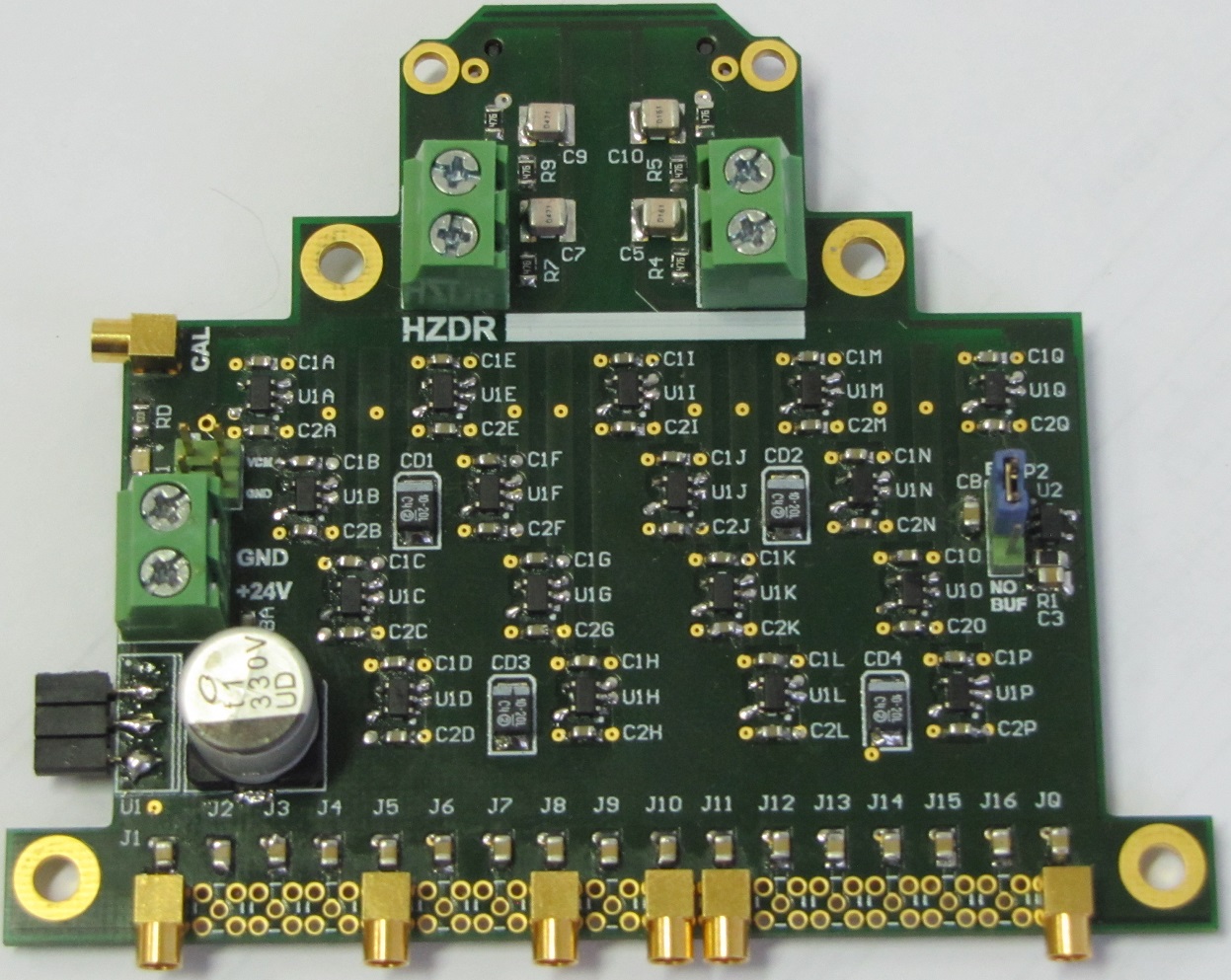}
\caption{The detector assembly with unpopulated readout boards (left) and a close-up of a populated printed circuit board with 17 charge-sensitive amplifiers (right).}
\label{fig_hardware}
\end{figure}

A major design guideline was the reduction of the total amount of components per readout channel. This could be achieved using a single operational amplifier to build the CSA without additional gain stages. Nevertheless, this circuit covers the desired measurement range and does not need any pulse shapers for its basic operation. In order to fulfill the requirements, the detector system must process energies up to $7\,\mathrm{MeV}$ and should be compatible with a digitizer system with an input voltage range of $2\,\mathrm{V}$. According to eq. \ref{eqn_moving_charge} and the steady-state charge-to-voltage gain of the CSA from eq.~\protect\ref{eqn_steadystate}, the feedback capacitance $C_\mathrm{F}$ should be at least $120\,\mathrm{fF}$. A higher gain is acceptable, since an attenuation of the pulse amplitude can be made with less effort. Since the feedback capacitance and the electrical characteristics of the detector are fixed design parameters, the operational amplifier and the feedback resistance $R_\mathrm{F}$ are the remaining components for an optimization of the readout circuit. As a first step, the drift time $t_\mathrm{D}$ of the moving charge must be taken into account to choose an appropriate value of the time constant $\tau$ of the CSA. With regard to the induced currents through the anode and cathode shown in figure~\protect\ref{fig_current}, we expect rise times up to $450\,\mathrm{ns}$ for the cathode and up to $200\,\mathrm{ns}$ for the anode. Because the anode signals carry the information used for spectroscopy, the ballistic deficit from eq.~\protect\ref{eqn_ballistic_deficit} should be minimized. A value of $1\,\%$ is appropriate. Therefore, the ratio $\frac{\tau}{t_\mathrm{D}}$ must be greater than $50$, which corresponds to a value of $10\,\mathrm{\mu{s}}$ for the time constant $\tau$ or $83\,\mathrm{M\Omega}$ for the feedback resistor $R_\mathrm{F}$ (the nearest matching part has $82\,\mathrm{M\Omega}$). With a loose constraint of $5\,\%$ ($\frac{\tau}{t_\mathrm{D}} = 9.66$) for the ballistic deficit on the cathode signal, we selected a feedback resistor of $47\,\mathrm{M\Omega}$ for that readout channel. For the implementation of the CSA with a COTS operational amplifier, we establish four important features for the parametric selection. First, the operational amplifier must have a large input impedance. Thus, the voltage noise gain of the feedback network ($~\frac{1}{\beta}$) is reduced. Second, it must have a high open-loop voltage gain, so that the effective input impedance regarding eq.~\protect\ref{eqn_zi} is minimized. The highest input impedance is achieved by operational amplifiers with a field-effect transistor at their input terminals. The third feature is that the equivalent voltage and current noise must be very low, and the fourth is that the operational amplifiers must have a sufficient large gain-bandwidth product to satisfy the requirements of rise time. For our application, the achievable timing of the detector pulses is important, and must be further investigated. For an advanced analysis, the rise time should not be limited (large gain-bandwidth product). In table~\protect\ref{tab_opas}, we list some COTS operational amplifiers from different vendors, which match the selection criteria.
\begin{table}
\centering
\caption{A list of suitable commercial off-the-shelf operational amplifiers from different vendors. They are compared in terms of our selection criteria: gain-bandwidth product ($f_\mathrm{GBP}$), zero-frequency open-loop gain ($A_\mathrm{OL}$) and input impedance ($Z_\mathrm{I}$). $v_\mathrm{n}$ is the equivalent voltage noise source and $i_\mathrm{n}$ the current noise source. All values are extracted from the datasheets~\protect\cite{ada4817,ltc6268,ltc626810,opa657}.}
\label{tab_opas}
\begin{tabular}{|l|l|l|l|l|l|l|}
\hline
{Vendor}            & {Product}    & {$f_\mathrm{GBP}$\,/\,MHz} & {$A_\mathrm{OL}$\,/\,dB}  & {$Z_{\mathrm{I}}\,/\,\Omega\,||\,\mathrm{pF}$} & {$v_\mathrm{n} / \frac{\mathrm{nV}}{\sqrt{\mathrm{Hz}}}$} & {$i_\mathrm{n}  / \frac{\mathrm{fA}}{\sqrt{\mathrm{Hz}}}$}\\
\hline
{Analog Devices}    & {ADA4817}    & {1050}        & {65}                           & {$  0.5\,\mathrm{T}\,||\,1.4$}   & {4}  &  {2.5}\\
{Linear Technology} & {LTC6268}    & {500}         & {108}                          & {$  0.5\,\mathrm{T}\,||\,0.55$}  & {4.3}&  {5.5}\\
{Linear Technology} & {LTC6268-10} & {4000}        & {108}                          & {$  0.5\,\mathrm{T}\,||\,0.55$}  & {4.0}&  {7.0}\\
{Texas Instruments} & {OPA657}     & {1600}        & {70}                           & {$  0.5\,\mathrm{T}\,||\,5.2$}   & {4.8}&  {1.3}\\
\hline
\end{tabular}
\end{table}
All operational amplifiers fulfill the requirements of a high input impedance and a relatively large gain-bandwidth product. For the first tests, we choose the OPA657 because of its larger open-loop gain and bandwidth in comparison to the ADA4817. For the second part, we choose the LTC6268-10 because of its outstanding parameters and very low parasitic capacitance. Both operational amplifiers are available in an almost pin compatible package. Consequently, the evaluation procedure can be done with the same hardware. After a first attempt with the LTC6268, which runs on the OPA657 PCB layout, we decided to populate the hardware with the LTC6268-10; unfortunately, this circuit tends to oscillate, as also observed in~\protect\cite{ltjournal}. Thus the implementation of the CSA with the OPA657 shows the best performance for the first prototype (see figure~\protect\ref{fig_hardware}).

Despite the higher parasitic input capacitance, a great advantage of the OPA657 over the LTC6268 is its wide supply voltage range of $12\,\mathrm{V}$. This provides a larger headroom for pulse pile-ups until the output voltage of the amplifier saturates. Thus, this amplifier is best-suited for high count rates with high energies. Nevertheless, the count rate capability mainly depends on the signal processing system, which is limited by the input voltage range of the digitizer and the digital pulse processing system \cite{abbene_high_rate}. Moreover, any rate limitations must be determined by field experiments. The readout board contains 17 CSAs, where one anode channel is equipped with a test pulse input and another channel is dedicated to the readout of the cathode. All necessary functionality, including low voltage power supply, high voltage filters, biasing, and decoupling capacitors, is included on the PCB. Each CSA has a passive low-pass filter at its output. The bandwidth is limited to $21.654\,\mathrm{MHz}$ ($16.15\,\mathrm{ns}$ rise time), which is sufficient for the CZT detector signals. One channel of the readout board is used for the evaluation of the CSA and contains a test input circuit. This circuit consists of a termination resistor and a series capacitor of $0.3\,\mathrm{pF} \pm 0.05\,\mathrm{pF}$ for the charge injection in accordance with~\protect\cite{knoll}. The theoretical performance parameters of the CSA are listed in table~\protect\ref{tab_performance}. These values are estimates and will be verified with experimental results.
\begin{table}
\centering
\caption{Numerical estimation of the electrical characteristics of the different readout channels. The calculations refer to a CSA based on the OPA657. The value of the equivalent noise charge (ENC) is given in units of the elementary charge $\mathrm{e}$.}
\begin{tabular}{|c|c|c|c|}
\hline
{} & {Cathode} & {Anode (Pixel)} & {Test input}\\
\hline
{$R_\mathrm{F}$\,||\,$C_\mathrm{F}$}              &{$47\,\mathrm{M\Omega}$\,||\,$100\,\mathrm{fF}$} & {$82\,\mathrm{M\Omega}$\,||\,$100\,\mathrm{fF}$} & {$82\,\mathrm{M\Omega}$\,||\,$100\,\mathrm{fF}$}\\
{$\frac{C_\mathrm{D}}{C_\mathrm{F}}$}             &{119}                                            & {53.05}                                          & {55}\\
{Ballistic deficit}                               &{$95.36\,\%$}                                    & {$98.79\,\%$}                                    & {$99.9\,\%$}\\
{Charge fraction}                                 &{$96.34\,\%$}                                    & {$98.32\,\%$}                                    & {$98.26\,\%$}\\
{Steady-state gain}                               &{$9.19\,\frac{\mathrm{V}}{\mathrm{pC}}$}         & {$9.71\,\frac{\mathrm{V}}{\mathrm{pC}}$}         & {$9.82\,\frac{\mathrm{V}}{\mathrm{pC}}$}\\
{Cutoff frequency}                                &{$13.84\,\mathrm{MHz}$}                          & {$30.11\,\mathrm{MHz}$}                          & {$29.08\,\mathrm{MHz}$}\\
{Rise time}                                       &{$25.27\,\mathrm{ns}$}                           & {$11.61\,\mathrm{ns}$}                           & {$12.03\,\mathrm{ns}$}\\
{Noise level (rms)}                               &{$2.62\,\mathrm{mV}$}                            & {$1.82\,\mathrm{mV}$}                            & {$1.80\,\mathrm{mV}$}\\
{Noise level (rms), BW limited}                   &{$2.01\,\mathrm{mV}$}                            & {$1.26\,\mathrm{mV}$}                            & {$1.21\,\mathrm{mV}$}\\
{ENC (rms), BW limited}                           &{$1365\,\mathrm{e}$}                             & {$810\,\mathrm{e}$}                              & {$769\,\mathrm{e}$}\\
{Peak amplitude at $511\,\mathrm{keV}$ in CZT}     &{$162.11\,\mathrm{mV}$}                          & {$171.38\,\mathrm{mV}$}                          & {$173.33\,\mathrm{mV}$}\\
\hline
\end{tabular}
\label{tab_performance}
\end{table}

\section{Results}
The readout board is evaluated with the test pulse input and with the detector shown in figure~\protect\ref{fig_detector}. The test pulse input is sourced by a signal generator with a step voltage input. According to the current-voltage relation of a capacitor, the step voltage applied to test input capacitor generates a current flowing into the CSA. With a varying shape of the voltage signal, arbitrary detector signals can be synthesized.
\subsection{Test pulse input}
\begin{figure}[ht]
\centering
\includegraphics[width=0.49\textwidth]{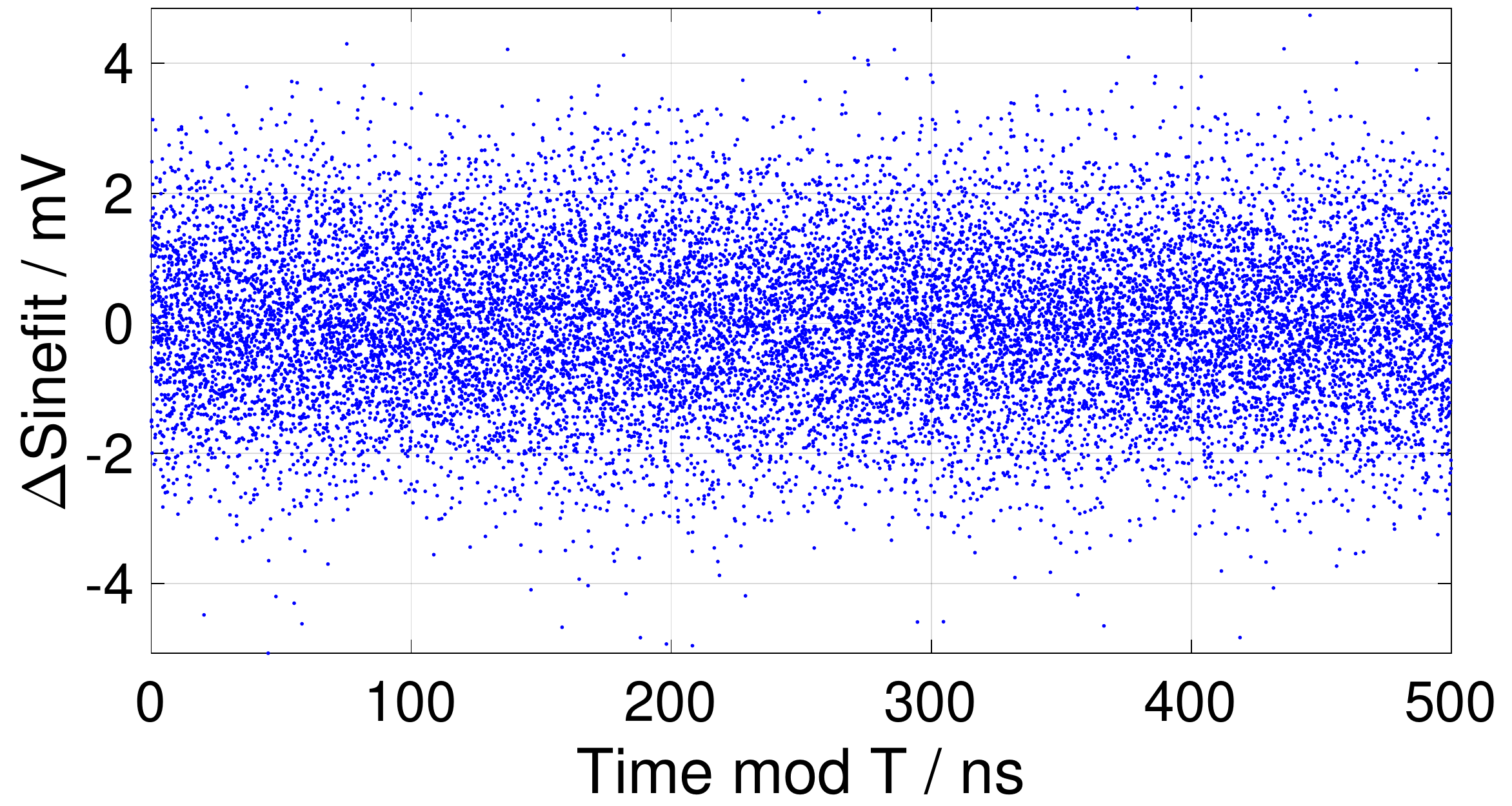}
\includegraphics[width=0.49\textwidth]{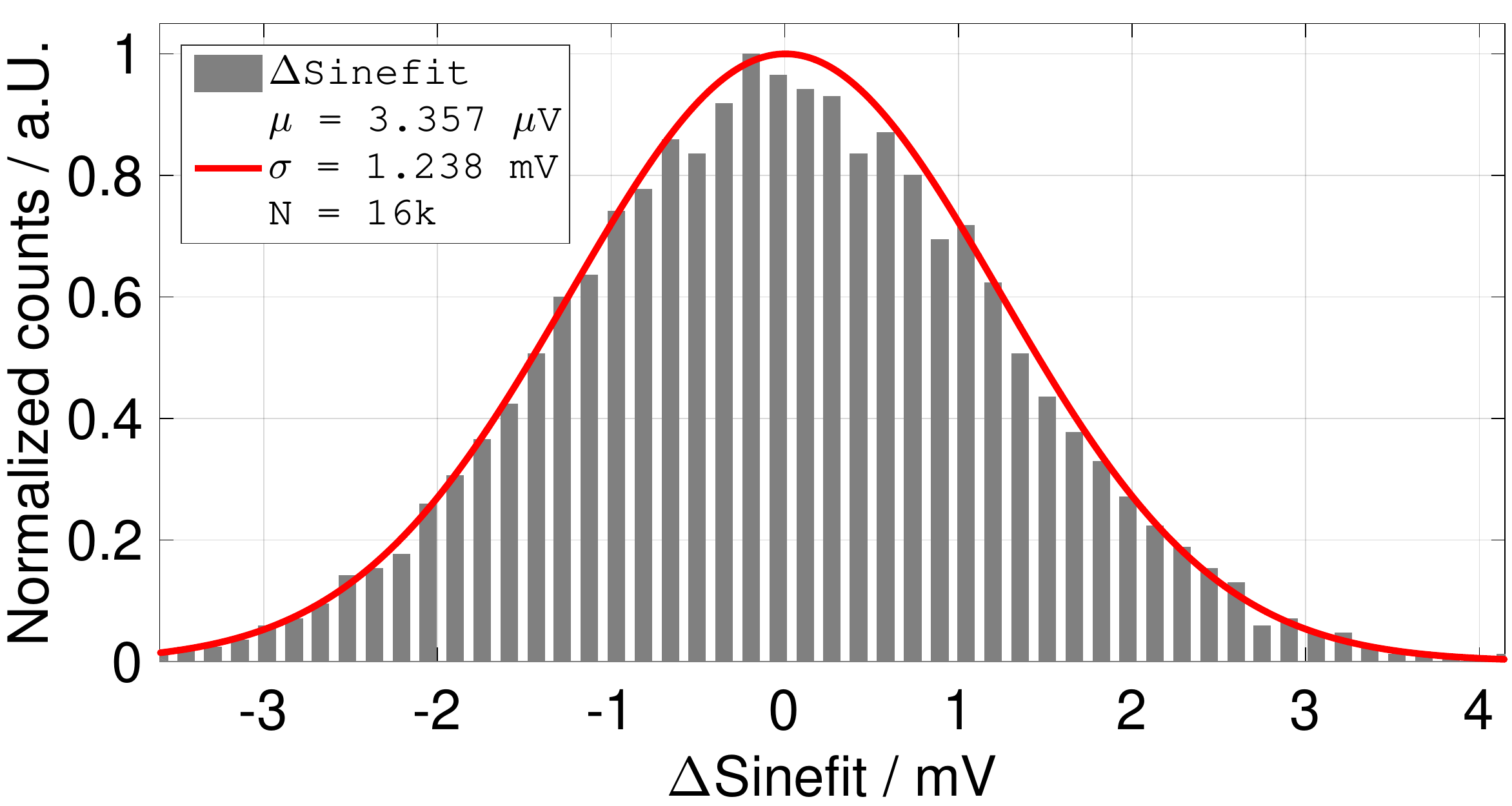}
\caption{Noise measurement of the charge-sensitive amplifier with a sine-wave test. A $2\,\mathrm{MHz}$ sine-wave was applied to the test input and the difference between the output signal and a four-parameter sine-wave fit ($\mathrm{{\Delta}Sinefit}$) is plotted against the period $T$ of the sine-wave (left). All harmonic distortions were eliminated by the test procedure, revealing the noise level (standard deviation $\sigma$ of $\mathrm{{\Delta}Sinefit}$, right).}
\label{fig_sinewavetest}
\end{figure}

The most important feature of the front-end electronics is the noise performance. There are various methods to analyze the noise of a linear time-invariant (LTI) system like the CSA. A common method is described in the IEEE Std 1241-201, referred to as "Sine-wave testing and fitting"~\protect\cite{ieee_adctest}. It is known that the response of an LTI system to a pure sine-wave is a sine-wave with the same frequency, but potentially different amplitude and phase. Tests with a sine-wave have the advantage that the waveform can be generated very accurately and the interpretation is done by standardized instruments and tools. If a sine-wave is applied to the input, the output depends on the transfer function of the system and is superimposed with noise. The input sine-wave is derived a posteriori by a four-parameter sine-wave fit, and the residual is the noise level, as described in~\protect\cite{ieee_adctest}. The waveform is captured by our digitizer with $100\,\mathrm{MSPS}$ and $14\,\mathrm{bit}$ resolution. The digitizer board provides an SNR of $71.3\,\mathrm{dB}$ at around $-1\,\mathrm{dB}$ of full scale input ($2.3\,\mathrm{V_{pp}}$). This corresponds to an rms~noise~level of $201\,\mathrm{{\mu}V}$. These values are measured at an input frequency of $2\,\mathrm{MHz}$. With the same setup, the sine-wave test with the test input of the CSA results in an rms noise level of $1.24\,\mathrm{mV}$. The result is shown in figure~\protect\ref{fig_sinewavetest}.
This is in accordance with the predicted values in table~\protect\ref{tab_performance} for the test input. 

\begin{figure}[ht]
\centering
\includegraphics[width=0.49\textwidth]{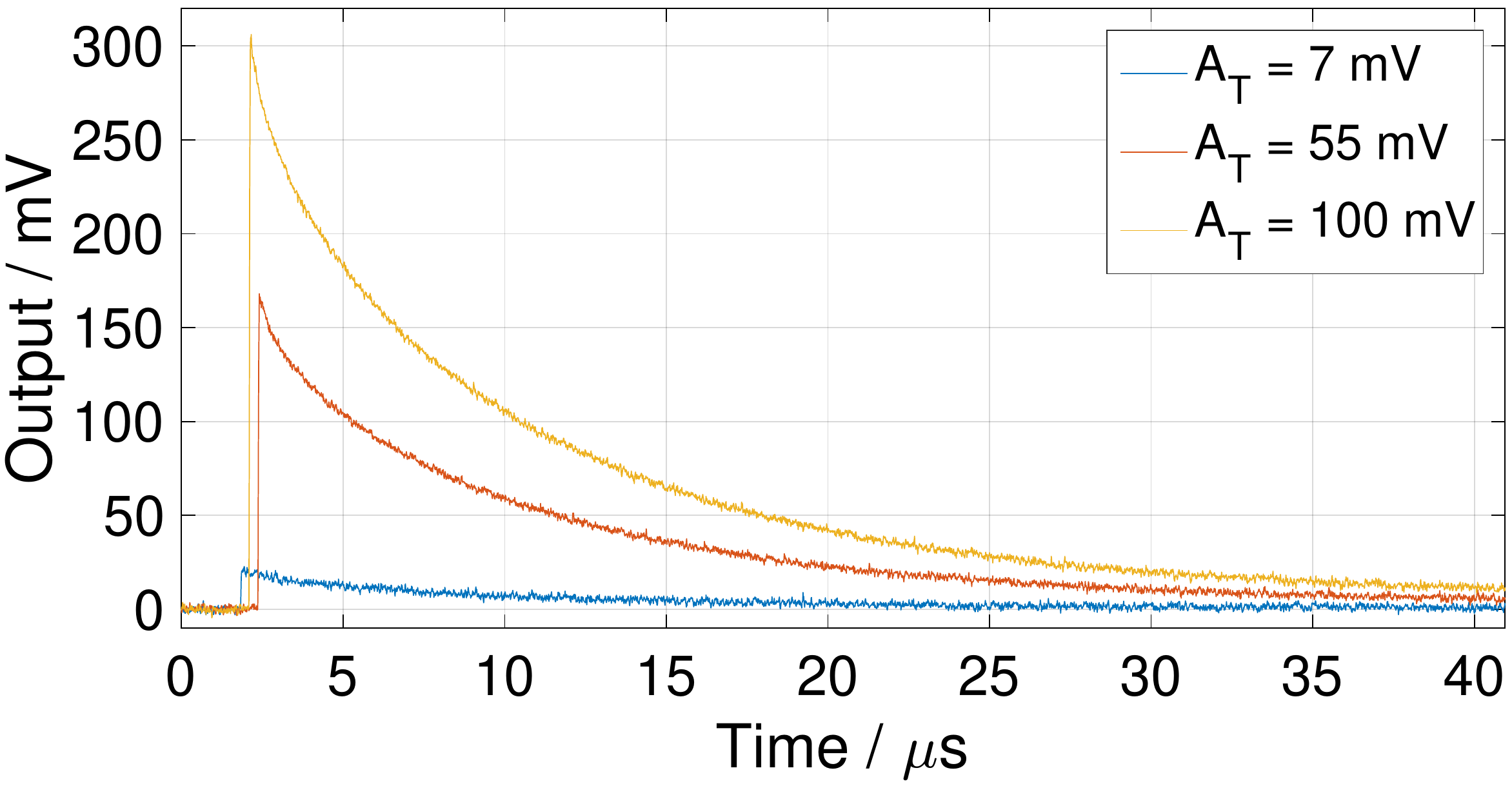}
\includegraphics[width=0.49\textwidth]{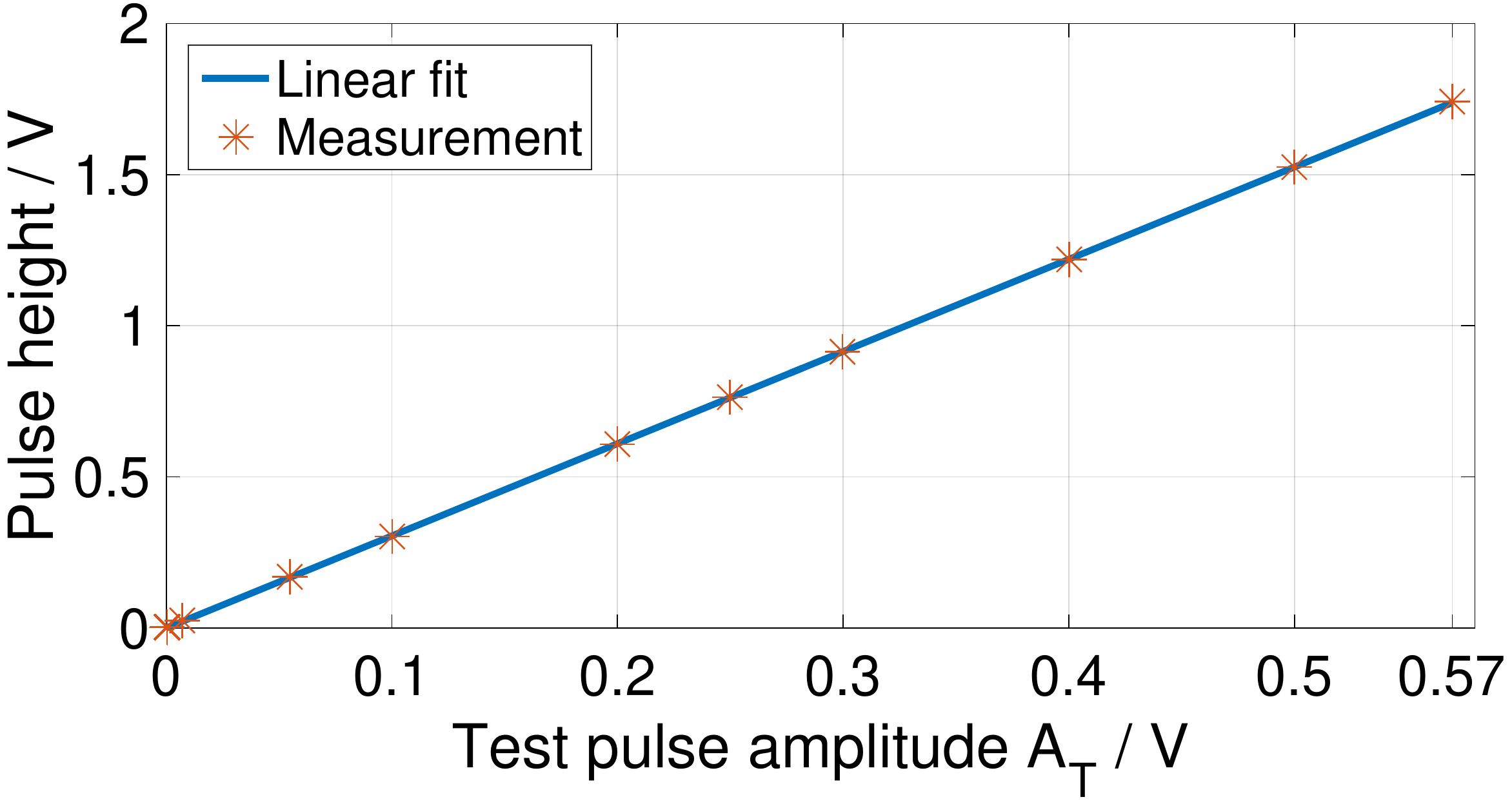}
\caption{A measurement with different rectangular pulses with amplitudes $A_\mathrm{T}$ injected to the test input capacitance. The output signal was recorded with the $100\,\mathrm{MSPS}$ digitizer (left). Each pulse height is the rms~value of $1\,k$ events (right). The linearity error is in the range from $-2\,\mathrm{mV}$ to $1.5\,\mathrm{mV}$.}
\label{fig_pulses}
\end{figure}

Moreover, other tests were made to validate the pulse shape at the output of the CSA. The signal generator was set up to generate a step input with a rise time of approximately $8\,\mathrm{ns}$. Figure~\protect\ref{fig_pulses} shows the output signals for a variation of the input amplitude $A_\mathrm{T}$.
The fit of the peak heights and the input shows an excellent linearity, with a maximum deviation of $2\,\mathrm{mV}$. The linear fit results in a gain of $3.05$, which correlates with the ratio of test input capacitance over feedback capacitance.

\begin{figure}[ht]
\centering
\includegraphics[width=0.49\textwidth]{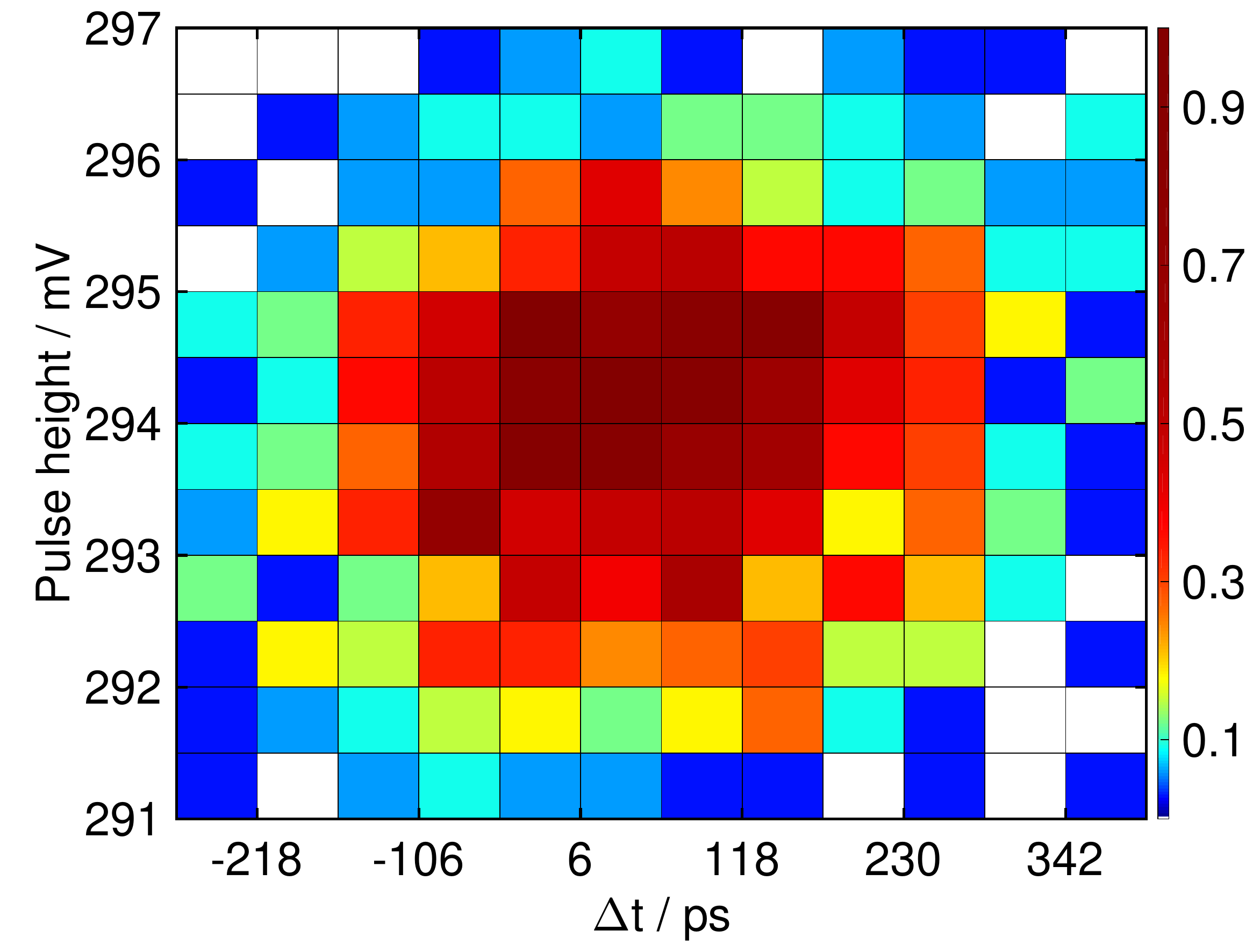}
\includegraphics[width=0.49\textwidth]{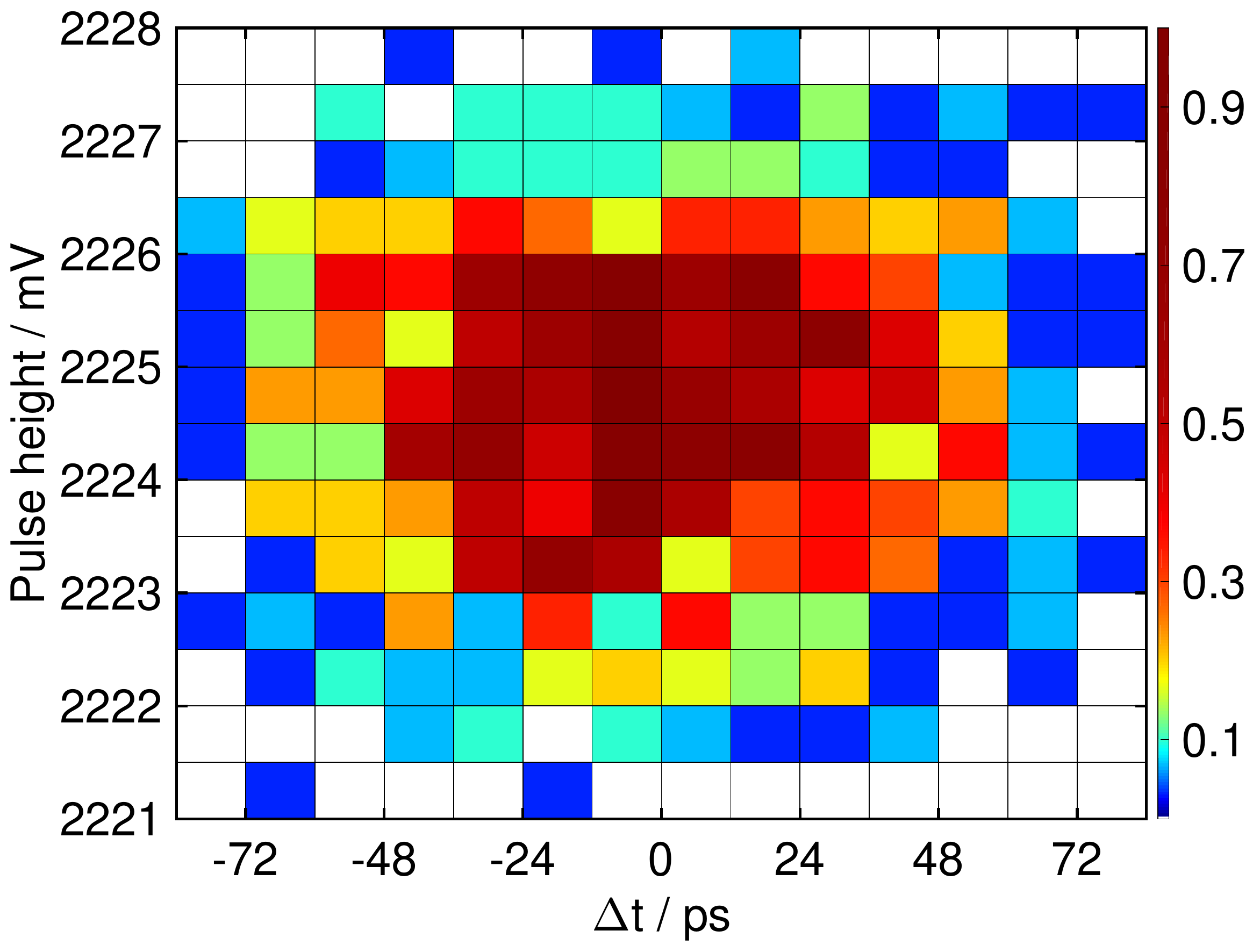}
\caption{Results of about $1\,k$ timing measurements with the $100\,\mathrm{MSPS}$ digitizer. The pulse height is the maximum value of the output from the charge-sensitive amplifier. The time $\Delta{t}$ is the difference between the pulse trigger and the timing reference from the signal generator. The standard deviation of the measurements are $\sigma_x = 131\,\mathrm{ps}$ and $\sigma_y = 1.15\,\mathrm{mV}$ for the left plot and $\sigma_x = 32\,\mathrm{ps}$ and $\sigma_y = 1.21\,\mathrm{mV}$ for the right plot. The colormap is in logarithmic scale.}
\label{fig_timing_energy_testpulse}
\end{figure}

The test pulse input was also used to evaluate the timing capabilities of the CSA. As before, the signals were captured with the digitizer and processed offline. The test pulses were synchronized to a known timing reference signal (sine-wave signal). The timing performance was obtained from the difference between the zero crossing point of the sine-wave and a digital constant fraction trigger ($\mathrm{fraction} = 0.2$) on the output signal of the CSA. Both timestamps were calculated by software. The results of the timing measurement are shown in figure~\protect\ref{fig_timing_energy_testpulse}.
At high signal amplitudes, the timing performance is in the range of several picoseconds ($32\,\mathrm{ps}$ standard deviation). This value increases with lower signal amplitudes, because the SNR decreases as well. This rough estimate of timing performance shows that the CSA does not limit the timing performance of the CZT detector, which is estimated to be in a range of some nanoseconds~\protect\cite{meng}.

\subsection{Pixel detector} 
The performance of the CSA is finally evaluated with the Redlen pixel detector and the hardware shown in figure~\protect\ref{fig_hardware}. The detector is biased at $-600\,\mathrm{V}$ and the signals are captured with the digitizer. The waveforms of a single pixel and the cathode are stored for an offline analysis. The first test should measure the energy on both electrodes dependent on the depth of interaction (DOI). As shown in figure~\protect\ref{fig_weighting_potentials}, the relationship between the DOI and the weighting potentials should be experimentally verified with the data from the detector. The DOI is measured either by the ratio of cathode-over-anode energy \protect\cite{wli} or by a direct measurement of the drift time of the moving charge. The drift time of charge correlates with the rise time of the CSA \protect\cite{verger}. For our analysis, we calculate the rise time as the time difference between two thresholds of the rising edge of the signal. On both the cathode and the anode signal, we set the thresholds to $10\,\%$ and $90\,\%$ of the peak amplitude. The data was collected with a ${}^{22}\mathrm{Na}$ radioactive source and an arbitrary selected pixel of the $2\,\times\,2$ center array. The anode was used to trigger an event while the cathode signal was captured at the same time. 
\begin{figure}[ht]
\centering
\includegraphics[width=0.49\textwidth]{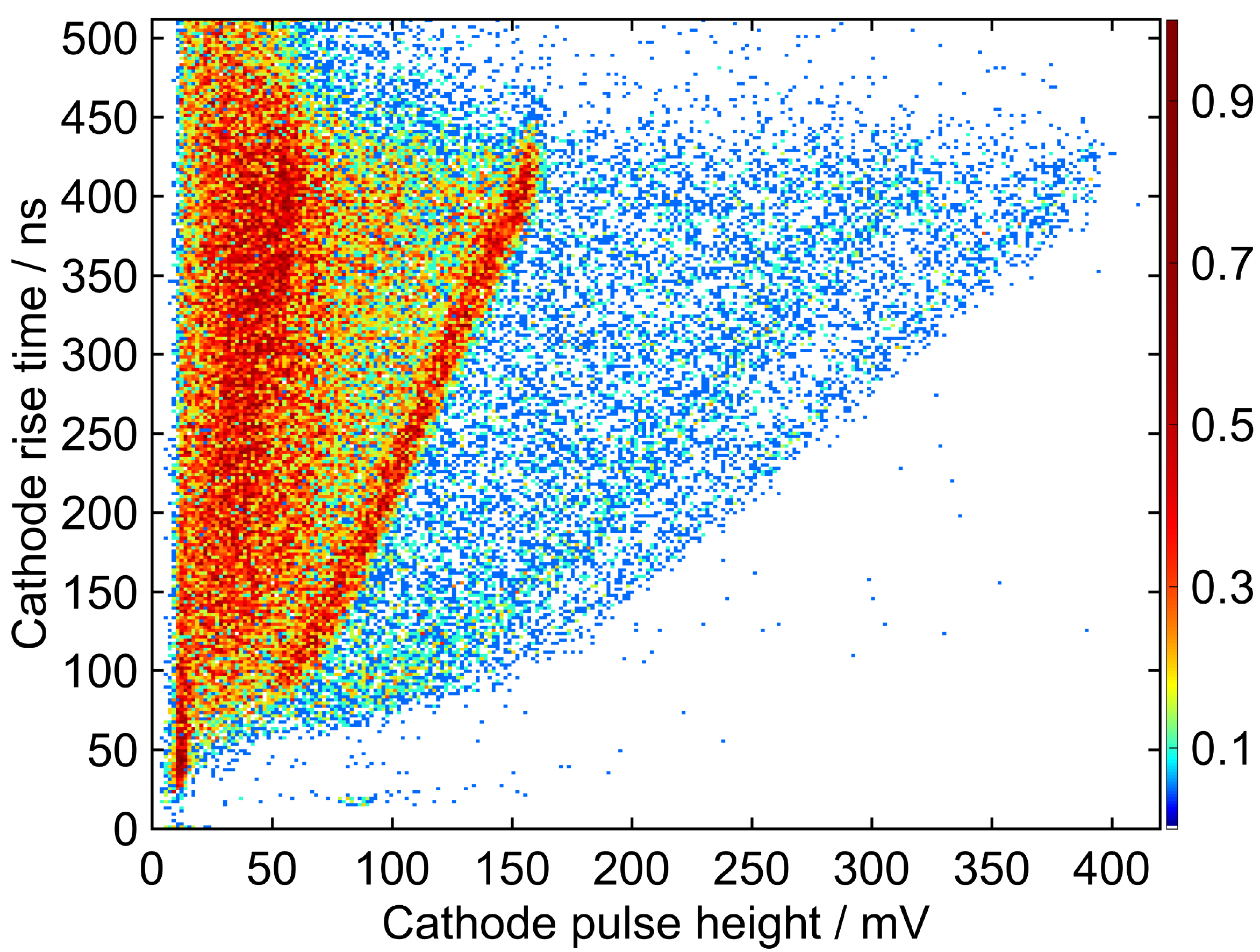}
\includegraphics[width=0.49\textwidth]{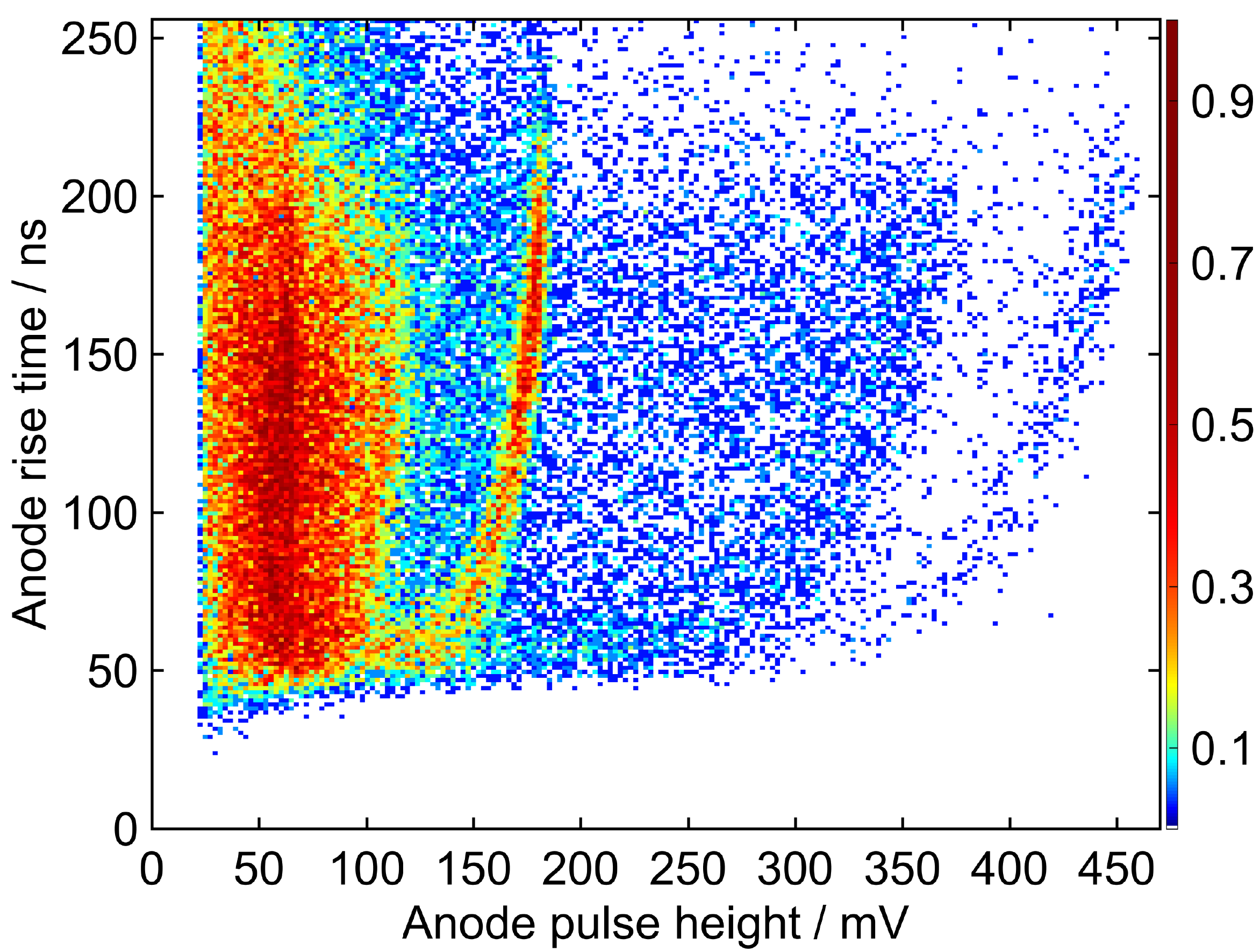}
\caption{Measurement of a ${}^{22}\mathrm{Na}$ radioactive source with the Redlen CZT detector and the charge-sensitive amplifier (CSA) readout board. The peak amplitudes of the output pulses from the CSA are plotted against the $10\,\%\,-\,90\,\%$\,rise times. The weighting potentials of the cathode (left) and an anode (right) are clearly visible along the $511\,\mathrm{keV}$ photopeak ($158\,\mathrm{mV}$ for the cathode and $182\,\mathrm{mV}$ for the anode at rise times of $410\,\mathrm{ns}$ and $200\,\mathrm{ns}$).}
\label{fig_drifttimes}
\end{figure}
This measurement shows the expected relationship of the DOI and the measured cathode and anode energy. The peak amplitude of the $511\,\mathrm{keV}$ photopeak of the cathode decreases linearly in correlation with the rise time. Events near the cathode cause the long rise times. The linear shape matches the expected weighting potential of the cathode. The pulse height of the anode signal also decreases dependent on the calculated rise time. The DOI is clearly visible along the predicted shape of the weighting potential, as shown in figure~\protect\ref{fig_weighting_potentials}. The expected peak amplitudes for the photopeak are in accordance with the calculated values in table~\protect\ref{tab_performance}. A summary of the measured key metrics for the CSA is given in table~\ref{tab_meas_params}.
\begin{table}
\centering
\caption{Measured parameters of the readout electronics. All channels are bandwidth-limited to about $21.654\,\mathrm{MHz}$ by a passive $RC$ low-pass filter. If appropriate, the standard deviation $\sigma$ is noted for the measured value.}
\label{tab_meas_params}
\begin{tabular}{|l|p{2.5cm}|l|p{5.0cm}|}
\hline
{Parameter}                                                             & {Measured value}                                               & {Channel}    & {Comment}\\
\hline
{Rise time $10\%-90\%$}                                                 & {$14.86\,\mathrm{ns}$,\newline $\sigma=371\,\mathrm{ps}$}      & {Test input} & {Input pulse: $3.0\,\mathrm{ns}$ fall time,\newline $V_\mathrm{out}: 2\,\mathrm{V_{pp}}$}\\
{Noise level (rms)}                                                     & {$1.24\,\mathrm{mV}$}                                          & {Test input} & {Sine-wave test}\\ 
{$\tau_\mathrm{1}$ ({$82\,\mathrm{M\Omega}$\,||\,$100\,\mathrm{fF}$})}  & {$8.091\,\mathrm{{\mu}s}$,\newline $\sigma=86\,\mathrm{ns}$}   & {Cathode}    & {Waveform fit: $f(t)=\mathrm{e}^{-t/\tau_1}$}\\
{$\tau_\mathrm{2}$ ({$47\,\mathrm{M\Omega}$\,||\,$100\,\mathrm{fF}$})}  & {$4.964\,\mathrm{{\mu}s}$,\newline $\sigma=181\,\mathrm{ns}$}  & {Anode}      & {Waveform fit: $f(t)=\mathrm{e}^{-t/\tau_2}$}\\
{Peak amplitude at $511\,\mathrm{keV}$}                                  & {$158\,\mathrm{mV}$}                                          & {Cathode}    & {Pulse height at longest drift time}\\
{Peak amplitude at $511\,\mathrm{keV}$}                                  & {$182\,\mathrm{mV}$}                                          & {Anode}      & {Pulse height at longest drift time}\\
{Steady-state gain}                                                     & {$8.95\,\frac{\mathrm{V}}{\mathrm{pC}}$}                       & {Cathode}    & {at $511\,\mathrm{keV}$ with $E_i=4.64\,\mathrm{eV}$}\\
{Steady-state gain}                                                     & {$10.31\,\frac{\mathrm{V}}{\mathrm{pC}}$}                      & {Anode}      & {at $511\,\mathrm{keV}$ with $E_i=4.64\,\mathrm{eV}$}\\
\hline
\end{tabular}
\end{table}

\begin{figure}[ht]
\centering
\includegraphics[width=0.49\textwidth]{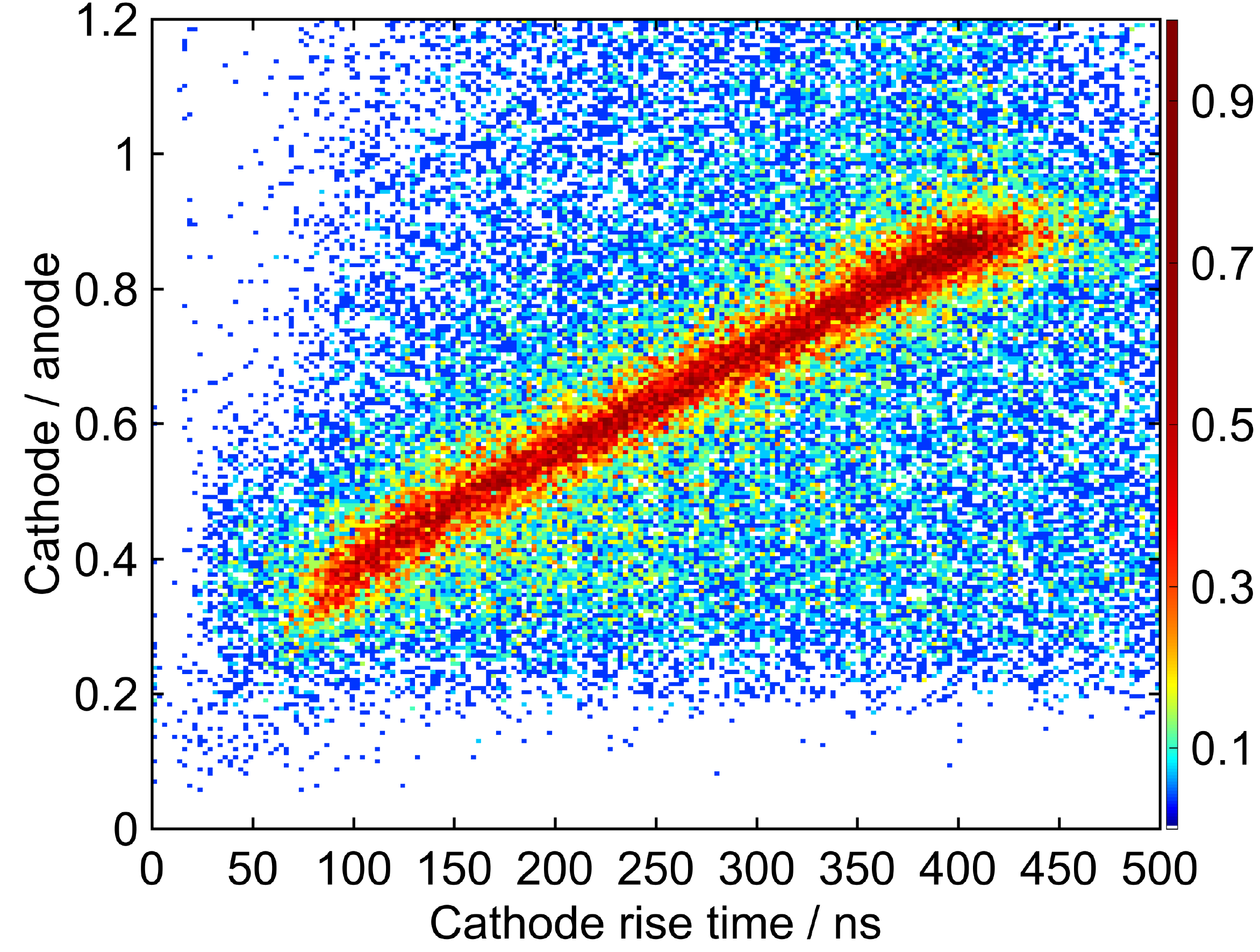}
\includegraphics[width=0.49\textwidth]{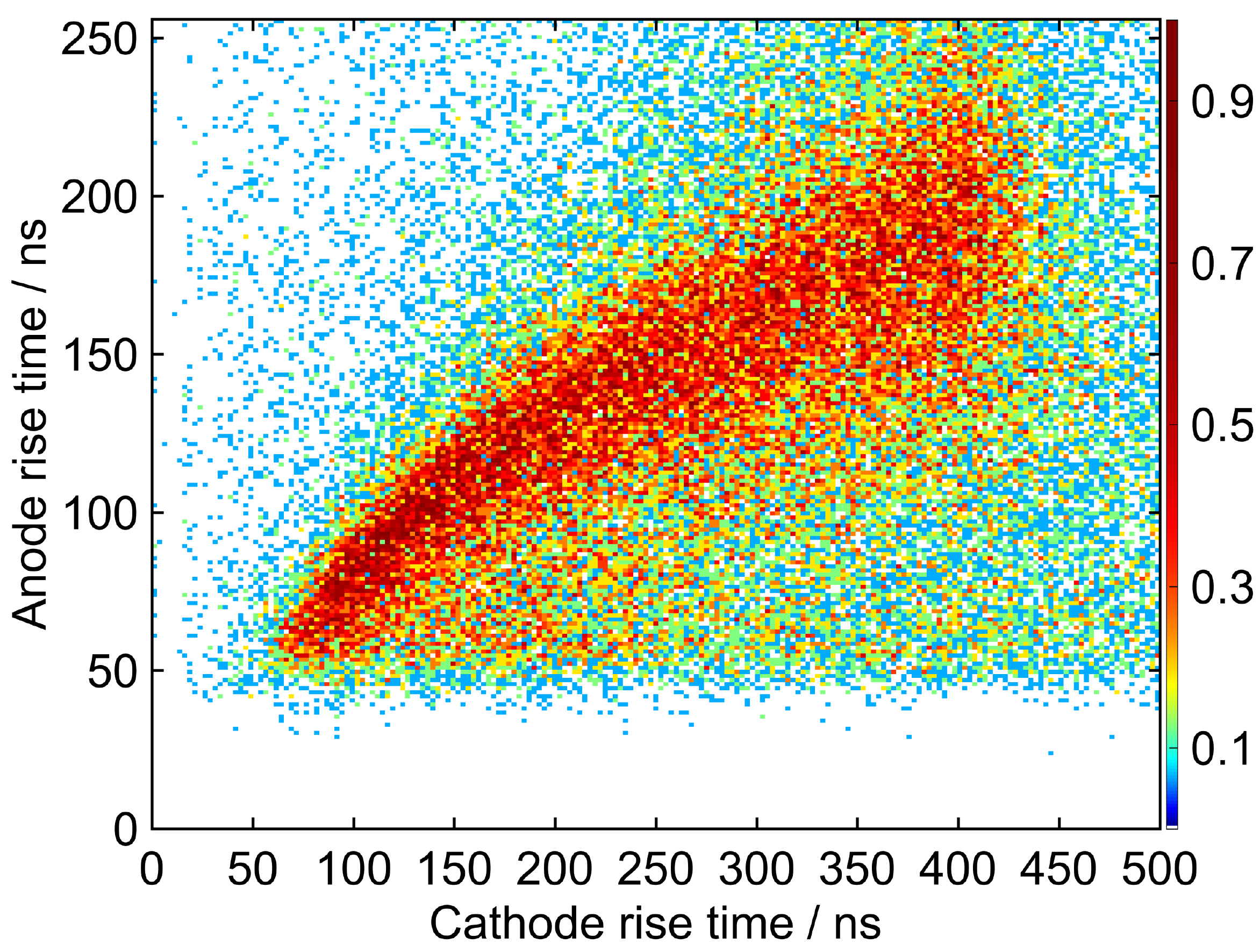}
\caption{Correlation of the cathode-over-anode ratio (CAR) with the measured rise times of the cathode pulses from the ${}^{22}\mathrm{Na}$ radioactive source measurement (left). The rise time has a strong correlation to the CAR and therefore to the depth of interaction (DOI) in the detector volume. The correlation of the anode rise time to the DOI is smeared out (right).}
\label{fig_drift_correlation}
\end{figure}

The correlation between the rise time of the cathode signal and the cathode-over-anode ratio is shown in figure~\protect\ref{fig_drift_correlation}. Both values carry information about the DOI. In contrast, the correlation between the cathode and anode rise time is smeared out. This is caused by uncertainties in the crossing of the low-level threshold, because of the slow rising component of the anode signal at the beginning. The low-level trigger is difficult to hit exactly without additional effort. 
However, the measurements show that the depth of interaction can be calculated by the rise time of the cathode signal or the ratio of cathode-over-anode energy. In conclusion, the measurement of the cathode rise time is more precise than the anode rise time, because the cathode signal rises with a nearly linear slope. An advantage of the rise time estimation is that the information is derived from one detector signal instead of two. Additionally, the calculation of cathode-over-anode ratio is error prone due to charge-sharing events.

\begin{figure}[ht]
\centering
\includegraphics[width=0.49\textwidth]{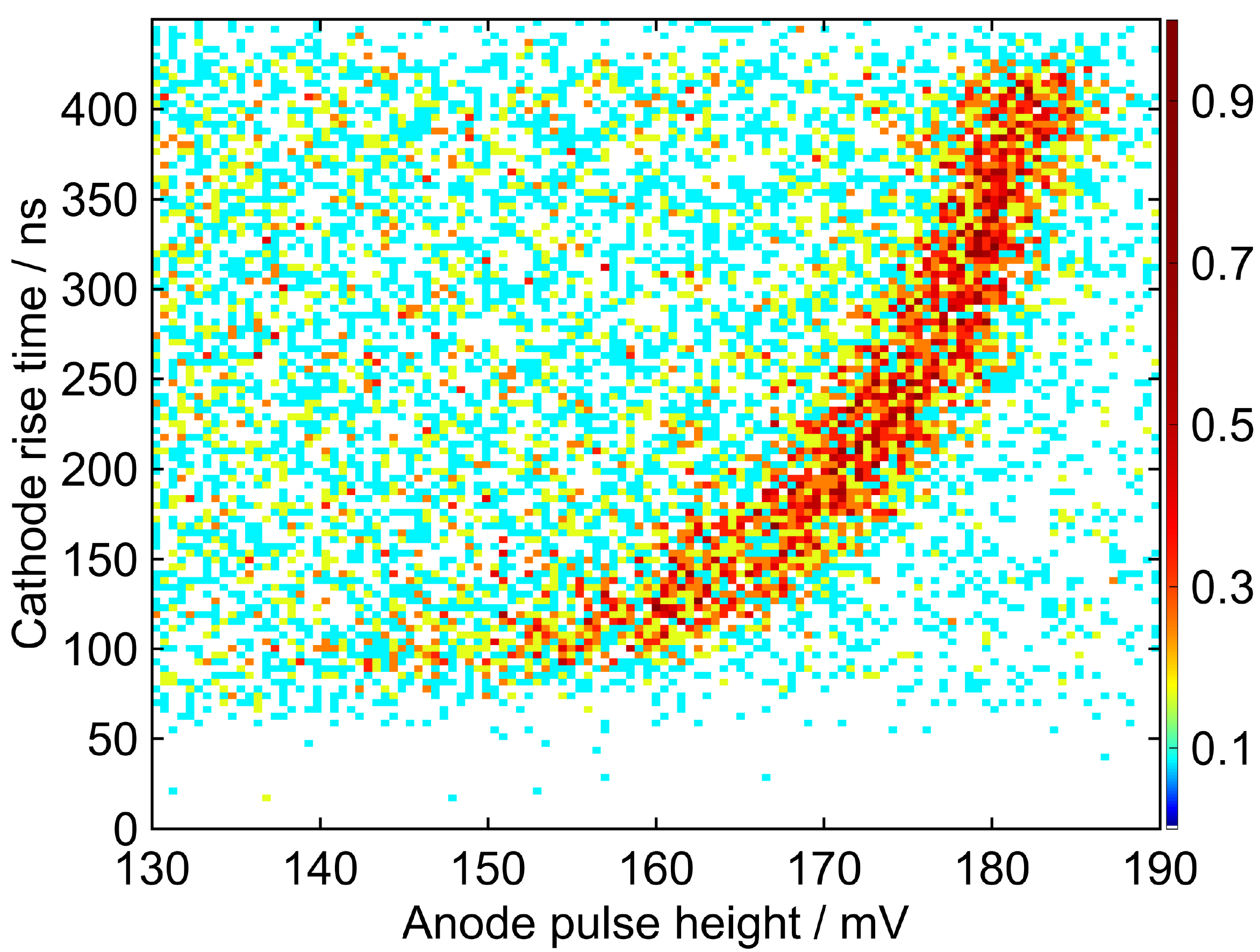}
\includegraphics[width=0.49\textwidth]{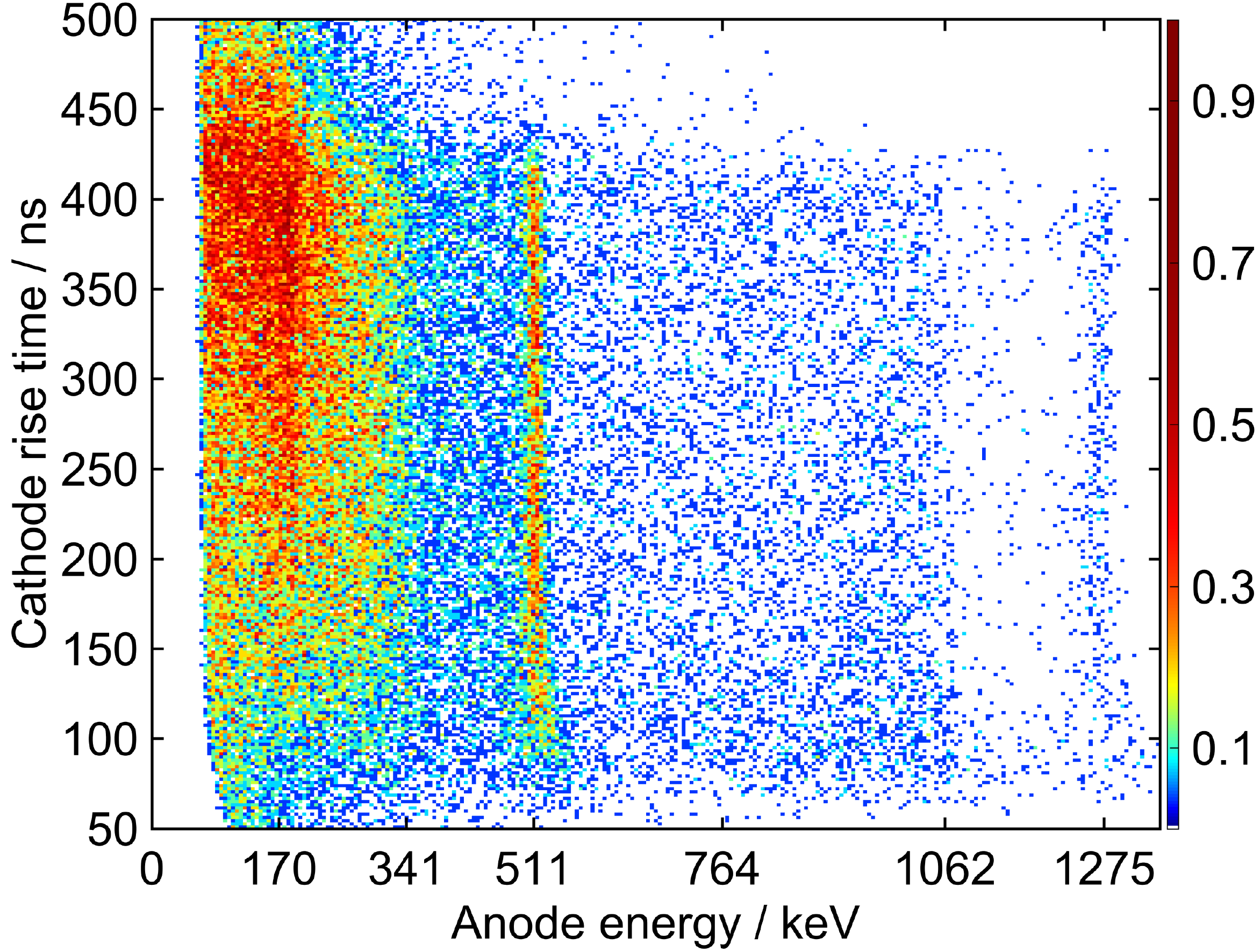}
\includegraphics[width=0.49\textwidth]{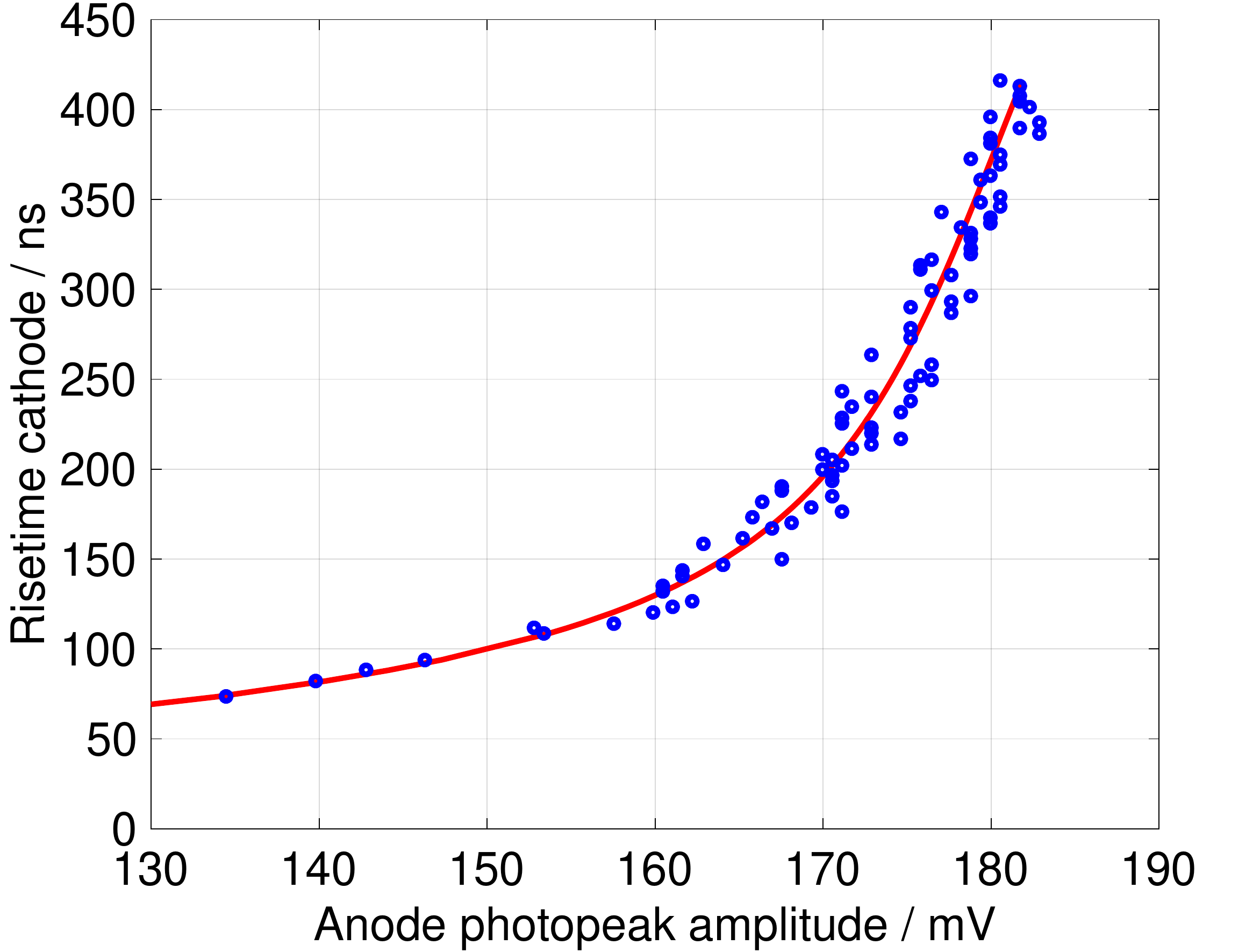}
\includegraphics[width=0.49\textwidth]{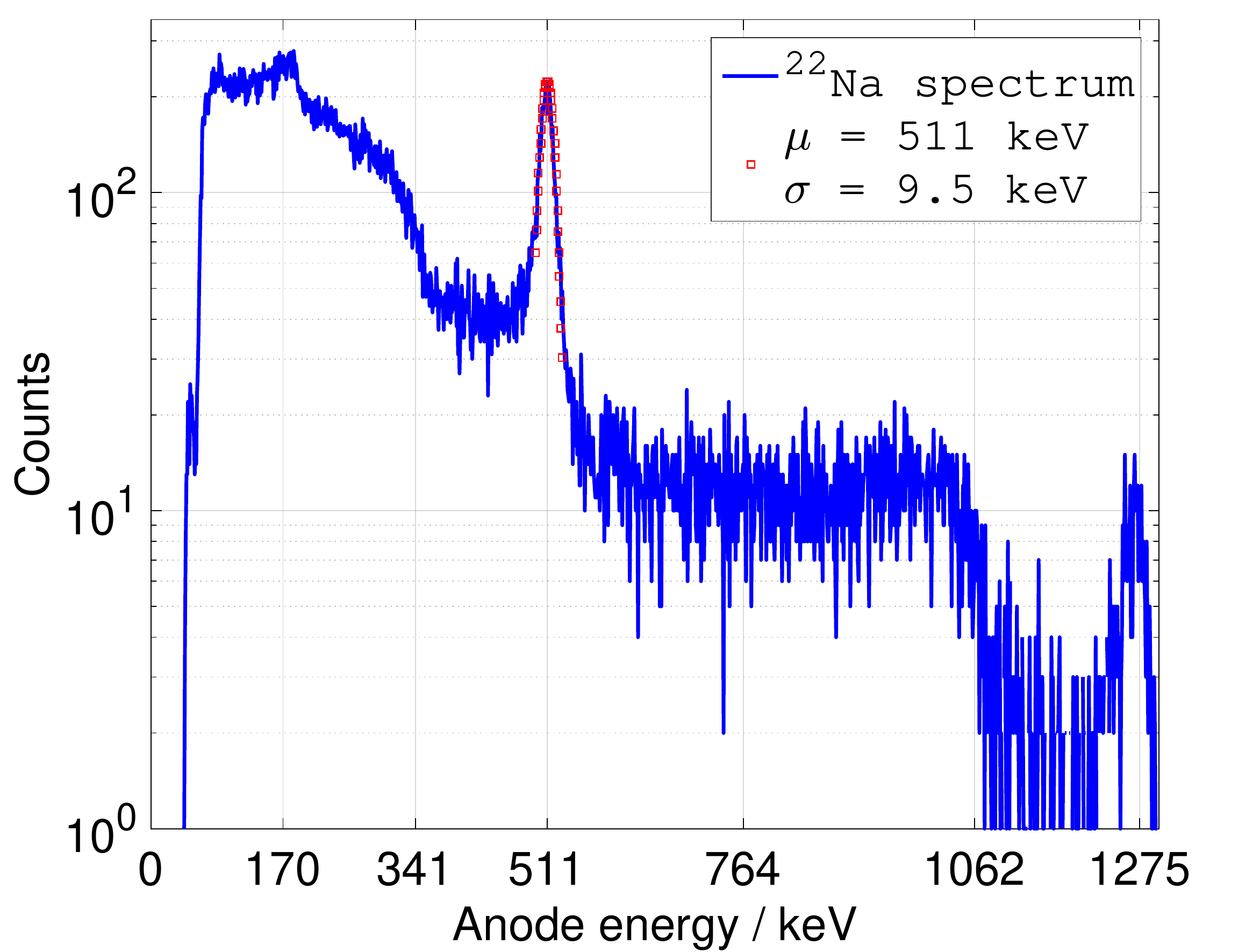}
\caption{The $511\,\mathrm{keV}$ photopeak along the signal rise time at the cathode with the corresponding data points used for the fit of the weighting potential (left). The corrected spectrum of the anode energy is independent of the depth of interaction (right). The selected pixel has an energy resolution of $9.5\,\mathrm{keV}$ (standard deviation) for the photopeak, which corresponds to $4.3\,\%\,\mathrm{FWHM}$. The spectrum is recorded without any additional pulse shapers.}
\label{fig_weightingfit}
\end{figure}

\begin{figure}[ht]
\centering
\includegraphics[width=0.75\textwidth]{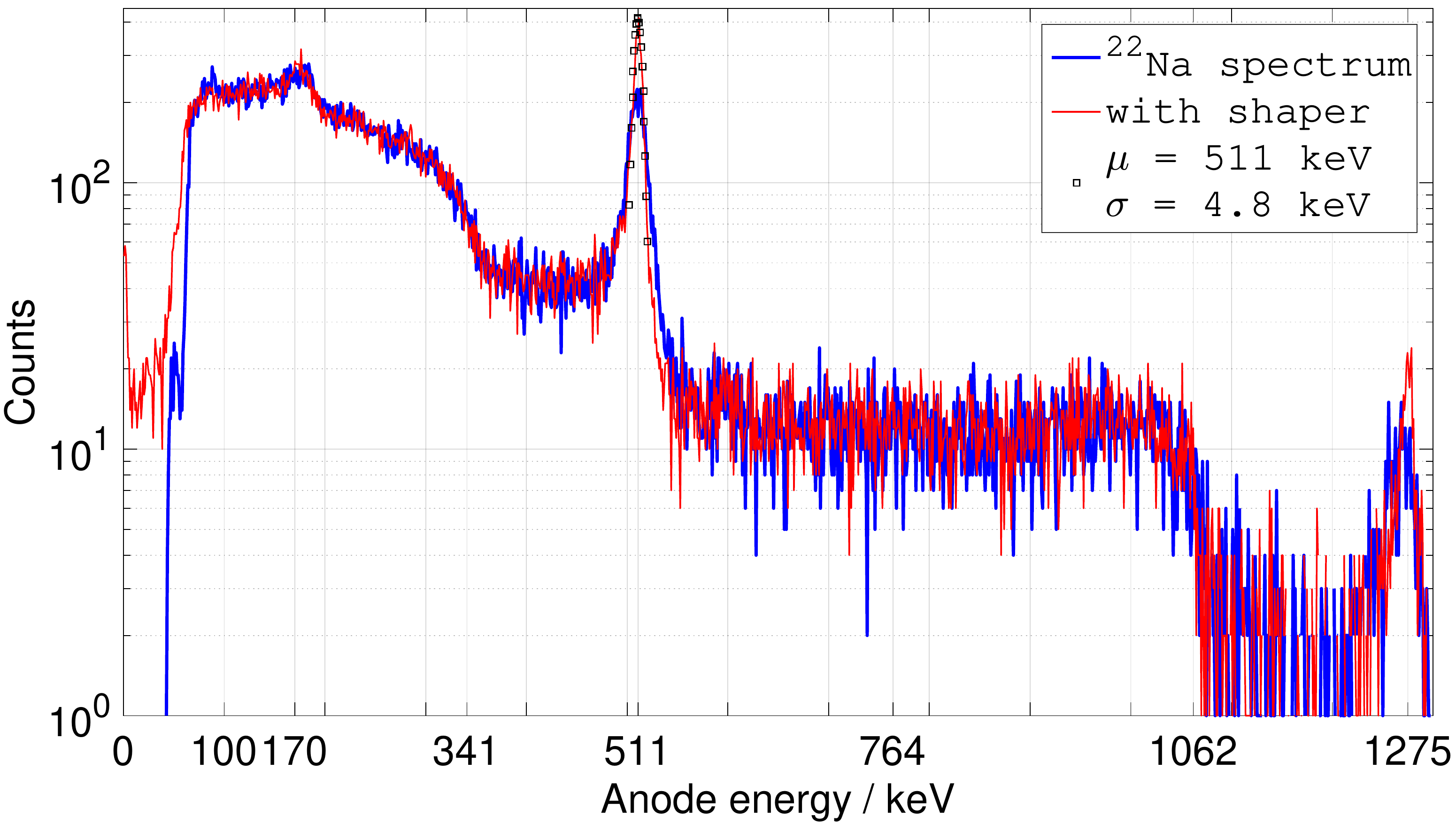}
\caption{The results of a measurement with our charge-sensitive amplifier and the Redlen CZT detector. The use of an additional digital pulse shaper improves the energy resolution of the $511\,\mathrm{keV}$ photopeak to $4.8\,\mathrm{keV}$ (standard deviation). This corresponds to $11\,\mathrm{keV}$ FWHM ($2.2\,\%$). All events of one pixel are shown.}
\label{fig_22naresult}
\end{figure}

It is evident that the anode energy has to be corrected dependent on the DOI. There are several approaches for the depth correction, which are all carried out empirically. The results are achieved with best-fit functions, based on e.g.\ polynomial~\protect\cite{donmez} or exponential~\protect\cite{hong},~\protect\cite{cho2} equations. Our approach is based on the weighting potential of a pixel, as this is the cause for the incorrect measurement. For depth correction, we select the events of the $511\,\mathrm{keV}$ photopeak along its cluster with reduced peak height and rise time, as shown in figure~\protect\ref{fig_weightingfit}.
A fit of the weighting potential according to eq.~\protect\ref{eqn_weightingpixel} matches the data points. Thus, the derived mathematical relationship is used to correct the anode energy, which is shown in figure~\protect\ref{fig_weightingfit}. We also evaluated polynomial equations for the fitting function, resulting in comparable results with a fourth-degree polynomial. The presented measurements were processes without any additional pulse shapers. This results in an energy resolution of $4.3\,\%$ (FWHM) for the $511\,\mathrm{keV}$ photopeak. With the use of additional digital pulse shapers~\cite{foedischnim}, the energy resolution could be improved to $2.2\,\%$ (see figure~\protect\ref{fig_22naresult}).

\section{Summary}
This paper has presented a circuit design and implementation of the analog front-end electronics for a cadmium zinc telluride (CdZnTe, CZT) pixel detector. Starting from the electrical equivalent circuit of the CZT, its electrical characteristics were discussed and a connection scheme presented. A short summary of the signal formation in a CZT pixel detector was provided, and the weighting potentials of the electrodes shown. Finally, these equations were applied for an analysis of the detector signals. After a comparison of different readout circuits, we focused our investigation on the charge-sensitive amplifier for the readout of the detector signals. As an ASIC-based solution is not available for the application with high gamma-ray energies and count rates, we designed the front-end electronics for the $8\,\times\,8$ pixel detector with commercial off-the-shelf operational amplifiers.

We have shown a detailed analysis of the charge-sensitive amplifier in conjunction with the electrical model of the CZT detector. The limits of the design in terms of gain, bandwidth, and noise were given with exact equations and numerical values. The performance of the readout electronics was measured with synthesized detector signals from a test pulser and with a pixel detector of size $20\,\times\,20\,\times\,5\,\mathrm{mm}^{3}$ from Redlen. The measurements with the test pulse showed that the rms~noise~level of $1.24\,\mathrm{mV}$ is below the nominal intrinsic resolution of about $8\,\mathrm{keV}$ (FWHM at $122\,\mathrm{keV}$) of the detector. Furthermore, we have shown that the signal-to-noise ratio is also sufficient for a timing far below $1\,\mathrm{ns}$, which outstrips the expected CZT performance. All the results have been achieved without an additional pulse shaper. We also investigated the performance of the charge-sensitive amplifiers with the detector from Redlen, and were able to verify that the claimed depth dependence is in accordance with the calculated weighting potential of a pixel. We presented a measurement of a ${}^{22}\mathrm{Na}$ radioactive source and showed a correction of the measured energy dependent on the depth of interaction. After depth correction, we obtain an energy resolution of about $22\,\mathrm{keV}$ ($4.3\,\%$ FWHM at $511\,\mathrm{keV}$). This result could be improved to $11\,\mathrm{keV}$ ($2.2\,\% \mathrm{FWHM}$) by the use of additional digital pulse shapers. Finally, the front-end electronics fulfill our requirements and operate from several $\mathrm{keV}$ up to $7\,\mathrm{MeV}$ with sub-nanosecond timing capabilities. The charge-sensitive front-end electronics were used in a multichannel digital signal processing system for a Compton camera prototype.

\end{document}